\documentstyle[aps,rmp,harvard,psfig,floats,twocolumn]{revtex}

                \def\F{{\bf F}}
 \def\G{{\bf G}}
                
\def\k{{\bf k}}
\def\q{{\bf q}}
\def\r{{\bf r}} \def\R{{\bf R}}
                \def\Z{{Z^\star}}
\def\U{{\bf u}}

\def \boldtau{\pmb{$\tau$}}

\def \SCF{{\scriptscriptstyle SCF}}
\def \KS{{\scriptscriptstyle KS}}
\def \NL{{\scriptscriptstyle NL}}
\def\ealla#1{{\rm e}^{#1}}

\def\eqref#1{(\ref{eq:#1})}
\def\secref#1{\ref{sec:#1}}

\def\E{\textsf{E}}
\def\EE{\textbf{\textsf{E}}}
\def\DD{\textbf{\textsf{D}}}
\def\P{\textsf{P}}
\def\PP{\textbf{\textsf{P}}}

\newdimen\dimensega

\def\Angstrom{\leavevmode\setbox0=\hbox{h}\dimensega=\ht0 
 \advance\dimensega by-1ex
 \rlap{\raise.67\dimensega\hbox{\char'27}}A}

\def\pmb#1{\setbox0=\hbox{#1}%
 \hbox{\kern-.025em\copy0\kern-\wd0
 \kern.05em\copy0\kern-\wd0
 \kern-0.025em\raise.0433em\box0} }

\def\psib{{\pmb{$\psi$}}}

\begin{document}

\title{Phonons and related properties of extended systems \\
 from density-functional perturbation theory}
\author{Stefano Baroni, Stefano de Gironcoli, Andrea Dal Corso}
\address{SISSA -- Scuola Internazionale Superiore di Studi Avanzati
 and \\ INFM -- Istituto Nazionale di Fisica della Materia \\
 via Beirut 2-4, 34014 Trieste, Italy}
\author{Paolo Giannozzi$^*$}
\address{Chemistry Department, Princeton University,
Princeton, New Jersey 08544}

\maketitle
\begin{abstract}
This article reviews the current status of lattice-dynamical
calculations in crystals, using density-functional perturbation
theory, with emphasis on the plane-wave pseudo-potential
method. Several specialized topics are treated, including the
implementation for metals, the calculation of the response to
macroscopic electric fields and their relevance to long wave-length
vibrations in polar materials, the response to strain deformations,
and higher-order responses. The success of this methodology is
demonstrated with a number of applications existing in the literature.
\end{abstract}
\narrowtext

\tableofcontents

\section{Introduction} \label{sec:intro}

The theory of lattice vibrations is one of the best established
chapters of modern solid state physics and very few of the astonishing
successes met by the latter could have been achieved without a strong
foundation of the former. A wide variety of physical properties of
solids depend on their lattice-dynamical behavior: infrared, Raman,
and neutron-diffraction spectra; specific heats, thermal expansion,
and heat conduction; phenomena related to the electron-phonon
interaction such as the resistivity of metals, super-conductivity, and
the temperature dependence of optical spectra, are just a few of
them. As a matter of fact, their understanding in terms of {\it
phonons} is considered to be one of the most convincing evidences that
our current quantum picture of solids is correct.

The basic theory of lattice vibrations dates back to the thirties and
the treatise of \citeasnoun{Born-Huang} is still considered today to
be the reference textbook in this field. These early formulations were
mainly concerned with establishing the general properties of the
dynamical matrices---such as e.g. their symmetry and/or analytical
properties---without even considering their connections with the
electronic properties which actually determine them. A systematic study
of these connections was not performed until the seventies
\cite{DeCicco,PCM}. The relationship between the electronic and the
lattice-dynamical properties of a system is important not only as a
matter of principle, but also because (and perhaps mainly because) it
is only exploiting these relations that it is possible to calculate the
lattice-dynamical properties of specific systems. 

The state of the art
of theoretical condensed-matter physics and of computational materials
science is such that it is nowadays possible to calculate specific
properties of specific (simple) materials using {\it ab initio}
quantum mechanical techniques whose only input information is the
chemical composition of the materials. In the specific case of
lattice-dynamical properties, a large number of {\it ab initio}
calculations based on the linear-response theory of lattice vibrations
\cite{DeCicco,PCM} have been made possible over the past 10 years by
the achievements of density-functional theory \cite{HK,KS}, and
by the development of density-functional perturbation theory
\cite{Zein1,BGT} which is a method to apply the former within the general
theoretical framework provided by the latter. Thanks to these
theoretical and algorithmic advances, it is now possible to obtain
accurate phonon dispersions on a fine grid of wave-vectors covering
the entire Brillouin zone, which compare directly with neutron
diffraction data, and from which several physical properties of the
system (such as heat capacities, thermal expansion coefficients,
temperature dependence of the band gap, and so on) can be calculated.

The purpose of the present paper is to illustrate in some detail the
theoretical framework of density-functional perturbation theory,
including several technical details which are useful for its
implementation within the plane-wave pseudo-potential scheme. We will
also provide a representative, though necessarily incomplete, choice
of significant applications to the physics of insulators and metals,
including their surfaces, alloys, and micro-structures.
\section{Density-Functional Perturbation Theory} \label{sec:dfpt}
\subsection{Lattice dynamics from electronic-structure theory}
\label{sec:ld-from-est}

The basic approximation which allows one to decouple the vibrational
from the electronic degrees of freedom in a solid is the {\it adiabatic
approximation} by Born and Oppenheimer (BO) \cite{Born-Oppenheimer}.
Within this approximation, the lattice-dynamical properties of a system
are determined by the eigenvalues ${\cal E}$ and eigenfunctions 
$\Phi$ of the
Schr\"odinger equation: \begin{equation} \left ( -\sum_I {\hbar^2
\over 2M_I} {\partial^2 \over \partial \R_I^2} + E(\R) \right )
\Phi(\R) = {\cal E} \Phi(\R), \label{eq:nucl-schr}
\end{equation} where $\R_I$ is the coordinate of the $I$-th nucleus,
$M_I$ its mass, $\R\equiv \{\R_I\}$ indicates the set of all the nuclear
coordinates, and $E(\R)$ is the clamped-ion energy of the system,
which is often referred to as the {\it Born-Oppenheimer energy surface}. In
practice, $E(\R)$ is the ground-state energy of a system of
interacting electrons moving in the field of fixed nuclei, whose
Hamiltonian---which acts onto the electronic variables and
depends parametrically upon $\R$---reads: \refstepcounter{equation} 
 \label{eq:el-hamilt} $$\displaylines{\quad 
H_{BO}(\R) = -{\hbar^2 \over 2m} \sum_i
{\partial^2 \over \partial \r_i^2} + {e^2 \over 2} \sum_{i\ne j}
{1\over |\r_i - \r_j|} \hfill \cr \hfill - 
\sum_{iI} {Z_I e^2 \over |\r_i - \R_I|} + E_N(\R), \quad
(\theequation) } $$
where $Z_I$ is the charge of the $I$-th nucleus, $-e$ is the electron charge,
and
 $E_N(\R)$ is the electrostatic interaction between
different nuclei:
\begin{equation} 
\label{eq:ion-ion}
E_N(\R)= {e^2 \over 2} \sum_{I\ne J}
{Z_I Z_J \over |\R_I - \R_J | }.
\end{equation}
The equilibrium geometry of the system is given by the condition that
the forces acting on individual nuclei vanish: \begin{equation} \F_I
\equiv -{\partial E(\R) \over \partial \R_I} = 0, \end{equation}
whereas the vibrational frequencies, $\omega$, are determined by the
eigenvalues of the Hessian of the BO energy, scaled by the nuclear
masses: \begin{equation} {\rm det} \left | {1\over \sqrt{M_I M_J}} 
{\partial^2 E(\R) \over \partial \R_I \partial \R_J} -\omega^2 
\right | = 0. \label{eq:secular-frequency} \end{equation}

The calculation of the equilibrium geometry and of the vibrational
properties of a system thus amounts to computing the first
and second derivatives of its BO energy surface. The
basic tool to accomplish this goal is the Hellmann-Feynman (HF)
theorem \cite{Hellmann,Feynman} which states that the first derivative
of the eigenvalues of a Hamiltonian that depends on a parameter
$\lambda$, $H_\lambda$, is given by the expectation value of the
derivative of the Hamiltonian: 
\begin{equation}
{\partial E_\lambda \over \partial \lambda}
= \left \langle \Psi_\lambda \left | {\partial H_\lambda \over
\partial \lambda} \right | \Psi_\lambda \right \rangle, 
\end{equation} where $\Psi_\lambda$ is the eigenfunction of $H_\lambda$
corresponding to the $E_\lambda$ eigenvalue: $H_\lambda \Psi_\lambda =
E_\lambda \Psi_\lambda$. In the BO
approximation, nuclear coordinates act as parameters in the
electronic Hamiltonian, Eq.\ \eqref{el-hamilt}. The
force acting on the $I$-th nucleus in the electronic ground state is
thus: 
\begin{equation}
\F_I = -{\partial E(\R)
\over \partial \R_I} =
- \left \langle \Psi(\R) \left | {\partial
H_{BO}(\R) \over \partial \R_I } \right | \Psi(\R) \right \rangle,
\end{equation}
where $\Psi(\r,\R) $ is the electronic ground-state wave-function of
the BO Hamiltonian. The BO Hamiltonian depends on $\R$ via the
electron-ion interaction that couples to the electronic degrees of
freedom only through the electron charge density. The HF theorem
states in this case that: \begin{equation} \F_I = - \int n_\R(\r)
{\partial V_\R(\r) \over \partial \R_I} d\r - {\partial E_N(\R)
\over \partial \R_I}, \label{eq:HF-force} \end{equation} where $
V_\R(\r)$ is the 
electron-nucleus interaction,
\begin{equation} V_\R(\r) = - \sum_{iI} {Z_I e^2 \over |\r_i - \R_I|},
\end{equation} and $n_\R(\r)$ is the ground-state
electron charge density corresponding to the nuclear configuration
$\R$. The Hessian of the BO energy surface appearing in
Eq.\ \eqref{secular-frequency} is obtained by 
differentiating the HF forces with respect to nuclear coordinates:
\refstepcounter{equation} \label{eq:hessian} $$\displaylines{
\quad
{\partial^2 E(\R) \over \partial \R_I
\partial \R_J} 
\equiv - {\partial \F_I \over \partial \R_J} = 
 \int {\partial n_\R(\r) \over \partial \R_J}
{\partial V_\R(\r) \over \partial \R_I} d\r
 \hfill \cr \hfill
+ \int n_\R(\r)
{\partial^2 V_\R(\r) \over \partial \R_I
\partial \R_J} d\r + {\partial^2 E_N(\R) \over \partial \R_I
\partial \R_J}. \quad (\theequation) } $$
Eq.\ \eqref{hessian} states that the calculation  of the
Hessian of the BO energy surfaces requires the calculation
of the ground-state electron charge density, $n_\R(\r)$, as well as of
its 
{\it linear response} to a distortion of the nuclear geometry, 
${\partial n_\R(\r) / \partial \R_I} $. This fundamental result was
first stated in the late sixties by 
\citeasnoun{DeCicco} and \citeasnoun{PCM}.
The Hessian
matrix is 
usually called the matrix of the {\it inter-atomic force constants} (IFC).
\subsection{Density Functional Theory} \label{sec:dft}

According to the preceding discussion, the calculation
of the {\it derivatives} of the BO energy surface with respect to the
nuclear coordinates only requires the knowledge of the electronic
charge-density distribution. This fact is a special case of a much
more general property of the systems of interacting electrons, known
as the \citeasnoun{HK} theorem. According to this
theorem, no two different potentials acting on the electrons of a
given system can give rise to a same ground-state electronic charge
density. This property can be used in conjunction with the standard
Rayleigh-Ritz variational principle of quantum mechanics to show that
a {\it universal} functional,\footnote{By {\it universal} it is meant
here that the functional is independent of the external potential 
acting on the
electrons, though it obviously depends on the form of the
electron-electron interaction.} $F[n(\r)]$, of the electron charge density
exists such that the functional: \begin{equation} E[n(\r)] = F[n(\r)]
+ \int n(\r) V(\r) d\r \label{eq:hk-energy} \end{equation} is 
minimized by the electron
charge density of the ground state corresponding to the external
potential $V(\r)$, under the constraint that the integral of $n(\r)$
equals the total number of electrons. Furthermore, the value of the
minimum coincides with 
the ground-state energy. This theorem provides the foundation of what
is currently known as {\it density-functional theory} (DFT)
\cite{DFT1,DFT2}. It allows an enormous conceptual
simplification of the quantum-mechanical problem of the search of the
ground-state properties of a system of interacting electrons, for it
replaces the traditional description based on wave-functions (which
depend on $3N$ independent variables, $N$ being the number of
electrons) with a much more tractable description in terms of the
charge density, which only depends on 3 variables. Two major problems
hamper a straightforward application of this remarkably simple result:
the fact that the form of the $F$ functional is unknown, and that the
conditions to be fulfilled for a function $n(\r)$ to be considered an
acceptable ground-state charge distribution (and hence the domain of
the functional $F$)
are poorly characterized. The second problem is hardly addressed at
all, and one usually contents oneself to impose the proper
normalization of 
the charge density by the use of a Lagrange multiplier. The first
problem can be handled by mapping the system onto an auxiliary system
of non interacting electrons \cite{KS} and by making appropriate
approximations along the lines described in the next subsection.

\subsubsection{The Kohn-Sham equations} \label{sec:Kohn-Sham}

The Hohenberg and Kohn theorem states that all the physical 
properties of a system of
interacting electrons are uniquely determined by its ground-state
charge-density distribution. This property holds independently of the
precise form of the electron-electron interaction. In particular,
when the strength of the electron-electron interaction vanishes,
$F[n(\r)]$ defines the
ground-state kinetic energy of a system of non-interacting electrons
as a functional of its
ground-state charge-density
distribution, $T_0[n(\r)]$.
This
fact has been used by \citeasnoun{KS}
to map the problem of a system of interacting electrons onto an
equivalent non-interacting problem. To this end, the unknown
functional $F[n(\r)]$ is cast into the form: \refstepcounter{equation} 
\label{eq:KS-funct} 
$$
\displaylines{\quad
F[n(\r)] = T_0[n(\r)] + {e^2\over 2} \int {n(\r)n(\r') \over |\r-\r'|}
d\r d\r' \hfill \cr \hfill + E_{xc}[n(\r)], \quad (\theequation) }
$$ where the
second term is the classical electrostatic self-interaction of the
electron charge-density distribution, and the so-called {\it
exchange-correlation energy}, $E_{xc}$, is {\it defined} by
Eq.\ \eqref{KS-funct}.\footnote{The exchange-correlation energy is
the name we give to the part of the energy functional that we do not
know how to calculate otherwise. For this reason, it has been named
{\it stupidity energy} by \citeasnoun{Feynman-StatPhys}. Whether or
not this is an useful concept depends on its magnitude with respect to
the total functional and on the quality of the approximations one can
find for it.} Variation of the energy functional with respect to
$n(\r)$ with the constraint that the number of electrons is kept fixed
leads formally to the same equation that would hold for a system
of non-interacting electrons subject to an effective potential (also
called {\it self-consistent}, SCF, potential) whose form is:
\begin{equation} V_{\SCF}(\r) = V(\r) + e^2 \int {n(\r')\over|\r-\r'|}
d\r' + v_{xc}(\r), \label{eq:Vscf} \end{equation} where \begin{equation}
v_{xc}(\r) \equiv {\delta E_{xc}\over \delta n(\r)} \label{eq:Vxc}
\end{equation} is the functional derivative of the
exchange-correlation energy, also called the {\it
exchange-correlation potential}.

The power of this trick is that, if one knew the
effective potential $V_{\SCF}(\r)$, the problem for non-interacting
electrons could be trivially solved without knowing the form of the
non-interacting kinetic-energy functional, $T_0$. To this end, one
should simply solve the one-electron Schr\"odinger equation:
\begin{equation} \left ( -{\hbar^2 \over 2m} {\partial^2 \over
\partial \r^2} + 
V_{\SCF}(\r) \right ) \psi_n(\r) = \epsilon_n
\psi_n(\r). \label{eq:ks-eq} \end{equation} The ground-state
charge-density distribution and non-interacting kinetic energy
functional would then be given in terms of the {\it auxiliary}
Kohn-Sham (KS) orbitals, $\psi_n(\r)$: \begin{eqnarray} 
n(\r)&=&  2 \sum_{n=1}^{N/2} |\psi_n(\r)|^2 \label{eq:n_of_r} \\
T_0[n(\r)] &=&  - 2 {\hbar^2\over 2m} \sum_{n=1}^{N/2} 
\int \psi^*_n(\r){\partial^2 \psi_n(\r) \over \partial \r^2} d\r ,
\label{eq:T0} \end{eqnarray} where $N$ is the number of electrons,
and the system is supposed to be non-magnetic,
so that each one of the $N/2$ lowest-lying orbital states accommodates
2 electrons of opposite spin. In periodic systems the index $n$
running over occupied states can be thought as a double label,
$n\equiv \{v,\k\}$, where $v$ indicates the set of valence bands, and
$\k$ is a wave-vector belonging to the first Brillouin zone (BZ).

The ground-state energy
given by Eqs.\ (\ref{eq:hk-energy},\ref{eq:KS-funct}) can be equivalently
expressed in terms of the KS eigenvalues: \refstepcounter{equation} 
\label{eq:toten-eigv}
$$\displaylines{\quad
E[n(\r)] = 2 \sum_{n=1}^{N/2}\epsilon_n - {e^2\over 2} \int
  {n(\r)n(\r')\over|\r-\r'|}d\r d\r' \hfill \cr \hfill
  + E_{xc}[n(\r)] - \int n(\r) v_{xc}(\r) d\r. \quad(\theequation) }
$$

Eq.\ \eqref{ks-eq} has the form of a non-linear Schr\"odinger equation
whose potential depends on its own eigenfunctions through the electron
charge-density distribution. Once an explicit form for the
exchange-correlation energy is available, this equation can be solved in a
self-consistent way using a variety of methods.

\subsubsection{Local-Density Approximation and beyond} \label{sec:LDA}

The Kohn-Sham scheme constitutes a practical way to implement DFT,
provided an accurate and reasonably easy-to-use approximation is
available for
the exchange-correlation energy,
$E_{xc}[n(\r)]$. In their original paper, \citeasnoun{KS} proposed to
assume that each small volume of the system (so small that the charge
density can be thought to be constant therein) contributes the same
exchange-correlation energy as an equal volume of a homogeneous
electron gas at the same density. With this
assumption, the exchange-correlation energy functional and potential
read: \begin{eqnarray}
E_{xc}[n(\r)] &=& \int \epsilon_{xc}(n(\r))n(\r)d\r, \\
v_{xc}(n(\r)) &=& 
\left(\epsilon_{xc}(n)+n{d\epsilon_{xc}(n)\over dn}\right)_{n=n(\r)},
\label{eq:LDA}
\end{eqnarray}
where $\epsilon_{xc}(n)$ is the exchange-correlation energy per
particle in an homogeneous electron gas at density $n$. This
approximation is known as the {\it local density approximation}
(LDA).
Approximate forms for $\epsilon_{xc}(n)$ have been known for a long time.
Numerical results from nearly exact Monte-Carlo calculations for the
homogeneous electron gas
by \citeasnoun{CA} have been parameterized 
by \citeasnoun{PZ} with a simple analytical form.
More accurate parameterizations have been recently
proposed by \citeasnoun{OrtizLDA}.
All these different forms are very similar in the range of electron
densities relevant to condensed-matter applications and yield
very similar results.

The LDA is {\it exact} in the limit of high density or of a slowly
varying charge-density distribution \cite{KS}.
LDA has turned out to be much more successful than originally
expected (see for instance \citeasnoun{DFTgeneral}), in spite of its
extreme simplicity. For weakly correlated materials, such
as semiconductors and simple metals, LDA accurately describes
structural and vibrational properties: the correct structure 
is usually found to have the lowest energy, while bond lengths, 
bulk moduli, and phonon frequencies are accurate within a few percent.

LDA also has some well-known drawbacks. 
A large overestimate ($\sim 20$\%) of the crystal cohesive and molecular
binding energies is possibly the worst failure of this approximation,
together with 
its inability to properly describe strongly correlated systems, such
as {\em e.g.} transition-metal oxides. Much effort has been put in 
the search for better functionals than LDA (see for instance
\citeasnoun{MetaGGA}). The use of {\em gradient corrections} 
\cite{Becke,PBE} to LDA has become widespread in recent years.
Gradient corrections are generally found to 
improve the account of electron correlations in finite or
semi-infinite systems, such as molecules or surfaces;
less so in infinite solids. 

In general, DFT is a ground-state theory and KS eigenvalues and 
eigenvector do not have a well defined physical meaning. 
Nevertheless, in the lack of better and equally general methods,
 KS eigenvalues are often used to estimate excitation
energies. The general features of the low-lying energy bands in solids
obtained in this way are generally considered to be at least
qualitatively correct, in spite of the fact that the LDA is known to
substantially underestimate the optical gaps in insulators.
\subsection{Linear response} \label{sec:linear-response}

In Sec. \secref{ld-from-est} we have seen that the electron-density
linear response of a system determines the matrix of its IFC's,
Eq.\ \eqref{hessian}. Let us see now how this response 
can be obtained within DFT. The procedure described in the following
is usually referred to as {\em density-functional perturbation theory}
(DFPT) \cite{Zein1,BGT,Gonze2}. 

In order to simplify the notation and make
the argument more general, we assume that the external potential
acting on the electrons is a differentiable function of a set of
parameters, ${\boldmath{\lambda}} \equiv \{\lambda_i \}$ ($\lambda_i
\equiv \R_I$ in the case of lattice dynamics). According to the HF
theorem, the first and second derivatives of the ground-state energy
read: \begin{eqnarray} 
{\partial E\over \partial \lambda_i} &=& \int
{\partial V_{\boldmath \lambda}(\r) \over\partial\lambda_i} n_
{ \boldmath \lambda}(\r)d\r, \label{eq:dE1} \\ {\partial^2 E\over
\partial\lambda_i \partial\lambda_j} & = & \int {\partial^2 
V_{\boldmath \lambda}(\r) \over \partial \lambda_i
\partial\lambda_j}n_{\boldmath \lambda}(\r)d\r \nonumber \\ &&
\quad \quad \quad \mbox{} + \int {\partial n_{\boldmath \lambda}(\r)
\over\partial\lambda_i} {\partial V_{\boldmath \lambda}(\r) \over
\partial \lambda_j} d\r.\label{eq:dE2}
\end{eqnarray}
The electron-density response, $ {\partial n_{\boldmath \lambda}(\r)
/ \partial\lambda_i} $, appearing in Eq.\ \eqref{dE2} can be
evaluated by linearizing Eqs.\ \eqref{n_of_r}, \eqref{ks-eq}, and
\eqref{Vscf} with respect to wave-function, density, and potential 
variations. Linearization of Eq.\ \eqref{n_of_r} leads to: 
\begin{equation} \Delta n(\r) = 4~ {\rm Re} \sum_{n=1}^{N/2}
\psi_n^*(\r) 
\Delta \psi_n(\r) , \label{eq:dn} \end{equation} 
where the finite-difference operator $\Delta^\lambda$ is defined as:
\begin{equation} \Delta^\lambda F = \sum_i {\partial
F_\lambda \over \partial \lambda_i} \Delta \lambda_i.
\label{eq:Delta_lambda} \end{equation}  
The
super-script $\lambda$ has been omitted in Eq.\ \eqref{dn}, as well as
in any subsequent 
formulas where such an omission does not give rise to ambiguities.
Since the external
potential (both unperturbed and perturbed) is real, each KS
eigenfunction and its complex conjugate are degenerate. As a
consequence, the imaginary part of the sum appearing in Eq.\ \eqref{dn}
vanishes, so that the prescription to keep only the real part can be
dropped.

The variation of the KS orbitals, $\Delta \psi_n(\r)$, is
obtained by standard first-order perturbation theory \cite{messiah}:
\begin{equation} \label{eq:linear} (H_{\SCF}-\epsilon_n ) |
\Delta\psi_n \rangle  = - (\Delta V_{\SCF} - 
\Delta \epsilon_n )| \psi_n
\rangle, \end{equation} where 
\begin{equation}
H_{\SCF}=  -{\hbar^2 \over 2 m} {\partial^2 \over \partial \r^2} +
V_{\SCF}(\r) \end{equation}
is the unperturbed KS Hamiltonian,
\refstepcounter{equation} \label{eq:dVscf} $$\displaylines{\quad
\Delta V_{\SCF}(\r) = \Delta V(\r) + e^2 
\int {\Delta n(\r') \over
|\r-\r'|} d\r' \hfill \cr \hfill + \left . {d v_{xc}(n) \over d n}
\right |_{n=n(\r)} \Delta n(\r), \quad (\theequation) } $$
is the first-order correction to the self-consistent potential, and
$\Delta \epsilon_n = \langle \psi_n | \Delta V_{\SCF} | 
\psi_n
\rangle $ is the first order variation of the KS eigenvalue,
$\epsilon_n$. 

In the atomic physics literature, an equation
analogous to Eq.\ 
\eqref{linear} is known as the Sternheimer equation, after the work in
which it was first used to calculate atomic polarizabilities
\cite{Sternheimer}. A self-consistent version of the Sternheimer
equation was introduced by \citeasnoun{Mahan} to calculate atomic
polarizabilities within DFT in the LDA. Similar methods are known in
the quantum chemistry literature under the generic name of {\em
analytic evaluation of second-order energy derivatives}
\cite{CHF,Amos}. In the 
specific context of the Hartree-Fock approximation, the resulting
algorithm is called the {\em coupled Hartree-Fock} method \cite{CHF}.

Equations (\ref{eq:dn}--\ref{eq:dVscf}) form a set of self-consistent
equations for the perturbed system completely analogous to the KS
equations in the unperturbed case---Eqs.\ \eqref{Vscf}, \eqref{ks-eq},
and \eqref{n_of_r}---with the KS eigenvalue equation, Eq.\ \eqref{ks-eq},
being replaced by the solution of a linear system, Eq.\
\eqref{linear}.
In the present case, the self-consistency requirement manifests itself
in the dependence of the right-hand side upon the solution of the
linear system. As $\Delta V_{\SCF}(\r)$ is a linear functional of
$\Delta n(\r)$ which in turn depends linearly on the $\Delta \psi$'s,
the whole self-consistent calculation can be cast in terms of a
generalized linear problem. Note however that the right-hand side of
Eq.\ \eqref{linear} for $\Delta\psi_n$ depends through $\Delta n$
on the solution of all
the similar equations holding for the $\Delta\psi_m$ ($m\ne
n$). Hence, all the $N$ equations, Eq.\ \eqref{linear}, are linearly
coupled to each other, and the set of all the $\Delta\psi$'s is the
solution of a linear problem whose dimension is ($NM/2 \times NM/2$), $M$
being the size of the basis set used to describe the
$\psi$'s. The explicit form of this big linear equation can be worked
out directly from Eqs.\ (\ref{eq:dn}--\ref{eq:dVscf}), or it can be
equivalently derived from a variational principle, as explained in
Sec. \secref{variational}. Whether this big linear system is better
solved directly by iterative methods or by the self-consistent
solution of the smaller linear systems given by Eq.\ \eqref{linear} 
is a matter of computational strategy.

The first-order correction to a given eigen-function of the
Schr\"odinger equation, given by Eq.\ \eqref{linear}, is often
expressed in terms of a sum over the spectrum of the unperturbed
Hamiltonian, 
\begin{equation} \Delta \psi_n(\r) = \sum_{m\ne n} \psi_m(\r) {\langle
\psi_m | \Delta V_{\SCF} | \psi_n \rangle \over \epsilon_n-\epsilon_m}, 
\label{eq:pert_th} \end{equation} running over all
the states of the system, occupied and empty, with the exception of
the state being considered, for which the energy denominator would
vanish. Using Eq.\ \eqref{pert_th}, the electron charge-density
response, Eq.\ \eqref{dn}, can be cast into the form: \begin{equation}
\Delta n(\r) = 4 \sum_{n=1}^{N/2} \sum_{m\ne n}\psi_n^*(\r) 
\psi_m(\r) {\langle \psi_m | \Delta V_{\SCF} | \psi_n \rangle 
\over \epsilon_n-\epsilon_m}. \label{eq:dn-2} \end{equation}
Eq.\ \eqref{dn-2} shows that the contributions to the electron-density
response coming from products of occupied states cancel each other, so
that the $m$ index can be thought as running onto conduction states
only. This is equivalent to say that the electron-density distribution
does not respond to a perturbation which only acts onto the
occupied-state manifold (or, more generally, to the component of any
perturbation which couples occupied states among each other).

The explicit evaluation of $\Delta \psi_n(\r)$ from
Eq.\ \eqref{pert_th} would require the knowledge of the full spectrum
of the KS Hamiltonian and extensive summations over conduction bands.
In Eq.\ \eqref{linear}, instead, only the knowledge of the occupied
states of the system is needed to construct the right-hand side of the
equation, and efficient iterative algorithms---such as conjugate
gradient 
\cite{NumRec,CG,Payne} 
or minimal residual \cite{NumRec,GMRES}
methods---can be used 
for the solution of the linear system. In this way the 
computational cost of the
determination of the density response to a single
perturbation is
of the same order as that needed
for the calculation of the unperturbed ground-state density.

The left-hand side of Eq.\ \eqref{linear} is singular because the
linear operator appearing therein has a null eigenvalue. However, we
saw above that the response of the system to an external perturbation
only depends on the component of the perturbation which
couples the occupied-state manifold with the empty-state one. The
projection onto the empty-state manifold of the first-order correction
to occupied orbitals can be obtained from Eq.\ \eqref{linear} by
replacing its right-hand side with $-P_c \Delta V_{\SCF}
|\psi_n \rangle $, where $P_c$ is the projector onto the empty-state
manifold, and by adding to the linear operator on its
left-hand side, $H_{\SCF} - \epsilon_n$, a multiple of the projector
onto the occupied-state manifold, $P_v$, so as to make it
non-singular:
\begin{equation} ( H_{\SCF} + \alpha P_v - \epsilon_n )  | \Delta
\psi_n \rangle = - P_c \Delta V_{\SCF} | \psi_n \rangle.
\label{eq:linear-2} \end{equation} 
In practice, if the linear system is solved by  
the conjugate-gradient or any other iterative method and the trial
solution is chosen orthogonal to the occupied-state manifold,
orthogonality is maintained during iteration without having to care
about the extra $P_v$ term on the left-hand side of
Eq.\ \eqref{linear-2}.

The above discussion applies to insulators where the gap is finite. In
metals a finite density of states (DOS) occurs at the Fermi energy,
and a change in the orbital occupation number may occur upon the
application of an infinitesimal perturbation. The modifications of
DFPT needed to treat the linear response of metals have been discussed
by \citeasnoun{SdGmetalli} and will be presented in some detail in
Sec. \secref{metals}.

\subsubsection{Monochromatic perturbations} \label{sec:monochromatic} 

One of the greatest advantages of DFPT---as compared to other
non-perturbative methods to calculate the vibrational properties
of crystalline solids (such as, {\em e.g.} the frozen phonon or the
molecular-dynamics spectral analysis methods)---is that within DFPT the
responses to perturbations of different wave-lengths
are decoupled. This
feature allows one to calculate phonon frequencies at arbitrary
wave-vectors ${\bf q}$
avoiding the use of super-cells and with a workload which is essentially
independent of the phonon wave-length. To see this in some detail,
we first rewrite Eq.\ \eqref{linear-2} by explicitly indicating the
wave-vector $\k$ and band index $v$ of the unperturbed 
wave-function, $\psi_v^\k$,
and by projecting both sides of the equation over the manifold of states
of wave-vector $\k+\q$. 
Translational
invariance requires that the projector onto the $\k+\q$ manifold, 
$P^{\k+\q}$, commutes
with $H_{\SCF}$ and with the projectors onto the occupied- and
empty-state manifolds, $P_v$ and $P_c$. By indicating with
$P^{\k+\q} P_v = P^{\k+\q}_v$ and $P^{\k+\q} P_c =
P^{\k+\q}_c$ the projectors onto the occupied and empty states of
wave-vector $\k+\q$,
Eq.\ \eqref{linear-2} can be rewritten as:
\refstepcounter{equation} \label{eq:linear-q1} $$ \displaylines{\quad
\left(H_{\SCF} +\alpha P^{\k+\q}_v - \epsilon_v^\k
\right) | \Delta\psi_v^{\k+\q} \rangle = \hfill \cr \hfill - P^{\k+\q}_c
\Delta V_{\SCF} | \psi_v^\k \rangle, \quad (\theequation) } $$
where $ | \Delta\psi_v^{\k+\q} \rangle = P^{\k+\q} |
\Delta\psi_v^\k \rangle$.  By decomposing the perturbing
potential, $\Delta V_{\SCF}$, into Fourier components,
\begin{equation} \Delta V_{\SCF}(\r) = \sum_\q
\Delta v_{\SCF}^\q(\r) {\rm e}^{i\q\cdot\r}, 
\end{equation} 
where $\Delta v_{\SCF}^\q(\r)$ is a lattice-periodic function,
Eq.\ \eqref{linear-q1} can be cast into the form:
\refstepcounter{equation} \label{eq:linear-q2} 
$$\displaylines{
\quad 
\left(H_{\SCF}^{\k+\q} + \alpha \sum_{v^\prime} | u_{v^\prime}^{\k+\q}
\rangle
\langle u_{v^\prime}^{\k+\q} | - \epsilon_v^\k \right) 
|\Delta u_v^{\k+\q} \rangle = \hfill \cr \hfill
-\left[ 1 - \sum_{v^\prime} | u_{v^\prime}^{\k+\q} \rangle \langle
u_{v^\prime}^{\k+\q} |\right ] \;
\Delta v_{\SCF}^\q | u_v^\k  \rangle, \quad (\theequation) }
$$
where $v^\prime$ runs over the occupied states at $\k+\q$,
$u_v^\k$ and $\Delta u_v^{\k+\q}$ are the
periodic parts of the unperturbed wave-function and of the $\k+\q$
Fourier component of its first-order correction,
respectively, and the coordinate-representation kernel of the operator
$H_{\SCF}^{\k+\q}$,  
$h_{\SCF}^{\k+\q}(\r,\r')
= \langle \r | H_{\SCF}^{\k+\q} | \r' \rangle $, is defined in terms of
the kernel of the 
SCF hamiltonian, $h_{\SCF}^{(0)}(\r,\r')
= \langle \r | H_{\SCF} | \r' \rangle $, by the relation:
\begin{equation} h_{\SCF}^{\k+\q}(\r,\r') = \ealla{-i(\k+\q)\cdot\r}
h_{\SCF}^{0}(\r,\r') \ealla{i(\k+\q)\cdot\r'}. \end{equation}
Eq.\ \eqref{linear-q2} shows that the time-consuming step
of the self-consistent process, Eq.\ \eqref{linear-2}, can be carried on
working on lattice-periodic functions only, and the corresponding
numerical workload is therefore independent of the wave-length of the
perturbation. 

Let us now see how the other two steps of the
self-consistent process, Eqs.\ \eqref{dn} and \eqref{dVscf}, can be
carried on in a similar way by treating each Fourier component of the
perturbing potential and of the charge-density response independently. 
The Fourier components of any real function (such as $\Delta n$ and
$\Delta v$) with wave-vectors $\q$ and $-\q$ are complex conjugate of
each other: $\Delta n^{-\q}(\r) = (\Delta
n^\q(\r))^*$, and similarly for the potential. Because of
time-reversal symmetry, a similar results applies to wave-functions: 
$\Delta u^{\k+\q}_v(\r) = (\Delta u^{-\k-\q}_v(\r))^*$.
Taking into account these relations,
the Fourier component of the charge-density response at 
wave-vector $\q$ are obtained from Eq.\ \eqref{dn}:
\begin{equation} \Delta n^\q_v(\r) = 4 \sum_{\k v} u^{\k
*}_v(\r) \Delta u^{\k+\q}_v(\r). \label{eq:dnq} \end{equation}
Eq.\ \eqref{dVscf} is a {\it linear} relation between the 
self-consistent variation of the  potential and the variation
of the electron charge-density distribution. 
The Fourier component of the self-consistent potential 
response reads:
\begin{eqnarray} 
\Delta v_\SCF^\q(\r) = \Delta v^\q(\r) & + & e^2 
\int {\Delta n^\q(\r') \over
|\r-\r'|} \ealla{-i\q\cdot(\r-\r')}d\r' \nonumber \\
 & & + \left . {d v_{xc}(n) \over d n}
\right |_{n=n(\r)} \Delta n^\q(\r).
\label{eq:dVq} \end{eqnarray}

The sampling of the BZ needed for the evaluation of \eqref{dnq} is
analogous to that needed for the calculation of the unperturbed
electron charge density, \eqref{n_of_r}, and it requires in most cases
an equal number of discrete $\k$ points. An exception to this rule
occurs when calculating the response of insulators to
macroscopic electric fields---as discussed in Sec. \secref{homelfields}---and
for the calculation of phonons in metals in the presence of Kohn
anomalies---as discussed in Sec. \secref{metals}.

In conclusion, Eqs.\ \eqref{linear-q2}, \eqref{dnq}, and \eqref{dVq}
form a set of self-consistent relations for the charge-density and
wave-function linear response to a perturbation of a wave-vector,
$\q$, which can be solved in terms of lattice-periodic functions only,
and which is decoupled from all the other sets of similar equations
holding for other Fourier components of a same perturbation. Thus,
perturbations of different periodicity can be treated independently of
each other with a numerical workload which is, for each perturbation,
of the same order as that needed for the unperturbed system.

\subsubsection{Homogeneous electric fields} \label{sec:homelfields}

The electron density response to a homogeneous ({\em macroscopic})
electric field requires a special treatment. In fact, several
electrostatic properties of an infinite solid are strictly speaking
ill-defined in the long wave-length limit because the electrostatic
potential describing a homogeneous electric field $\EE$, ($V_{_\EE}(\r) =
e \EE\cdot \r$), is both unbound from below and 
incompatible with Born-von-K\'arm\'an periodic boundary
conditions. In the linear regime, however, these pathologies can
be efficiently treated in an elementary way. As in the harmonic
approximation
the lattice-dynamical properties of polar
insulators depend for long wave-lengths on the linear response
to a homogeneous electric field (see Sec.\secref{polar}), 
we will
limit here our analysis to the linear regime only and postpone the
discussion of non-linear electrostatic effects to Sec. \secref{higher}.

From a mathematical point of view, the main difficulty in treating
macroscopic electric fields stems from the fact that the position
operator, $\r$, is ill-defined in a periodic system and so are its
matrix elements between wave-functions satisfying Born-von-K\'arm\'an
boundary conditions. The wave-function response to a given
perturbation, Eq.\ \eqref{pert_th}, however, only depends on the {\em
off-diagonal} matrix elements of the perturbing potential between
eigen-functions of the unperturbed Hamiltonian. Such matrix elements
are indeed well defined even for a macroscopic electric field, as it
can be seen by re-writing them in
terms of the commutator between $\r$ and the unperturbed Hamiltonian,
which is a lattice-periodic operator: \begin{equation} \langle \psi_m
| \r | \psi_n \rangle = { \langle \psi_m | [H_\SCF,\r] | \psi_n
\rangle \over \epsilon_m -\epsilon_n },\quad \forall\ m \neq n. 
\label{eq:Commutator} \end{equation}
If the self-consistent potential acting on the electrons is local, the
above commutator is simply proportional to the momentum operator:
\begin{equation}
 \left[H_\SCF , \r \right] = 
     -{\hbar^2\over m} {\partial\over\partial \r}.
 \end{equation}
Otherwise, the
commutator will contain an explicit contribution from the non-local
part of the potential \cite{Hx1,BGT,Hx2}. 

When calculating the response of a crystal to an applied electric
field, $\EE_0$, one must consider that the {\em screened} field acting
on the electrons is: \begin{equation}  \EE=\EE_0-4\pi \PP, \label{eq:macro}
\end{equation}
where $\PP$ is the electronic polarization linearly induced by the
screened ({\em i.e.} self-consistent) field, $\EE$:
\begin{equation} {\PP} = - {e \over V}\int_V\r\Delta^\EE n(\r) d\r.
\label{eq:dP} \end{equation}
In addition to the {\it macroscopic}
screening, expressed by Eq.\ \eqref{macro}, 
the density response to an external macroscopic electric field, $\EE_0$,
also involves {\it microscopic} Fourier components,
\begin{equation} \Delta^\EE n(\r) = 4 \sum_{n=1}^{N/2} \psi_n^*(\r) 
\Delta^\EE\psi_n(\r), \label{eq:dnlf} \end{equation} 
responsible for the so called 
{\it local fields}, that must be taken into account in the
self-consistent procedure. 

Eq.\ \eqref{dP} is of course well defined
for any finite system. The electric polarization of a macroscopic
piece of matter, however, is ill defined in that it depends
on the details of the charge
distribution at the surface of the sample. Nevertheless, the
polarization {\it linearly} induced by a given perturbation is well
defined, and Eq.\ \eqref{dP} can in fact be recast into a
boundary-insensitive form \cite{Littlewood}. To see this, we use
Eq.\ \eqref{dn} and we obtain from Eq.\ \eqref{dP}: \begin{eqnarray}
\P_\alpha &=& -{4e\over V} \sum_{n=1}^{N/2} \langle \psi_n
|  r_\alpha| \Delta^\EE\psi_n \rangle \nonumber \\
&= & -{4e\over V}
\sum_{n=1}^{N/2} \sum_{m=N/2+1}^\infty 
{\langle \psi_n |[H_\SCF,r_\alpha]| |\psi_m\rangle \over
(\epsilon_n - \epsilon_m)} \langle \psi_m | \Delta^\EE \psi_n \rangle,
\nonumber \\
& &
\label{eq:dP1}
\end{eqnarray}
where the subscript $\alpha$ indicates Cartesian components.
Let us introduce the wave-function
$\bar{\psi}_n^\alpha (\r)$ defined as: 
\begin{equation} \bar{\psi}_n^\alpha (\r) = \sum_{m\ne n}
\psi_m(\r) {\langle \psi_m | [H_{\SCF},r_\alpha] | \psi_n
\rangle \over (\epsilon_m-\epsilon_n)}. \end{equation} 
$ \bar{\psi}$ satisfies an equation of the
same kind as Eq.\ \eqref{linear-2} with the perturbing potential on its
right-hand side replaced by $[H_{\SCF},r_\alpha]$:
\begin{equation} ( H_{\SCF} - \epsilon_n )  | 
\bar \psi_n^\alpha \rangle = P_c [H_{\SCF},r_\alpha] | \psi_n 
\rangle. \label{eq:psibar} \end{equation}
The induced polarization, Eq.\ \eqref{dP1}, can be recast into the form:
\begin{equation} \P_\alpha = -{4e\over V} \sum_{n=1}^{N/2}
\langle\bar{\psi}_n^\alpha |\Delta^\EE \psi_n \rangle,
\label{eq:dP2} \end{equation}
where the anti-hermitian character of the commutator has been used.

The first-order correction
to a crystal wave-function due to a perturbing homogeneous
electric field, $\EE$, is given by the response to the full, 
macroscopic {\em and} microscopic, perturbation: 
\refstepcounter{equation}
\label{eq:deltapsiElocf} $$ \displaylines{ \quad
(H_{\SCF} - \epsilon_n) |
\Delta^\EE \psi_n \rangle = -e \sum_\alpha \E_\alpha
|\bar{\psi}_n^\alpha \rangle 
\hfill \cr \hfill - P_c
\Delta V^{lf} 
| \psi_n \rangle, \quad(\theequation) } $$
where:
\refstepcounter{equation} \label{eq:dvlf} $$\displaylines{\quad
\Delta V^{lf}(\r)
= e^2 \int {\Delta^\EE n(\r') \over |\r-\r'|} d\r' \hfill \cr \hfill
+ \left . {d
v_{xc}(n) \over d n} \right |_{n=n(\r)} \Delta^\EE n(\r). \quad
(\theequation) } $$

The terms of the sum appearing in \eqref{dP2} implicitly behave as the
inverse of the difference between a conduction and a valence energy
eigenvalue to the third power, $\sim (\epsilon_c - \epsilon_v
)^{-3}$. One of the energy denominators results from the standard
first-order perturbation calculation of the perturbed wave function,
$\Delta^\EE \psi_n $, Eq. \eqref{pert_th}. A second energy
denominator comes from the term $\bar{\psi}$ in the right-hand side 
of Eq. \eqref{deltapsiElocf}, which requires by itself the solution 
of a linear equation analogous to that of
first-order perturbation theory, Eq. \eqref{psibar}. The third energy
denominator comes from the $\bar{\psi}$, which appears explicitly in
the bracket of Eq. \eqref{dP2}. This dependence of the terms of the
sum upon the direct gaps at different $\k$ points of the BZ may
require a rather fine sampling of the BZ when the fundamental gap is
small. In these cases, the number of $\k$ points needed to compute the
dielectric constant is substantially larger than that needed by a
standard unperturbed calculation \cite{raman,Hx1,piezoSdG}.

The self-consistent cycle defined by Eqs.\
(\ref{eq:macro}-\ref{eq:dvlf}) can be performed starting from a given
external macroscopic electric field, $\EE_0$, and updating at each
iteration the macroscopic and microscopic part of electronic density
response and the perturbing potential, via Eqs.\ \eqref{dP2},
\eqref{macro}, \eqref{dnlf}, and \eqref{dvlf}; then solve Eq.\
\eqref{deltapsiElocf} for $\Delta^\EE \psi$ and repeat. However, for
computational purposes, it is more convenient and simple to keep the
value of the screened electric field, $\EE$, fixed and let only the
microscopic components of the potential vary during the the iterative
procedure, that is reached simply iterating Eq.\
\eqref{deltapsiElocf}, \eqref{dnlf} and \eqref{dvlf}.  The macroscopic
polarization is then calculated only at the end when self-consistency
is achieved using Eq.\ \eqref{dP2}. Physically, this is equivalent to
calculating the polarization response to a given {\it screened}
electric field, \EE, instead of the response to the {\it bare}
electric field $\EE_0$.

The electronic contribution to the dielectric tensor,
$\epsilon^{\alpha\beta}_\infty$, for the general (low-symmetry)
case, can be derived from simple electrostatics. Using 
Eq.\ \eqref{macro} and the definition
of $\epsilon^{\alpha\beta}_\infty$, one has:
\begin{equation}
 \E_{0 \alpha} = (\E_\alpha + 4\pi \P_\alpha)
            = \sum_\beta \epsilon^{\alpha \beta}_\infty \E_\beta.
\end{equation}
Using Eq.\ \eqref{dP2} to calculate the polarization induced in the 
$\alpha$ direction when a field is applied in the $\beta$ direction,
one finally obtains:
\begin{equation}
\epsilon^{\alpha \beta}_\infty = \delta_{\alpha\beta} -
{16\pi e \over V\E_\beta} \sum_{n=1}^{N/2}
\langle\bar{\psi}_n^\alpha |\Delta^{\E_\beta} \psi_n \rangle .
\end{equation}

\subsubsection{Relation to the variational principle}
\label{sec:variational} The KS equations are the
Euler equations which solve the Hohenberg-Kohn variational
principle. The equations of DFPT introduced in Sec.
\secref{linear-response} can be seen as a set of equations which solve
{\em approximately} the variational principle when the external
potential is perturbed. Alternatively, these equations can be
considered as those which minimize {\em exactly} an approximate energy
functional \cite{GonzeCG,Gonze1,Gonze3}. To see this point in some
detail, let us consider the energy functional as depending explicitly
on the set of KS orbitals, $\psi \equiv \{\psi_n\}$
(assumed to be real)
and parametrically on the external
potential, $V(\r)$: \refstepcounter{equation}
\label{eq:ks-functional[phi]} $$ \displaylines{\quad
E[\psi;V] =
-2 {\hbar^2 \over 2m} \sum_{n=1}^{N/2} \int \psi_n(\r){\partial^2
\psi_n(\r) 
\over \partial \r^2} d\r \hfill \cr +\int V(\r) n(\r) d\r + {e^2 \over 2} \int
{n(\r)n(\r')\over|\r-\r'|}d\r d\r' \cr \hfill +
E_{xc}[n(\r)]. \quad(\theequation) } $$
The functional derivative of the above functional
with respect to $\psi_n$ is: 
\begin{equation} {\delta E \over \delta \psi_n(\r)} = 2
H_{\SCF} \psi_n(\r). \label{eq:dEoverdphi} \end{equation} The Euler
equations thus read: \begin{equation} H_{\SCF} |\psi_n\rangle = 
\sum_{m=1}^{N/2}
\Lambda_{nm} |\psi_m\rangle, \label{eq:kohn-sham-euler} \end{equation} where
the $\Lambda$'s are a set of Lagrange multipliers introduced so as to
enforce the ortho-normality of the $\psi$'s.  Eqs.\ \eqref{kohn-sham-euler} 
are invariant with respect to a unitary transformation within the manifold
of the $\psi$'s, so that the usual KS equation, Eq.\ \eqref{ks-eq}, is
recovered in the representation which diagonalizes the $\Lambda$
matrix.

Let us now indicate by $\psi^{(0)}$ the solutions of the KS equations
corresponding to a particular choice of the external potential,
$V^{(0)}(\r)$ (the {\em unperturbed} potential), and let us indicate by
$\Delta V$ and $\Delta\psi$ the differences between the actual
potential and orbitals and their unperturbed values. The energy
functional, Eq.\ \eqref{ks-functional[phi]}, can be equally
seen as depending on $\Delta\psi$ and $\Delta V$: $
E \equiv E[\{\psi^{(0)} + \Delta \psi\}; V^{(0)}+\Delta V]. $ We
now consider the {\em approximate} functional, $E^{(2)}$, which is
obtained from $E$ by truncating its Taylor expansion in terms of
$\Delta\psi$ and $\Delta V$ to second order: 
\refstepcounter{equation} \label{eq:E2} $$ \displaylines{\quad
E^{(2)}[\{\Delta\psi\}; \Delta V] = E[\{\psi^{(0)}\};V^{(0)}] + \int {\delta E
\over \delta V(\r)} \Delta V(\r) d\r  \hfill \cr
+ \sum_{n=1}^{N/2} \int {\delta E \over \delta
\psi_n(\r)} \Delta \psi_n(\r) d\r \cr + {1\over 2} \sum_{n=1}^{N/2} \int
{\delta^2 E \over \delta \psi_n(\r) \delta V(\r')} \Delta \psi_n(\r)
\Delta V(\r') d\r d\r' \cr \hfill + {1\over 2} \sum_{n,m=1}^{N/2} \int
{\delta^2 E \over 
\delta \psi_n(\r)\delta \psi_m(\r')} \Delta \psi_n(\r) \Delta
\psi_m(\r') d\r d\r', \hfill \llap{(\theequation)} } $$
where we have omitted the second derivative with
respect to $V$ because $E$ is linear in $V$. Besides
Eq.\ \eqref{dEoverdphi}, the required functional derivatives are:
\begin{eqnarray} {\delta E \over \delta V(\r)}\biggr|_{\psi=\psi^{(0)}} 
&=& n^{(0)}(\r) \\
{\delta^2 E \over \delta \psi_n(\r) \delta V(\r')} 
\biggr|_{\psi=\psi^{(0)}}
&=&
2\psi^{(0)}_n(\r) \delta(\r-\r') \\ {\delta^2 E \over \delta
\psi_n(\r) \delta \psi_m(\r')}\biggr|_{\psi=\psi^{(0)}}
 &=& 2h^{(0)}_{\SCF}(\r,\r') \nonumber
\\ 
 && + 4 K(\r,\r')
\psi^{(0)}_n(\r) \psi^{(0)}_m(\r'), \end{eqnarray} where: \begin{equation}
K(\r,\r') = {e^2\over |\r-\r'|} + {\delta^2 E_{xc} \over \delta n(\r)
\delta n(\r') }, \label{eq:Kappa} \end{equation} 
and $h^{(0)}_{\SCF}(\r,\r')$ is the
kernel of the unperturbed KS Hamiltonian.
The energy functional, Eq.\ \eqref{E2}, must be minimized under the
constraints that the resulting solutions lead to an orthonormal set of
occupied states: \begin{equation} \langle \psi^{(0)}_n + \Delta \psi_n |
\psi^{(0)}_m + \Delta \psi_m \rangle = \delta_{nm}. 
\label{eq:delta-constraints} \end{equation} This leads to the Euler
equations: \refstepcounter{equation} \label{eq:delta-euler} $$
\displaylines{ \quad H^{(0)}_{\SCF}|\Delta \psi_n\rangle - \sum_{m=1}^{N/2}
 \Lambda_{nm}
|\Delta\psi_m\rangle = \hfill \cr\hfill -\Delta V_{\SCF}|\psi^{(0)}_n\rangle
 + \sum_{m=1}^{N/2}(\Lambda_{nm}-\epsilon_n \delta_{nm})|\psi^{(0)}_m\rangle, 
\quad (\theequation) }
$$ where the $\Lambda$'s are a set of Lagrange multipliers introduced
so as to enforce the constraints, Eq.\ \eqref{delta-constraints}, and
\refstepcounter{equation} \label{eq:DeltaVvsDeltapsi} $$
\displaylines{\quad
\Delta V_{\SCF}(\r) = \Delta V(\r) \hfill \cr \hfill
+ 2 \sum_{n=1}^{N/2} \int
d\r' K(\r,\r') \psi^{(0)}_n(\r') \Delta \psi_n(\r') \quad(\theequation)} $$
is the
first-order variation of the self-consistent potential. We now project
both sides of Eq.\ \eqref{delta-euler} onto $\psi^{(0)}_k$. Taking
into account that by Eq.\ \eqref{delta-constraints} $\langle
\Delta\psi_n | \psi^{(0)}_k \rangle = {\cal O}(\Delta^2)$, to linear order
in $\Delta$ one obtains: \begin{equation} \Lambda_{mn} - \epsilon_n
\delta_{nm} = \langle \psi^{(0)}_m | \Delta V_{\SCF} | \psi^{(0)}_n \rangle
. \end{equation} To linear order in $\Delta$, Eq.\ \eqref{delta-euler}
can thus be cast into the form: \begin{eqnarray} (H^{(0)}_{\SCF} -
\epsilon_n ) |\Delta\psi_n\rangle &=& - \Delta V_{\SCF} |\psi^{(0)}_n\rangle 
\nonumber \\& & \mbox{} + \sum_{m=1}^{N/2}
|\psi^{(0)}_m\rangle \langle \psi^{(0)}_m | \Delta V_{\SCF} | \psi^{(0)}_n 
\rangle
\nonumber \\ &=& -P_c \Delta V_{\SCF} |\psi^{(0)}_n\rangle , \label{eq:linear-var}
\end{eqnarray} which is essentially the same as Eq.\ \eqref{linear-2}.
The set of perturbed orbitals, $\{\Delta\psi\}$, is the solution of $N$
coupled linear systems whose dimension is the size of the basis set,
$M$ (Eq.\ \eqref{linear-var} or Eq.\ \eqref{linear-2}, where the
coupling comes from the dependence of the self-consistent potential on
all the orbitals). Alternatively, it can be seen as a huge linear
system of dimension $M\times N/2$, which is obtained by inserting the
expression for $\Delta V_{\SCF}$, Eq.\ \eqref{DeltaVvsDeltapsi}, into
Eqs.\ \eqref{linear-var} or \eqref{linear-2}. The latter is naturally
derived from the minimization of the functional $E^{(2)}$, Eq.\ \eqref{E2},
which is quadratic.

This variational approach shows that the error on the
functional to be minimized, Eq.\ \eqref{E2}, is proportional to the 
{\em square} of the error on the minimization variables, 
$\Delta\psi$. This fact can be exploited in the calculation 
of the second-order mixed derivatives, Eq.\ \eqref{dE2}. 
It can be shown that a variational expression
can be constructed for mixed derivatives as well \cite{GonzeCG,Gonze3}.

\subsubsection{Metals}
\label{sec:metals} 

Density-functional perturbation theory, as presented above, is directly
applicable to metals, provided the (electronic) temperature vanishes, 
so that a clear cut separation between occupied and empty 
states is possible. In this case, however, the number of $k$ points
needed to correctly represent the effect of the Fermi surface would
be very large. Practical implementations of DFPT to metallic systems have
been discussed by \citeasnoun{Quong_LR} and by \citeasnoun{SdGmetalli},
in the pseudo-potential formalism, and by \citeasnoun{LMTO_LR1},
in the Linearized Muffin-Tin Orbital framework. In the following, we
will closely follow the formulation of \citeasnoun{SdGmetalli} which is
based on the smearing technique for dealing with Fermi-surface effects.

In the smearing approach each KS energy level is broadened by a smearing
function: \begin{equation}\delta_\sigma(\epsilon) = {1\over \sigma}
\tilde\delta(\epsilon/\sigma), \end{equation} where $\tilde\delta(x)$
is any function that integrates to 1---so that $\delta_\sigma(\epsilon)$
tends to the Dirac $\delta$ function in the limit of vanishing 
smearing width, $\sigma$.
Many kinds of smearing functions, $\tilde\delta(x)$, can be used:
Fermi-Dirac broadening,\footnote{In this case the broadening function
is the derivative of the Fermi-Dirac distribution function: $\tilde\delta(x) = 
{1\over 2} \left
(1 + \cosh (x)\right)^{-1}$.}
Lorentzian, Gaussian \cite{smearing}, Gaussian combined with polynomials
\cite{MP}, or {\it cold smearing} \cite{Nicola-coldsmearing} functions,
to recall just a few of those used in the literature. While the choice
of a given smearing function is to some extent a matter of personal
taste and computational convenience, the specific choice of
the Fermi-Dirac broadening allows one to explicitly account for the
effects of a finite temperature ($T=\sigma / k_B$), when needed.
The (local) density
of states resulting from the broadened energy levels will be the original
density of states, convoluted with the smearing function: \begin{equation}
n(\r,\epsilon) = \sum_n {1\over\sigma} \tilde\delta\left(\frac{\epsilon -
\epsilon_n}{\sigma}\right) |\psi_n(\r)|^2, \label{eq:ldos} \end{equation}
where the sum refers both to the discrete $\k$-vector index and to
band and spin indices, for all bands. 
From this basic quantity the electron density follows:
\begin{equation} n(\r) = \int_{-\infty}^{\epsilon_F} n(\r,\epsilon)
d\epsilon = \sum_n \tilde\theta\left({\epsilon_F - \epsilon_n \over
\sigma}\right) |\psi_n(\r)|^2, \label{eq:n-metalli}\end{equation} where
$\tilde\theta(x)= \int_{-\infty}^{x} \tilde\delta(y) dy$ is a smooth
approximation to the step function, and the Fermi energy is determined
by the normalization to the total number of electrons, \begin{equation}
N = \int_{-\infty}^{\epsilon_F} n(\epsilon) d\epsilon = \sum_n
\tilde\theta\left( {\epsilon_F - \epsilon_n\over \sigma}\right). \end{equation}
The advantage of this procedure is that after convolution the (modified)
local density of states can be computed accurately on a discrete grid
of points in the BZ, provided that the average separation
between neighboring eigenvalues is small with respect to the broadening
width, $\sigma$.

Given the definition of the local density of states, Eq.\ \eqref{ldos}, 
the consistent way to define
the auxiliary KS kinetic-energy functional---or its analogue in
the finite temperature theory \cite{mermin}---is through the Legendre
transform of the single-particle energy integral:
\begin{eqnarray} T_s[n]
&=& \int_{-\infty}^{\epsilon_F} \epsilon n(\epsilon) d\epsilon
- \int V_{\SCF}(\r) n(\r) d\r \nonumber \\
&=& \sum_n \left [ - \frac{\hbar^2}{2m}
\tilde\theta\left(\frac{\epsilon_F - \epsilon_n}{\sigma}\right)
\int \psi^*_n(\r){\partial^2 \psi_n(\r)  \over \partial \r^2} d\r
\right .\nonumber \\ &&
\quad\quad \quad\quad \quad\quad \quad\quad \quad\quad 
\left . + \sigma \tilde\theta_1
\left(\frac{\epsilon_F - \epsilon_n}{\sigma}\right) \right ],
\label{eq:Tmetalli} \end {eqnarray} where $\tilde\theta_1(x) =
\int_{-\infty}^{x} y \tilde \delta(y) dy$.  Note that when a finite
electronic temperature is considered the KS auxiliary functional,
$T_s[n]$, contains both the kinetic energy and the entropy contribution
to the electronic free energy of the independent particle system. These
two contributions appear separately in the last expression for $T_s[n]$
in Eq.\ \eqref{Tmetalli} where it can be verified that for Fermi-Dirac
broadening $\tilde\theta_1(x) = f {\log} (f) + (1-f) {\log}
(1-f)$, with $f=\tilde\theta(x)$ as required.  With the definitions
above, for any kind of smearing function, the usual KS equations follow
from the minimization of the total energy. The price to be paid for
the computational simplicity of the smearing approach is that the
computed total energy depends on the chosen broadening width and results
for finite broadening widths must be corrected for, unless the shape of the
smearing function is such as to reduce this dependence within acceptable
values. Various discussions of this issue can be found in the literature
\cite{MP,Nicola-coldsmearing,SdGmetalli,DeVita}.

Forces and other first-order energy derivatives can be computed in
the usual way from Eqs.\ \eqref{HF-force} and \eqref{dE1} which only
require the knowledge of the unperturbed electronic density. Similarly
second-order derivatives are computed from Eq.\ \eqref{dE2} where
also the first order variation of the density is needed. Direct
variation of Eq.\ \eqref{n-metalli} provides the required expression for
$\Delta n (\r)$: 
\refstepcounter{equation} \label{eq:dn-metalli} 
$$ \displaylines{ \quad
\Delta n (\r) = \sum_{n}
\tilde\theta_{F,n} \left[ \psi^*_n(\r) \Delta \psi_n(\r)
+ c.c. \right] \hfill \cr\hfill
+\sum_{n} |\psi_n(\r)|^2 \tilde\delta_{F,n} 
( \Delta \epsilon_F -\Delta \epsilon_n ), \quad (\theequation) } $$
where we have defined
$\tilde\theta_{n,m} = \tilde\theta((\epsilon_n-\epsilon_m)/\sigma)$
and $\tilde\delta_{n,m} = (1/\sigma) \tilde\delta
((\epsilon_n-\epsilon_m)/\sigma)$. The last term in Eq.\
\eqref{dn-metalli} accounts for possible changes in occupation numbers
induced by shifts 
in the single particle energies ($\Delta \epsilon_n=
\langle\psi_n|\Delta V_{\SCF}|\psi_n\rangle$), as well as
in the Fermi energy of the system. Whether this term is
present or not depends on the thermodynamic ensemble used: it is absent
if the chemical potential is kept fixed, whereas it might be present
if it is the
number of electrons which is fixed. Even in this last
case the Fermi energy is unaffected by the perturbation to linear order,
unless the perturbation is lattice-periodic (monochromatic with $\q=0$).
Let us neglect this term for the time being and come back to it at the
end of the section.

Substituting in Eq.\ \eqref{dn-metalli} the definition for $\Delta
\psi_n(\r)$ from standard perturbation theory, Eq.\ \eqref{pert_th},
and exploiting the symmetry between the two contributions in square
brackets, one obtains the following well known
expression for the first-order variation of the electronic density
in a metal: \refstepcounter{equation} \label{eq:deltan_metal-1}
$$\displaylines{\quad
\Delta n (\r) = \sum_{n,m}
\frac{\tilde\theta_{F,n} - \tilde\theta_{F,m} } {\epsilon_n -
\epsilon_m} \hfill \cr \hfill 
\times ~\psi^*_n (\r) \psi_m(\r) \langle \psi_m | \Delta
V_{\SCF} | \psi_n \rangle, \quad(\theequation) } $$
where the term in $\Delta
\epsilon_n$ in Eq.\ \eqref{dn-metalli} has become the $n=m$ term
in the above sum and the incremental ratio $(\tilde\theta_{F,n}
- \tilde\theta_{F,m} )/ (\epsilon_n - \epsilon_m) $ must be
substituted with its limit, $-\tilde\delta_{F,n}$, whenever $\epsilon_m
\rightarrow \epsilon_n$. This limit is always finite for any finite
broadening line-width, or temperature, and this expression is therefore
numerically stable even in the presence of vanishingly small virtual
excitation energies.

In order to avoid the double sum over occupied and unoccupied states we use
the relation $\tilde\theta(x) + \tilde\theta(-x) = 1$ and the symmetry
between $i$ and $j$ to get 
\refstepcounter{equation} \label{eq:deltan_metal-2} $$\displaylines{\quad
 \Delta n(\r) = 2
\sum_{n,m} \frac{\tilde\theta_{F,n} - \tilde\theta_{F,m} } {\epsilon_n
- \epsilon_m} \tilde \theta_{m,n}  \hfill \cr \hfill
\times ~\psi^*_n(\r) \psi_m(\r) \langle
\psi_m | \Delta V_{\SCF} | \psi_n \rangle, 
\quad(\theequation) } $$
where the sum over the first index can be limited to the 
states that have non negligible occupation. This expression can be further
simplified, avoiding the explicit sum over the second index, by re-writing
it as: \begin{equation} \Delta n(\r) = 2 \sum_n \psi_n^*(\r)
\Delta \psi_n(\r), \label{eq:dn-metalli-2} \end{equation}
where the $\Delta \psi_n$'s satisfy the equation
\begin{equation} \left[ H_{\SCF} + Q - \epsilon_n \right] \Delta \psi_n
= - \left [\tilde \theta _{F,n} - P_n \right ] \Delta V_{\SCF}
| \psi_n \rangle, \label{eq:linear-metalli} \end{equation}
with
$$
Q = \sum_k \alpha_k |\psi_k\rangle\langle\psi_k|, \quad\quad
P_n= \sum_m \beta_{n,m} |\psi_m\rangle\langle\psi_m|,$$
\begin{equation}
\beta_{n,m} = \tilde\theta_{F,n} \tilde\theta_{n,m} + 
              \tilde\theta_{F,m} \tilde\theta_{m,n} +
              \alpha_m \frac{\tilde\theta_{F,n}-\tilde\theta_{F,m}}
                       {\epsilon_n-\epsilon_m} \tilde\theta_{m,n}.
\label{eq:proiettori-metalli}
\end{equation}
In the above equations the $\alpha_k$'s are chosen in such a way that
the $Q$ operator makes the linear system, Eq.\ \eqref{linear-metalli},
nonsingular for all non vanishing $\Delta \psi_n$. A possible simple
choice is \begin{equation} \alpha_k = {\max} (\epsilon_F + \Delta
- \epsilon_k,0), \end{equation} 
with $\Delta \approx 3 \div 4\sigma$. Another, even simpler, choice is
to set 
$\alpha_k $ equal to the occupied bandwidth plus, say, $3 \sigma$
for all partially occupied states and zero when the state is totally
unoccupied.
It can be easily verified that since $\alpha_k$ vanishes when $\psi_k$
is unoccupied then also $\beta_{n,m}$ vanishes when any of its indices
refers to an unoccupied state. Therefore the $Q$ and $P$ operators
involve only 
the small number of partially filled bands and the first-order variation
of the wave functions and of the charge density can be computed
avoiding any
explicit reference to the unoccupied states, much in the same way as
for insulating materials. In fact, if the above scheme is
applied to an insulator using a smearing width much smaller
than its fundamental band gap, all the metallic equations, 
Eqs.\ (\ref{eq:dn-metalli-2}-\ref{eq:proiettori-metalli}), reduce
numerically to their insulating analogues, Eqs.\ \eqref{dn} and
\eqref{linear-2}. 

The expression for the charge density linearly induced by a given
perturbation, Eqs. \eqref{deltan_metal-1} and \eqref{deltan_metal-2},
involves an energy denominator which vanishes for metals. In one
dimension, this vanishing denominator gives rise to a divergence in
the screening of perturbations whose wave-vector is twice the Fermi
momentum, $2k_F$. This divergence is smeared in two dimensions and
suppressed in three dimensions by volume effects. However, if the
topology of the Fermi surface is such that two finite portions of it
are parallel and connected by a wave-vector which for convenience we
will name $2k_F$ ({\em nesting}), the screening to perturbations 
of wave-vector
$2k_F$ will diverge even in three dimension. This is the physical
mechanism giving rise to Kohn anomalies in the vibrational spectra of
certain metals. The sampling of the BZ necessary to evaluate the sum
over (partially) occupied states in Eqs. \eqref{deltan_metal-1} and
\eqref{deltan_metal-2} is in ordinary cases similar to that needed to
calculate the unperturbed charge-density distribution. Near a
Kohn anomaly, however, a fine sampling of the Fermi surface is
necessary and the number of needed points in the BZ is correspondingly
larger. 

Periodic ($\q = 0$) perturbations may induce a shift of the Fermi energy.
In this case Eq.\ \eqref{dn-metalli-2} must be  modified
as: \begin{equation} \Delta n(\r) = 2 \sum_n \psi_n^*(\r)
\Delta \psi_n(\r) + n(\r, \epsilon_F) \Delta \epsilon_F,
\label{eq:dn-metalli-3} \end{equation} all other technical details
remaining unchanged.
In order to find out the appropriate value of the Fermi energy shift
let us examine the perturbation in the $\q \rightarrow 0$ limit. 
Let us consider the Fourier transform of the self-consistent
perturbing potential $\Delta V_{\SCF}(\q)= 1/V 
\int \Delta V_{\SCF}(\r) e^{-i\q\cdot\r} d\r$.
Its macroscopic ($\q \approx 0$) component reads:
\begin{equation} \Delta V_{\SCF}(\q) = \Delta V(\q) +
\frac{4\pi e^2}{q^2} \Delta n(\q) + {d v_{xc} \over d n} \Delta
n(\q), \label{eq:deltaV-macro} \end{equation} where the last
term is the exchange-correlation contribution, $\Delta V(\q) =
-\frac{4\pi e^2}{q^2} \Delta n_{ext}(\q)$ is the macroscopic
electrostatic component of the  external perturbing potential, and 
$\Delta n_{ext}(\q)$ is finite in the $\q \rightarrow 0$ limit. 
On the other hand the macroscopic component of the density response is 
\begin{equation} \Delta n(\q) = - n(\epsilon_F)
\Delta V_{\SCF} (\q) + \Delta n^{lf}(\q),
\label{eq:deltan-macro} \end{equation} where $\Delta n^{lf}$ is the
density response to the non-macroscopic (local fields) components of
the self-consistent potential. As $\Delta n^{lf}$ and the DOS at the
Fermi energy, $n(\epsilon_F)$, are both finite, $\Delta V_{\SCF}$ and
$\Delta n$ must not diverge when $\q \rightarrow 0$ otherwise  Eqs.\
\eqref{deltaV-macro} and \eqref{deltan-macro} could not be satisfied at
the same time. 
This implies, from Eq.\ \eqref{deltaV-macro}, macroscopic 
charge neutrality for the perturbed system, that is $\Delta n_{ext}(\q) =
\Delta n(\q) + {\cal O}(q^2) $ for $\q\approx 0$; a condition that in turn
implies in Eq.\ \eqref{deltan-macro} 
\begin{equation} \Delta V_{\SCF}(\q) = -
\frac{ \Delta n_{ext}(\q) - \Delta n^{lf} (\q)}
     {n(\epsilon_F)} + {\cal O}(q^2).
\end{equation} The lattice periodic ($\q=0$) result can be obtained by
taking the $\q \rightarrow 0$ limit of the above equations. In this
case however, it is costumary to set arbitrarily to zero the macroscopic
electrostatic component of the self-consistent potential and the charge
neutrality condition is thus enforced by a compensating shift in the
Fermi energy, equal and opposite to the above result: \begin{equation}
\Delta \epsilon_F = \frac{ \Delta n_{ext}(\q=0) + \int n(\r,\epsilon_F)
\Delta V_{\SCF}(\r) d\r }{ n(\epsilon_F)}.  \end{equation}
\subsection{Phonons}
\label{sec:phonons}
\subsubsection{Vibrational states in crystalline solids}

In crystalline solids, the nuclear positions appearing in the
definition of the IFC's,
Eq.\ \eqref{secular-frequency}, are labelled by an index, $I$, which
indicates the unit cell to which a given atom belongs, $l$, and the
position of the atom within that unit cell, $s$: $I\equiv
\{l,s\}$. The position of the $I$-th atom is thus: \begin{equation}
\R_I \equiv \R_l + \boldtau_s + \U_{s}(l), \end{equation} where $\R_l$
is the position of the $l$-th unit cell in the Bravais lattice,
$\boldtau_s$ is the equilibrium position of the $s$-th atom in the
unit cell, and $\U_{s}(l)$ indicates the deviation from equilibrium of
the nuclear position. Because of translational invariance, the matrix
of the IFC's, Eq.\ \eqref{hessian},
depends on $l$ and $m$ only
through the difference $\R \equiv \R_l -\R_m$: \begin{equation}
C^{\alpha\beta}_{st}(l,m) \equiv {\partial^2 E \over \partial
u^\alpha_s(l) \partial u^\beta_t(m)} = C^{\alpha\beta}_{st}(\R_l-\R_m),
\end{equation}
where the greek super-scripts indicate
Cartesian components. The Fourier transform of $C^{\alpha \beta}_{s
t}(\R)$ with respect to $\R$, 
$\widetilde C^{\alpha \beta}_{st}(\q)$, 
can be seen as the second derivative of the BO
energy surface with respect to the amplitude of a lattice distortion
of definite wave-vector: \refstepcounter{equation} 
\label{eq:c_of_q} $$ \displaylines{ \quad \widetilde C^{\alpha \beta}
_{st}(\q) \equiv \sum_\R \ealla{-i\q\cdot\R}
C^{\alpha \beta}_{s t}(\R) \hfill \cr \hfill
= {1 \over N_c} {\partial^2 E \over \partial u^{* \alpha}_s(\q) \partial
u^\beta_t(\q)}, \quad(\theequation) } $$ 
where $N_c$ is the number of unit cells in the crystal, 
and the vector $\U_s(\q)$ is
defined by the distortion pattern: \begin{equation} \R_I[\U_s(\q)] =
\R_l + \boldtau_s + \U_s(\q) \ealla{i\q\cdot\R_l}. \label{eq:RIofuq}
\end{equation} Phonon frequencies, $\omega(\q)$, are solutions of the
secular equation: \begin{equation} \det \left | {1\over \sqrt{M_s
M_t}} \widetilde C^{\alpha\beta}_{s t}(\q) -\omega^2(\q) \right | = 0.
\label{eq:secular-frequency(q)} \end{equation} 

Translational invariance can be alternatively stated in this context
by saying that a lattice distortion of wave-vector $\q$ does not
induce a force response in the crystal at wave-vector $\q' \ne \q$, in
agreement with the analysis carried out in Sec. \secref{monochromatic}. 
Because of this property, IFC's are most easily calculated in
reciprocal space and, 
when they are needed in direct space, they can be readily obtained by
Fourier transform (see Sec. \secref{force-constants}).

The reciprocal-space expression for the matrix of IFC's,
Eq.\ \eqref{hessian}, is the sum of an electronic and of an 
ionic contribution: \begin{equation} \widetilde C^{\alpha \beta}_{s
t}(\q) =  ^{el}\negthinspace \widetilde C^{\alpha 
\beta}_{s t}(\q) + ^{ion}\negthinspace \widetilde C^{\alpha \beta}_{s
t}(\q), \end{equation} where: \refstepcounter{equation} 
\label{eq:forceconstants} $$\displaylines{ \quad
^{el} \negthinspace \widetilde C^{\alpha \beta}_{st}(\q) = 
{1 \over N_c} \left [ \int \left ( {\partial n(\r)\over \partial  
u^\alpha_s (\q)}\right )^* {\partial V_{ion}(\r) \over \partial
u^\beta_t(\q)} d\r \right. \hfill \cr \hfill \left. + \int n(\r)
{\partial^2 V_{ion}(\r) \over\partial u^{*\alpha}_s(\q) \partial u^\beta
_t(\q)}d\r \right] , \quad (\theequation) } $$
and 
\begin{equation} V_{ion}(\r) = \sum_{ls} v_s(\r-\R_l-\boldtau_s
-\U_s(l)), \label{eq:Vion} \end{equation} $v_s$ being the ionic
(pseudo-) potential corresponding to the $s$-th atomic species.
All the derivatives must be
calculated for $\U_s(\q)=0$.
The ionic contribution comes from the ion-ion interaction
energy (the last term of Eq.\ \eqref{hessian}) and does not
depend on the electronic structure. The explicit expression
of $ ^{ion}\negthinspace \widetilde C^{\alpha \beta}_{st}(\q)$
for periodic systems is given in the appendix.

Using Eqs.\ \eqref{RIofuq} and
\eqref{Vion}, the derivatives of the potential appearing in
Eq.\ \eqref{forceconstants} read: \begin{equation} {\partial
V_{ion}(\r) \over \partial u^\alpha_s(\q)} = -\sum_l{\partial
v_s(\r-\R_l-\boldtau_s) \over \partial \r} \ealla{i\q
\cdot \R_l}, \label{eq:dVion} \end{equation} while the corresponding
derivative of the electron charge-density distribution is given by
Eqs.\ \eqref{linear-q2} and \eqref{dnq}.

\subsubsection{Long wave-length vibrations in polar materials}
\label{sec:polar}

In polar semiconductors and insulators, the long-range character of
the Coulomb forces gives rise to macroscopic electric fields for
longitudinal optic (LO)
phonons in the long wave-length limit. At any finite wave-length,
polar semiconductors are dealt with in the same way as non polar
ones. In the long wave-length limit, however, phonons are coupled to
macroscopic electric fields which must be treated with some care
because the corresponding electronic potential, $V_{\EE}(\r) = e
\E\cdot\r$, is not lattice-periodic (see Sec. \secref{homelfields}). A
physically transparent picture of the coupling between zone-center
phonons and macroscopic electric fields is provided by the Huang's
phenomenological model \cite{Born-Huang}, which we briefly discuss in
the case of a
cubic (or tetrahedral) lattice with two atoms per unit cell. The most
general quadratic expression of the energy as a function of the phonon
optic coordinates, $\bf u$, and the {\it electrical degrees of freedom}
(i.e.  the field itself, $\EE$), is:
\begin{equation}
E({\bf u},\EE) = {1\over2} M\omega_0^2u^2 -
{\Omega\over 8\pi}\epsilon_\infty \E^2 - \Z {\bf u}\cdot\EE,
\label{eq:Huang}
\end{equation}
where $M$ is the nuclear reduced mass, $\Omega$ is the volume of the unit cell,
$\epsilon_\infty$ the electronic dielectric constant of the crystal (i.e. the
static dielectric constant with clamped nuclei, ${\bf u}=0$), and the
coupling between {\bf u} \ and \EE, $\Z$, is known as the {\it Born effective
charge} of the ions (see  {\em e.g.}, \citeasnoun{Boettger},
Sect. 1.5). The variables which are conjugate to {\bf u} and
\EE\ are the 
force {\bf F} acting on the ions and the electrical induction, \DD:
\begin{eqnarray}
{\bf F} & \equiv & -{\partial{ E} \over
\partial {\bf u}} = -M\omega_0^2 {\bf u} + \Z \EE \label{eq:F} \\
{\DD} & \equiv & -{4\pi\over\Omega}{\partial{ E}\over \partial \EE} =
{4\pi\over \Omega} \Z{\bf u} + \epsilon_\infty\EE. \label{eq:D}
\end{eqnarray}
In the absence of free external charges, Maxwell equations give: 
\begin{eqnarray}
\mbox{rot}~    \EE & \sim & i\q\times\EE  = 0  \label{eq:Maxwell-a} \\
\mbox{div}~{\DD}& \sim & i\q\cdot {\DD} = 0. \label{eq:Maxwell-b} 
\end{eqnarray}
For transverse modes ($\EE\perp\q$), Eq.\ \eqref{Maxwell-a} gives 
$\EE_T=0$, and Eq.\ \eqref{F} ${\bf F}_T= -M\omega_0^2{\bf u}$:
the transverse frequency is therefore $\omega_T=\omega_0$. For longitudinal
modes ($\EE\parallel\q$), Eq.\ \eqref{Maxwell-b} gives ${\DD}_L=0$ 
and Eq.\ \eqref{D} gives $\EE_L=-{4\pi
\Z \over \Omega\epsilon_\infty}{\bf u} $; 
Eq.\ \eqref{F} gives
${\bf F}_L=
-\left (M\omega_0^2 + {4\pi \Z \over \Omega\epsilon_\infty}\right ){\bf u}$:
the longitudinal frequency is therefore $\omega_L=\sqrt{\omega_0^2+{4\pi
\Z \over \Omega\epsilon_\infty M}}$. 
These results, which are {\it exact} in the case of cubic and tetrahedral
systems, can be easily generalized to crystals of arbitrary symmetry
\cite{Boettger}.

The first-principles calculation of $\epsilon_\infty$ and $\Z$ proceeds
from Eqs.\ \eqref{F} and \eqref{D}. Let us start for instance from 
Eq.\ \eqref{D}
which---expressed in terms of the macroscopic electric polarization of the
medium and generalized to the case of many atoms per cell---reads: 
\begin{equation}
{\PP} =
{1\over\Omega} \sum_s \Z_s{\bf u}_s + {\epsilon_\infty-1 \over 4\pi} \EE
.\label{eq:Polarization}
\end{equation}
In the general, low-symmetry case, Eq.\ \eqref{Polarization} must be
read as a tensor equation stating that the Born
effective-charge tensor of the $s$-th ion is 
the partial derivative of the macroscopic polarization
with respect to a periodic displacement of all the ions of the
$s$-species at zero macroscopic electric field: \begin{equation}
\Z^{\alpha \beta}_s = \Omega \left . {\partial \P_\alpha \over \partial
u^\beta_s(\q=0)} \right |_{\EE=0}; \end{equation} while the electronic
dielectric-constant tensor is the derivative of the polarization with
respect to the macroscopic electric field at clamped nuclei:
\begin{equation} \epsilon^{\alpha \beta}_\infty = \delta_{\alpha\beta}
+ 4\pi \left . {\partial \P_\alpha \over \partial \E_\beta} \right |
_{\U_s(\q=0) = 0} . \end{equation}

In the long wave-length limit, the matrix of the force constants can
be split into the sum of an analytic and a non analytic contribution
\cite{Born-Huang,Cochran}: \begin{equation} \widetilde C^{\alpha
\beta}_{st} =  ^{an} \negthinspace \widetilde C^{\alpha \beta}_{st} + 
^{na} \negthinspace \widetilde C^{\alpha \beta}_{st}, \label{eq:DX}
\end{equation} where the analytic part, $^{an} \negthinspace
\widetilde C$, is the matrix obtained from the response to a
zone-center phonon, calculated at zero macroscopic 
electric field. The non analytic part has the general form
\cite{Cochran}: \begin{eqnarray} ^{na} \negthinspace
\widetilde C^{\alpha\beta}_{st} &=& {4 \pi \over\Omega} {{\sum_\gamma
\Z^{\gamma \alpha }_s q_\gamma} {\sum_\nu \Z^{\nu\beta }_t q_\nu}
\over \sum_{\gamma,\nu} q_\gamma \epsilon_\infty^{\gamma \nu} q_\nu }
\nonumber \\ &=& {4 \pi \over \Omega} { \left (\q\cdot {\bf \Z}_s
\right )_\alpha \left (\q\cdot{\bf \Z}_t \right )_\beta \over \q \cdot
\pmb{$\epsilon$}^\infty \cdot\q }. \label{eq:CINA} \end{eqnarray}
Eq.\ \eqref{CINA} shows that all the information necessary to deal with
the non analytic part of the dynamical matrix is contained in the
macroscopic dielectric constant of the system, and in the Born
effective charges $\Z$, whereas the analytic contribution can be
calculated just ignoring any macroscopic polarization associated to
the phonon. All these quantities can be easily obtained within DFPT
\cite{bulkoni}.

It is worth mentioning that effective charges can be
calculated  using an approach to
the electrostatics of quantum dielectrics based on topological
concepts, the {\em Berry's phase} approach to macroscopic
polarization \cite{BerryPhase,P}. When used at the same level 
of accuracy, the linear-response and Berry's phase approaches 
yield the same results within numerical uncertainties.

\subsubsection{Inter-atomic force constants}
\label{sec:force-constants} 

The considerations developed so far in the present section allow in
principle to calculate the vibrational frequencies at any (finite or
infinite) phonon wave-length. Phonon frequencies are usually rather
smooth functions of the wave-vector, so that suitable interpolation
techniques can be used when complete dispersions are needed. Simple
concepts from (discrete) Fourier analysis show that the smoother the
phonon dispersions ({\em 
i.e.} the smoother the matrix elements of $\widetilde C$ as functions
of $\q$), the shorter is the range of real-space IFC's: 
\begin{equation} C^{\alpha \beta}_{st}(\R) = 
{1 \over N_c} \sum_\q e^{i\q\cdot \R} \widetilde C^{\alpha \beta}_{st} (\q),
\label{Cq} \end{equation} {\em i.e.} the smaller the number of their
non-vanishing values (to any given accuracy). Real-space IFC's
can thus be readily obtained by Fourier analyzing
a set of force-constant matrices calculated and tabulated over a
uniform grid of points in reciprocal space. The most efficient way of
calculating all these Fourier transforms numerically is the {\em fast
Fourier transform} (FFT) technique (see {\em e.g.} \citeasnoun{NumRec}). 
Once real-space IFC's have been thus obtained, dynamical matrices in 
reciprocal-space (and, hence, vibrational frequencies) can be obtained
at {\em any} wave-vector (not necessarily contained in the original
grid) by FFT. The shorter the range of
real-space force constants, the coarser will be the reciprocal-space
grid needed for such Fourier interpolation. In practice, the size of the
reciprocal-space grid will be assessed a posteriori by verifying that
it yields vanishing real-space constants (to within a given accuracy)
beyond some cutoff radius. A simple rule of thumb is to include in 
the FFT grid enough points in the BZ so as to reach neighbor
interactions extending up to 2-3 bond-lengths, and check the accuracy of
the interpolation against the full calculation on some points {\it not}
included in the grid. 

The above considerations apply as such to metals, away from Kohn
anomalies, and to non-polar insulators. The presence of Kohn
anomalies in metals is associated to long-range IFC's propagating along
the direction of the wave-vector of the anomaly. Catching the details
of the anomaly with a regular grid of wave-vectors in the BZ would be
very impractical. In these cases, once the position of the anomaly has
been located, it is much simpler to refine the grid locally just
around the anomaly.

In polar materials, instead, real-space IFC's are long-ranged in all
directions, as a consequence of the long-range dipole-dipole
interaction between ionic effective charges. For this reason, Fourier
interpolation would be inefficient in this case. The
dipole-dipole interaction is precisely the physical origin of the
non-analytic behavior of the reciprocal-space dynamical matrices in
the long wave-length limit, whose form is however known exactly in
terms of the ionic effective charges (Eq.\ \eqref{CINA}). Eq.\
\eqref{CINA} expresses the long wave-length limit of the
reciprocal-space force constants of {\em any} system whose atoms carry
a charge equal to $\Z_s$. If the force constants of a system of
point charges $\Z_s$ is subtracted from those
calculated for the physical system under consideration, the resulting
difference will be analytic in the long wave-length limit, and its
Fourier transform short ranged. For polar materials, Fourier
interpolation is thus efficiently applicable to the analytic
contribution to the reciprocal-space force constants, whereas the full
non-analytic behavior can be easily restored by adding the force
constants of a suitable point-charge model \cite{bulkoni}.  A
description of the technical details necessary to implement the
Fourier interpolation of reciprocal-space force constants in the
general case of materials with anisotropic effective charges can be found in
\citeasnoun{GonzeFC}.
  \subsection{Homogeneous deformations}
    \subsubsection{Elastic properties}

Elastic constants can be viewed as force constants
associated to homogeneous strains, i.e. to {\it macroscopic} distortions
of the crystal. In any finite system, there is no conceptual difference
between a strain and a microscopic distortion, and linear-response
techniques are straightforwardly applicable in both cases. 
In an infinite system, on the contrary, one cannot apply directly
linear-response techniques, because a homogeneous strain changes the
boundary conditions of the Hamiltonian.
The use of perturbation theory requires instead the existence of
a common basis set for the perturbed and unperturbed systems.
In order to use perturbation theory for homogeneous deformations, it
has been suggested that
one can introduce an intermediate fictitious Hamiltonian which is
related to the unperturbed one by an unitary transformation, 
and obeys the same boundary conditions as the strained Hamiltonian
\cite{BGTel}. 

Let us consider for simplicity a system under an isotropic strain 
(dilatation) of amplitude $\alpha$, $\{\R\} \rightarrow \{\alpha\R\}$,
whose corresponding elastic constant is 
the bulk modulus. The KS Hamiltonian, $H^{\alpha}$, for the strained
crystal is given by
\begin{eqnarray}
   H^{\alpha}_{\SCF} & =& -{\hbar^2\over 2m} {\partial^2\over\partial \r^2} +
   V^{\alpha}_{ion}(\r) \nonumber \\
   & & \mbox{} + e^2 \int {n^{\alpha}(\r')
                   \over{\mid\r-\r' \mid } } d\r' +
   v_{xc}\bigl ( n^{\alpha}(\r) \bigr ) ,
\end{eqnarray}
where the electron-ion potential, $V_{ion}^{\alpha}$, is
\begin{equation}
V_{ion}^{\alpha}(\r) = \sum_{ls} 
v_{s}(\r-\alpha{\bf R}_l-\alpha\boldtau_s). \label{eq:vstrained}
\end{equation}

The intermediate (fictitious) strained Hamiltonian,
$\tilde H^{\alpha}$, is obtained from
the unperturbed one, $H$, 
through a scale transformation: 
\begin{equation}
\tilde H^{\alpha}(\r, \partial / \partial \r) =
           H_{\SCF}(\r / \alpha ,\alpha \partial / \partial \r).
\end{equation}
$\tilde H^{\alpha}$ obeys the same boundary conditions as the physical
strained Hamiltonian $H^{\alpha}$, and hence perturbation theory can
be used to calculate the relative energy difference. At the same time
$\tilde H^{\alpha}$ and 
$H$ differ by a unitary transformation and their spectra
are trivially related:
\begin{eqnarray}
\tilde \epsilon^{\alpha}_{n} & = & \epsilon_{n}
\nonumber \\
\tilde \psi^{\alpha}_{n}(\r) & = &
   \alpha^{-{3 \over 2}} \psi_{n}(\r / \alpha) \\
\tilde n^{\alpha}(\r) & = &\alpha^{-3}n(\r / \alpha).
\nonumber
\end{eqnarray}

The energy change due to a strain can thus be computed in two steps: 
first one calculates the energy difference between the unperturbed
crystal and the fictitious strained one described by 
$\tilde H^{\alpha}$; one then computes the energy difference
between the latter and the physical strained system using perturbation
theory. The first step is trivial. The second step is less so,
because $\tilde H^{\alpha}$ is not a proper Kohn-Sham Hamiltonian:
the Hartree and XC terms in $\tilde H^{\alpha}$ are {\it not} the
Hartree and XC potentials generated by $\tilde n^{\alpha}$. One can
write $\tilde H^{\alpha}$ as a genuine Kohn-Sham Hamiltonian, 
provided that the definition of the external potential is changed:
\begin{eqnarray}
   \tilde H^{\alpha}& =& -{\hbar^2\alpha^{2}\over 2m}
                         {\partial^2\over\partial \r^2} +
   \tilde V^{\alpha}_{ext}(\r) \nonumber \\
   & &\mbox{} + e^2 \int {\tilde n^{\alpha}(\r')
                   \over{\mid\r-\r' \mid } } d\r' +
   v_{xc}\bigl ( \tilde n^{\alpha}(\r) \bigr ) ,
\end{eqnarray}
where
\begin{eqnarray}
   \tilde V^{\alpha}_{ext}(\r) & = &
   \sum_{ls} v_{s}(\r/\alpha-\R_l-\boldtau_s)\nonumber\\
    & & \mbox{} + e^2\left(1-{1\over \alpha}\right)
   \int {n(\r')\over{\mid\r/ \alpha -\r' \mid }} d\r' \nonumber \\
    & & \mbox{} +
   v_{xc}\bigl ( \alpha^{-3} n(\r/ \alpha) \bigr ) +
   v_{xc}\bigl ( n(\r/ \alpha) \bigr ) .
\end{eqnarray}
The energy difference between the fictitious and the real strained
systems, $E^{\alpha}- \tilde E^{\alpha}$, can now be calculated by
perturbation. The
details are given in \citeasnoun{BGTel}.
This method can be straightforwardly extended to generic elastic
constants. The algebra involved is however quite heavy. 

It must be
remarked that what is calculated in this way is only a {\it bare}
elastic constants: in Eq.\ \eqref{vstrained} the coordinates of
different atoms within a same unit cell are assumed to undergo the
same homogeneous scale transformation as the positions of different
unit cells. This is not true in general, and atoms rearrange themselves
within each unit cell, so as to minimize the total energy as a
function of the applied strain. To see how the elastic constants are
affected by the internal relaxation of the atomic positions, let us
write
the more general second-order expression of the crystal energy per 
unit cell as a second-order polynomial of the macroscopic (strain) 
and microscopic (phonon) modes: 
\begin{eqnarray}
  E &=&{\Omega\over 2} \sum_{\alpha\beta,\gamma\delta}
                   \lambda^0_{\alpha\beta,\gamma\delta}
                   \; e_{\alpha\beta} \; e_{\gamma\delta} 
    \nonumber \\ & & \mbox{}
     + {1\over 2} \sum_{s\alpha,t\beta}
                   C_{st}^{\alpha\beta}
                  \; u_s^\alpha \; u_t^\beta
     +       \Omega \sum_{s\alpha\gamma\delta}
                   \zeta_{s\alpha\gamma\delta} 
                  \; u_s^\alpha \; e_{\gamma\delta}
\end{eqnarray}
where $\Omega$ is the unit cell volume, $\lambda^0$ is the (bare) 
elastic constant matrix, $e$ the strain tensor, $C$ the zone-center 
reciprocal-space matrix of force constants, $u$ atomic displacements 
in the unit cell, $\zeta$ the coupling between macroscopic strain and 
atomic displacements (the matrix of the so-called {\em internal-strain 
parameters}). Crystal symmetry determines the number of independent 
non-vanishing terms in $\lambda^0$, $C$, and $\zeta$. If we allow the 
atoms to relax for a given strain state, minimization of the energy 
with respect to the $u$'s yields 
\begin{equation}
   E = {\Omega\over 2} \sum_{\alpha\beta,\gamma\delta}
                   \lambda_{\alpha\beta,\gamma\delta}
                   \; e_{\alpha\beta} \; e_{\gamma\delta},
\end{equation}
where
\begin{equation}
     \lambda_{\alpha\beta,\gamma\delta} = 
            \lambda^0_{\alpha\beta,\gamma\delta} - \Omega \sum_{s\mu,t\nu}
                   \zeta_{s\mu\alpha\beta} 
                   (C^{-1})_{st}^{\mu\nu}
                   \zeta_{t\nu\gamma\delta}.
\end{equation}
Both the force constants and the coupling between macroscopic strain
and atomic displacements can be readily calculated within DFPT.

\subsubsection{Piezo-electric properties}

The piezo-electric constants form a third-order tensor,
$\gamma_{\alpha,\gamma\delta}$, defined as the derivative of the
macroscopic electric
polarization with respect to a homogeneous strain, at vanishing macroscopic
field. This quantity---which has been demonstrated to be independent
of surface effects \cite{piezo}, {\it i.e.} independent on surface
termination---could in principle be evaluated as the electric-polarization
response to a given applied strain. Alternatively, and somewhat more
conveniently, it can also be calculated as the stress linearly induced by
an electric field at zero strain. The latter definition was used by 
\citeasnoun{piezoSdG} to calculate the piezo-electric tensor in III-V 
semiconductor compounds. The two definitions are equivalent
and can be deduced expanding the energy of the system to second order
in the macroscopic perturbations (strain, $e_{\alpha\beta}$, and electric
field, $\E_{\alpha}$):
\begin{eqnarray}
   E &=& {\Omega\over 2} \sum_{\alpha\beta,\gamma\delta}
                   \lambda_{\alpha\beta,\gamma\delta}
                   \; e_{\alpha\beta} \; e_{\gamma\delta}
   \nonumber \\ & & \mbox{}
                - \Omega \sum_{\alpha,\gamma\delta} 
                  \gamma_{\alpha,\gamma\delta}
                  \; \E_{\alpha} \; e_{\gamma\delta} -
                  {\Omega \over 8 \pi} \sum_{\alpha,\beta}
                   \epsilon_{0}^{\alpha\beta} \; \E_{\alpha}  \; \E_{\beta}. 
\label{eq:macropiezo}
\end{eqnarray}
The link with the microscopic description is provided, 
by expanding the crystal energy to second-order in the external
(strain and electric field) and internal (phonon) degrees of freedom:
\begin{eqnarray}
   E & = & {\Omega\over 2} \sum_{\alpha\beta,\gamma\delta}
                   \lambda^0_{\alpha\beta,\gamma\delta}
                   \; e_{\alpha\beta} \; e_{\gamma\delta} -
                 \Omega  \sum_{\alpha,\gamma\delta} 
                  \gamma_{\alpha,\gamma\delta}
                  \; \E_{\alpha} \; e_{\gamma\delta} 
   \nonumber \\ & & \mbox{}
             - {\Omega \over 8 \pi} \sum_{\alpha,\beta}
               \epsilon_{\infty}^{\alpha\beta} \; \E_{\alpha}  \; \E_{\beta} 
     + {1\over 2} \sum_{s\alpha,t\beta}
                   C_{st}^{\alpha\beta}
                  \; u_s^\alpha \; u_t^\beta
    \nonumber \\ & & \mbox{}
     +            \sum_{s\alpha\gamma\delta}
                   \zeta_{s\alpha\gamma\delta} 
                  \; u_s^\alpha \; e_{\gamma\delta}
     -    \Omega \sum_{s\alpha\beta} \Z_s^{\alpha\beta} u_s^\alpha \E_\beta.
\label{eq:micropiezo}
\end{eqnarray}
where $\lambda^0$, $\gamma^0$, and $\epsilon_\infty$ are the purely
electronic ({\em i.e.} clamped-ion) contributions to the elastic,
piezoelectric, and dielectric, constants respectively, and the
coupling between atomic displacements and macroscopic variables
(electric field and strain), are expressed by the effective charges,
$\Z$, and internal strain parameters, $\zeta$.  Once the macroscopic
variables, $\E$ and $e$, are fixed, The equilibrium values of the
internal degrees of freedom are given by the condition that the
derivatives of Eq. \eqref{micropiezo} with respect to the atomic
displacements vanish. When these equilibrium atomic positions as
functions of the macroscopic electric field and strain are inserted in
\eqref{micropiezo}, the resulting expression defines the total
piezoelectric constants as:
\begin{equation}
   \gamma_{\alpha, \gamma\delta} =   \gamma^0_{\alpha, \gamma\delta}
                 + \sum_{s\mu,t\nu} \Z_s^{\mu\alpha} (C^{-1})_{st}^{\mu\nu}
                   \zeta_{t\nu\gamma\delta}.
\end{equation}
The two resulting contributions to the piezoelectric constants are
often of opposite sign and close in absolute value, so that a well
converged calculation is needed in order to extract a reliable value
for their sum \cite{piezoSdG}.

The problem of a proper definition of piezoelectric properties in a
crystal displaying a spontaneous macroscopic polarization has been
raised recently by \citeasnoun{properpiezo1} and further discussed by
\citeasnoun{properpiezo2}.
\subsection{Higher-order responses}
\label{sec:higher}
    \subsubsection{The {\it 2n+1} theorem}

In Sec.~\secref{linear-response} we have seen that 
the knowledge of the first-order derivatives of the wave-functions is 
enough to calculate the second-order derivatives of the total energy.
This a special case of a very general theorem, known as the
{\it $2n+1$ theorem}, which states that the knowledge of the 
derivatives of the wave-functions up to order $n$ allows 
the calculation of the derivatives of the energy up to
order $2n+1$. This theorem, well known in quantum mechanics 
since many years, is a consequence of the variational principle and
it is valid also in DFT. In this context, its usefulness derives 
from the fact that the third-order derivatives of the total energy 
can be obtained from the first-order derivatives of the wave-functions. 
This opens the possibility to study phenomena which depends 
upon third order anharmonic terms in the energy expansion---such as
phonon line widths, Raman scattering cross sections, or nonlinear
optical responses---with a computational effort of the same order as
for harmonic properties (because the time-consuming step is 
the calculation of the first-order derivatives of
the wave-functions).

Several proofs of the $2n+1$ theorem can be found in the literature. 
In a DFT framework this theorem was first proved by
~\citeasnoun{2n+1}. 
Explicit expressions of the energy derivatives up to fourth order 
have been worked out by \citeasnoun{Gonze2}. The scope of this theorem
is much more general, as it concerns the properties of the extrema
of any functional depending upon some parameters \cite{Epstein}.
Let
$E[{\psib},\lambda]$ be a generic functional of $\psib$ which, for
$\lambda=0$, 
has an extremum at $\psib^{(0)}$: ${\delta E[\psib^{(0)},0] /
\delta \psib} = 0$. The position of the extremum will depend on
the value of the parameter $\lambda$: $\psib(\lambda) = \psib^{(0)} +
\Delta \psib(\lambda)$. The value of the functional at the extremum
will be:
\begin{equation}
E_{min}(\lambda) = E[{\psib}^{(0)}+\Delta {\psib}(\lambda), \lambda],
\end{equation} where $\Delta\psib(\lambda)$ is determined, for any
given value
of $\lambda$, by the extremum condition:
\begin{equation}
{\partial E[{\psib}^{(0)}+\Delta {\psib}, 
\lambda] \over
\partial \Delta {\psib} }= 0.
\end{equation} When $\lambda$ is small, both $\Delta\psib(\lambda)$
and 
$E_{min}(\lambda)$ will be well approximated by their Taylor expansion
in powers of $\lambda$:
\begin{eqnarray}
\Delta {\psib}(\lambda) &=&  
\sum_{l=1}^\infty {1\over l!} {d^l {\psib} \over d \lambda^l}  \lambda^l
\equiv \sum_{l=1}^\infty \Delta {\psib}^{(l)} \lambda^l,
\label{eq:qua1} \\
E_{min}(\lambda) &=&  
\sum_{l=0}^\infty {1\over l!} {d^l {E_{min}} \over d \lambda^l}  \lambda^l
\equiv \sum_{l=0}^\infty E^{(l)} \lambda^l.
\label{eq:qua2} \end{eqnarray} The $2n+1$ theorem states that the
knowledge of $\Delta\psib^{(l)}$ up to order $n$ is enough to
determine $E^{(l)}$ up to order $2n+1$. To demonstrate this, 
it is
convenient to first expand $E[{\psib}^{(0)}
+\Delta {\psib}, \lambda]$ into a 
Taylor series treating $\Delta {\psib}$ and $\lambda$
as independent variables:
\begin{eqnarray}
E[{\psib}^{(0)} &&+ \Delta {\psib}, \lambda]= \nonumber \\  
&& \sum_{p=0}^\infty \sum_{k=0}^\infty {1\over k! p!}
{\delta^{k+p}  E[{\psib}^{(0)}, 0]
\over \delta {\psib}^k  \delta \lambda^p}
{(\Delta {\psib})^k} {\lambda^p},\label{eq:due}
\end{eqnarray}
where we use the notation:
\begin{equation}
 {\delta^k E \over \delta {\psib}^k }(\Delta {\psib})^k =
\big( \sum_i \Delta \psi_i {\partial \over \partial \psi_i} \big)^k E.
\end{equation}
Variation of Eq.\ \eqref{due} with respect to $\Delta\psib$ leads to
the extremum condition:
\begin{eqnarray}
{\bf f} &=& {\partial E[{\psib}^{(0)}+\Delta {\psib}, 
\lambda] \over
\partial \Delta {\psib} } \nonumber \\
&=& \sum_{p=0}^\infty \sum_{k=1}^\infty {1\over (k-1)! p!}
{\delta^{k+p} E[{\psib}^{(0)}, 0] \over
\delta {\psib}^{k} \delta \lambda^p }
{(\Delta {\psib})^{k-1}  } { \lambda^p} \nonumber \\
&=& 0. \label{eq:tre} 
\end{eqnarray}
Introducing Eq.\ \eqref{qua1} into Eq.\ \eqref{tre} allows to formally
expand $\bf f$ as a power series of $\lambda$ only. The resulting
expression for $\bf f$ must vanish identically order by order in
$\lambda$. By equating the coefficients of $\lambda^n$ in the Taylor
expansion to zero, Eq.\ \eqref{tre}
yields an infinite number of equations, ${\bf
f}^{(n)}=0$, which determine $\Delta\psib^{(n)}$.

Introducing Eq.\ \eqref{qua1} into Eq.\ \eqref{due} allows to formally
expand $E$ as a power series of $\lambda$ only and therefore to
calculate the explicit form of $E^{(l)}$.
Since a term quadratic in ${\Delta \psib}^{(l)} \lambda^l$ is of order
$\lambda^{2l}$, $E^{(2n+1)}$ could depend only linearly upon
${\Delta \psib}^{(l)} \lambda^l $, if $l>n$.
Showing that for $l>n$ the coefficient of ${\Delta \psib}^{(l)}
\lambda^l$ in $E^{(2n+1)}$ is ${\bf f}^{(2n+1-l)}$ 
we prove the $2n+1$ theorem, since this force is zero by the minimum 
condition. 
We start by extracting ${\Delta \psib}^{(l)}\lambda^l$
from the product $(\Delta { \psib})^{k}$ appearing in Eq.\ \eqref{due},
using the relationship:
\begin{eqnarray}
\label{eq:nuova}
&(\Delta {\psib})^k& = (\Delta {\psib}-{\Delta \psib}^{(l)} \lambda^l
+ {\Delta \psib}^{(l)}\lambda^l)^k=  \\
&\ & k {\Delta \psib}^{(l)}\lambda^l (\Delta { \psib})^{k-1} +
(\Delta {\psib}-{\Delta \psib}^{(l)}\lambda^l)^k+{\cal O}\big(\lambda^{2n+1}
\big),
\nonumber
\end{eqnarray}
valid for $l>n$. The only term which is linear in 
${\Delta \psib}^{(l)}\lambda^l$
is the first term of the r.h.s. of Eq.\ \eqref{nuova}.
Inserting this term in Eq.\ \eqref{due} and recalling the
definition of ${\bf f}$, Eq.\ \eqref{tre}, we can write
$E^{(2n+1)}$ as:
\begin{eqnarray}
\label{eq:nuova1}
E^{(2n+1)}&=&{\Delta \psib}^{(2n+1)}\lambda^{2n+1} {\bf f}^{(0)}+\ldots
\\
&+&
{\Delta \psib}^{(l)}\lambda^l{\bf f}^{(2n+1-l)}+
\ldots \nonumber \\ 
&+&{\Delta \psib}^{(n+1)}\lambda^{(n+1)}{\bf f}^{(n)}+ P^{(2n+1)}
({\psib}^{(1)},\ldots,{\psib}^{(n)}),
\nonumber
\end{eqnarray}
where $P^{(2n+1)}$ is a polynomial of degree $2n+1$. From the condition
 ${\bf f}^{(i)}=0$ for every $i$, we get:
\begin{equation}
E^{(2n+1)}=
 P^{(2n+1)}({\psib}^{(1)},\ldots,{\psib}^{(n)}),
\label{eq:teorema}
\end{equation}
that proves the $2n+1$ theorem.

In order to apply this theorem to DFT we can take as
${\psib}$ a vector whose elements are the
coefficients of all the occupied wave-functions $\{\psi_i\}$ on a given basis
and $\lambda$ as a parameter measuring the magnitude of 
the perturbation. The orthogonality constraint can be dealt with  
as shown for instance by \citeasnoun{Mauri}, writing the total 
energy functional of DFT for non ortho-normalized orbitals.
The demonstration presented here, applied to this functional,
provides high-order derivatives of the DFT energy.
Note that the ${\psib}$ are arbitrary linear combinations of the 
occupied eigenstates of the Hamiltonian, which are only required 
to minimize the total energy functional. For instance, 
in a crystalline solid ${\psib}$ could represent 
Wannier functions if they are used instead of Bloch functions 
to describe the electronic states. 
In alternative to the path followed here, the $2n+1$ theorem 
can be demonstrated also for constrained functionals,
with Lagrange multipliers used to impose the orthogonality
of the orbitals~\cite{Gonze1}. 

\subsubsection{Nonlinear susceptibilities}

Within DFT, the third-order derivatives of the energy,
Eq.\ \eqref{hk-energy}, are \cite{2n+1}: 
\refstepcounter{equation} \label{eq:etre} 
$$\displaylines{
\quad
E^{(3)}= 2\sum_{n=1}^{N/2} \langle \Delta \psi_n | 
\Delta V_{\SCF}-\Delta
\epsilon_n | \Delta \psi_n \rangle \hfill \cr \mbox{} + 
2\sum_{n=1}^{N/2} \left( \left \langle \Delta \psi_n \left | 
{\partial^2 V \over \partial \lambda^2} 
\right |\psi_n \right \rangle+c.c.\right) \cr
\mbox{} + 2\sum_{n=1}^{N/2} \left \langle \psi_n \left | 
{\partial^3 V \over \partial \lambda^3} 
\right |\psi_n \right \rangle + 
{1\over 6} \int K^{(3)}({\bf r}_1,{\bf r}_2,{\bf r}_3) \cr 
\hfill\times \Delta n({\bf r}_1) \Delta n({\bf r}_2) 
\Delta n({\bf r}_3) d{\bf r}_1 d{\bf r}_2 d{\bf r}_3,
\quad (\theequation)
\label{eq:dft2n1}} $$
where $K^{(3)}$ is the third order functional derivative of the 
exchange and correlation energy
with respect to the density. In this section, 
we have dropped the indication of the order in $\lambda$ from 
$\Delta \psi_n$ which is always the first order term.

The solution of Eq.\ \eqref{linear-2} yields the projection on 
the conduction manifold of the first order change of the wave-functions.
Therefore we need to recast Eq.\ \eqref{dft2n1} in a 
form which does not depend on the projection 
of $| \Delta \psi_i \rangle $ on the valence manifold.
This expression for $E^{(3)}$ exists
since the total energy functional 
is invariant with respect to a unitary transformation within the 
manifold of the occupied orbitals.
The required transformation  has been carried out 
by~\citeasnoun{anharm} and by \citeasnoun{nonlinear}.
After some algebra one obtains: \refstepcounter{equation}
\label{eq:etretr} $$\displaylines{
\quad E^{(3)}= 2\sum_{n=1}^{N/2} \langle \Delta \psi_n |P_c
\Delta V_{\SCF} P_c| \Delta \psi_n \rangle  \hfill \cr \mbox{}
- 2\sum_{n,m=1}^{N/2} \langle \Delta \psi_n |P_c| \Delta \psi_m \rangle
\langle \psi_m| \Delta V_{\SCF}|\psi_n \rangle \cr \mbox{}
+ 2\sum_{n=1}^{N/2} \left( \left \langle \Delta \psi_n \left | P_c
{\partial^2 V \over \partial \lambda^2}
\right |\psi_n \right \rangle+ c.c.\right) \cr \mbox{} +
2\sum_{n=1}^{N/2} \left \langle \psi_n \left | {\partial^3 V \over \partial
\lambda^3} \right
|\psi_n \right \rangle +
{1\over 6} \int K^{(3)}({\bf r}_1,{\bf r}_2,{\bf r}_3) \cr
\hfill \times \Delta n({\bf r}_1) \Delta n({\bf r}_2) 
\Delta n({\bf r}_3)  d{\bf r}_1 d{\bf r}_2 d{\bf
r}_3. \quad(\theequation) } $$
When the perturbation $\lambda$ is an atomic displacement,
as in Sec.~\secref{phonons}, $E^{(3)}$ gives the first anharmonic 
corrections to the energy. These corrections are responsible,
for instance, for the decay 
of the phonon modes into vibrations of lower frequency.
The line width of the phonon lines in Raman scattering, 
after subtraction of isotopic and inhomogeneous 
broadening, is proportional to $E^{(3)}$ if higher-order
processes are neglected. The comparison of theoretical and 
experimental values is described in Sec.~\secref{anharmonic}.
The generalization of Eq.\ \eqref{etretr} to metals can be done by the
techniques introduced in Sec.~\secref{metals}.
The first application has been recently presented~\cite{lazzeri}.

When the perturbation $\lambda$ is an electric field,
as in Sec.~\secref{homelfields}, $E^{(3)}$ is proportional to
the nonlinear optical susceptibilities of a material at low frequency.
Unfortunately, in this case, Eq.\ \eqref{etretr} cannot be directly
used to compute $E^{(3)}$.
In fact it contains a term $\langle \psi_m| \Delta^{\EE}
V_{\SCF}|\psi_n\rangle$ that becomes ill defined when $n=m$
(Eq.\ \eqref{Commutator} cannot be applied: the energy denominator 
vanishes).
The same problem arises also when Eq.\ \eqref{etretr} is generalized
to mixed third order derivatives of the energy and
one perturbation is an electric field. These derivatives
allows one to account, theoretically, for 
the intensities of non-resonant Raman lines~\cite{raman}
or for nonlinear infrared absorption.

A solution to this problem has been proposed by~\citeasnoun{nonlinear},
switching to a Wannier representation for the electronic orbitals,
and applying the $2n+1$ theorem to a total energy functional
for a periodic insulating solid in a finite electric field originally
proposed by \citeasnoun{Nunes}. This functional,
which exploits the properties of localized orbitals, is written as:
\refstepcounter{equation} \label{eq:uno} $$\displaylines{\quad
E\left[\{w_{0,m}\}, \E_0 \right] 
= \sum_l \sum_{m,n=1}^N \langle w_{0,m} | H +
e \E_0 x | w_{l,n} \rangle \hfill \cr \hfill \times
\bigl ( 2 \delta_{0m,ln} - \langle w_{l,n} | w_{0,m} \rangle
\bigr ), 
\quad (\theequation) } $$
where $H$ is the unperturbed Hamiltonian of the solid,
$\E_0$ is the electric field, $x$ is the position operator, $e$ is the
electron charge, and $\{w_{l,m}\}$ are functions---in
general non-orthonormal---localized around
the unit cell identified by the Bravais lattice vector, $R_l$.
The function $w_{l,m}$ is obtained by translating
the function centered at the origin by a vector $R_l$: $w_{l,m}(x)= 
w_{0,m}(x-R_l)$. In practical applications, 
$w_{0,m}$ is constrained to vanish outside
a localization region of radius $R_c$ centered
at the origin. For simplicity in Eq.\ \eqref{uno} we have limited
ourselves to one-dimensional systems of
non interacting electrons. The electron-electron interaction---when
dealt with within any self-consistent-field scheme---does not yield
any additional problems.
We stress here that the expectation value of $x$ is well defined for 
any finite cut-off radius $R_c$.
Furthermore we note that even if no orthogonality constraints are
imposed on the $w_{l,m}$'s, at the minimum they become 
approximately orthonormal as shown by~\citeasnoun{Mauri}.

In analogy with Eq.\ \eqref{etretr} one can show that the
nonlinear optical susceptibility $\chi^{(2)}$ is given by:
\refstepcounter{equation} \label{eq:chi2p} 
$$\displaylines{
\quad\chi^{(2)} = -3 E^{(3)}/\E_0^3= \hfill \cr
-{3\over \E_0^2} \bigg( \sum_{m=1}^N e\langle w^{(1)}_{0,m}|P_c x P_c
|w^{(1)}_{0,m}\rangle \cr \hfill 
- \sum_{m,n=1}^N \sum_l e \langle w_{0,m}| x
|w_{l,n}\rangle\langle w^{(1)}_{l,n}|P_c|w^{(1)}_{0,m}\rangle\bigg),
\quad (\theequation) } $$
the $w^{(1)}$'s are solutions of a linear system similar to 
Eq.\ \eqref{deltapsiElocf} which can be obtained from the condition 
$\delta E^{(2)}/ \delta w^{(1)}=0$:
\refstepcounter{equation} $$\displaylines{\quad
- P_c e \E_0 x |w_{0,m}\rangle =  H P_c |w^{(1)}_{0,m}\rangle 
\label{perturbazione} \hfill \cr \hfill
-
\sum_{n=1}^N\sum_l
P_c |w^{(1)}_{l,n}\rangle\langle w_{l,n}| H | w_{0,m}\rangle,
\quad (\theequation) } $$
where $P_c = 1 - \sum_{n=1}^N \sum_l| w_{l,n}\rangle
\langle w_{l,n}| $ is the projector on the conduction bands in
the Wannier representation.

Eq.\ \eqref{chi2p} is difficult to implement since it requires a 
electronic structure code which calculates localized Wannier functions.
However, Eq.\ \eqref{chi2p} can be rewritten in terms of Bloch functions.
We recall that the Wannier functions are defined in terms of the
Bloch functions $\psi^k_n(x)$ as:
\begin{equation}
w_{0,n}(x) = { \Omega \over 2 \pi } \int_{BZ} dk\ \psi^k_n(x),
\label{fromBtoW}
\end{equation}
where the integral is done over the first BZ,
$\Omega$ is the dimension of the unit cell, the Bloch functions are
normalized on the unit cell, and $\psi^{k+G}_n(x) = \psi^k_n(x) $;
here $G$ is a reciprocal lattice vector. Inserting this definition in
Eq.\ \eqref{chi2p} and using the relationship:
$x \psi^k_n(x) = - {i\partial\over \partial k} \psi^k_n(x)  + e^{ikx}
{i \partial \over \partial k} u^k_n(x) $,
where $u^k_n(x) = e^{-ikx} \psi^k_n(x) $ are the periodic parts of
the Bloch wave-functions, one finally obtains:
\refstepcounter{equation} 
\label{chi2v} $$ \displaylines{
\quad\chi^{(2)} = 
3 i {e \Omega \over 2 \pi \E_0^2}
\hfill \cr \hfill 
\times \sum_{m,n=1}^N
\int_{BZ} dk \left \langle u^k_m \left | {\partial \over
\partial k} 
\Bigl( |u^k_n
\rangle \langle \tilde u^{k(1)}_{n} | \Bigr) \right |\tilde
u^{k(1)}_{m} \right \rangle, \hfill (\theequation) } $$
where $\tilde u^{k(1)}_{n}=P_c  u^{k(1)}_{n}$.

The formalism, as introduced in this paper, allows one to access 
nonlinear phenomena in a frequency range where one can assume
that the electrons are
in the ground state. The generalization to finite frequencies, 
starting from time dependent DFT has been explored by~\citeasnoun{chi2DFT}. 
\section{Implementations} \label{sec:implem}
\subsection{Plane waves and pseudo-potentials}
\label{sec:implem_pw}

The first---and still today by far the most numerous---implementations
of DFPT were based on the plane-wave (PW) pseudo-potential (PP) method
\cite{Pickett}. 
PW's have many attractive features: they are simple to use,
orthonormal by construction, unbiased by the atomic
positions. Contrary to what
happens with localized (atomic-like) basis sets, it is very 
simple to check for convergence, by just increasing the 
size of the basis set, as given by the
kinetic-energy cutoff.
The FFT algorithm allows to quickly go back and forth from reciprocal 
to real space.
An especially important advantage of PW's is the 
absence of \citeasnoun{Pulay} terms in the calculation 
of energy derivatives. As a consequence the HF
expressions 
for forces and for force constants are valid without any
correction when a PW basis set is used.

PW's are used in conjunction with PP's.
A PP is a fictitious electron-ion interaction
potential, acting on valence electrons only, that mimics the
interaction with the inner electrons---which are supposed to be frozen
in the core---as well as
the effective repulsion exerted by the latter on the former,
due to their mutual orthogonality. Modern {\em norm-conserving} PP's
\cite{HSC} are determined uniquely from the properties of the isolated
atom, while the requirement of {\em norm conservation} ensures
an optimal {\em transferability}. By the latter expression one
indicates the 
ability of the PP's to provide results whose
quality is to a
large extent independent of the local
chemical environment of the individual atoms. Norm-conserving PP's are
angular-momentum dependent ({\em i.e.} they are {\em non-local}
operators) and a special care must be taken to ensure that the atomic
valence (pseudo-) wave-functions associated with them are sufficiently
smooth in the atomic (pseudo-) core, so that they can be efficiently
dealt with using a PW basis set.
Experience has shown that the use of PP's is practically
equivalent to the frozen core approximation within an all-electron
approach.
The pseudo-potential approximation implicitly assumes that the energy
functional is linear with respect to the partition of the total charge
into core and valence contributions.
In some atoms (such as, {\em e.g.}, alkali atoms) the loss of accuracy
due to the neglect of nonlinearity in the exchange-correlation energy
functional can be intolerably high. For such cases the 
{\em non-linear core correction} of \citeasnoun{CoreCorr}
turns out to be very useful.

From a computational point of view, it is very convenient to recast the
angular-momentum dependent
part of PP's into a sum over a few projectors \cite{KB}. 
This is called the {\em separable} form of PP's.
The use of PW's and of separable PP's, together with FFT and iterative 
diagonalization or minimization techniques, 
allows a fast and efficient solution of the KS equations
for systems containing up to hundreds of atoms in the unit cell. 
The technical aspects of the implementation of the KS equations in a
PW-PP framework have been extensively described in the literature 
(see e.g. \citeasnoun{Pickett}, \citeasnoun{Payne}, 
\citeasnoun{Australia}). 

The implementation of DFPT in a PW-PP framework is a 
straightforward extension of  the implementation of the KS 
equations. 
The only modifications to the theory expounded 
in the preceding chapter are related to the non-local 
character of PP's.
In DFT calculations this is accounted for by modifying 
the electron-ion interaction term in the energy functional, 
Eq.\ \eqref{hk-energy}, as follows :
\begin{eqnarray}
E[n(\r)] & = & F[n(\r)] \nonumber \\
&\mbox{} & + 2 \sum_{n=1}^{N/2} \int \psi^*_n(\r) V(\r,\r')\psi_n(\r')d\r d\r',
\end{eqnarray}
where $V(\r,\r')$ is a sum of atomic nonlocal PP's.
The results of Sec.\secref{phonons} must be generalized accordingly.
Eqs.\ \eqref{dE1} and \eqref{dE2} become:
\begin{eqnarray}
 {\partial E \over \partial \lambda_i } = 2 \sum_{n=1}^{N/2}
\left\langle\psi_n \left | {\partial V_\lambda \over \partial\lambda_i }
\right |\psi_n\right\rangle
\end{eqnarray}
and
\begin{eqnarray}
 {\partial^2 E \over \partial \lambda_i \partial \lambda_j} & = &
2 \sum_{n=1}^{N/2}
\Biggl (\left\langle 
    \psi_n\left |
         {\partial^2 V_\lambda\over\partial\lambda_i \partial\lambda_j}
    \right|\psi_n
 \right\rangle\nonumber  \\ & &  \mbox{} +
\left\langle 
    {\partial\psi_n\over \partial \lambda_i}\left |
        {\partial V_\lambda\over\partial\lambda_j}
    \right| \psi_n\right\rangle 
+ c.c.\Biggr ).
\end{eqnarray}

The expression \eqref{forceconstants} for the electronic
contribution to force constants is modified as follows:
\begin{eqnarray}
^{el}\negthinspace\widetilde C^{\alpha\beta}_{st}(\q) & = &
{2 \over N_c} \sum_{n=1}^{N/2} \Biggl (\left\langle 
    {\partial\psi_n\over\partial u^\alpha_s(\q)}\left|
       {\partial V_{ion}\over\partial u^\beta_t(\q)}
    \right|\psi_n
\right\rangle + c.c. \nonumber  \\& & \mbox{} + 
\left\langle 
   \psi_n \left|
       {\partial^2 V_{ion}\over
        \partial u^{*\alpha}_s(\q)\partial u^\beta_t(\q)}
   \right|\psi_n
\right\rangle
\Biggl )
\end{eqnarray}

The procedure outlined above can be easily extended to 
the case of PP's with non-linear core corrections \cite{II-VI}.
In this case,
the
exchange-correlation functional
$ E_{xc}[n(\r)]$
must be replaced
in Eq.\ \eqref{KS-funct},
by $E_{xc}[n(\r)+n_c(\r)]$,
where $n(\r)$ is the atomic valence charge density, and
$n_c(\r)$ is the core charge, or a suitable smooth 
approximation to it \cite{CoreCorr}.
The exchange-correlation potential $ v_{xc}(\r)$ 
of Eq.\ \eqref{Vscf} is replaced by $ v_{xc}(n(\r)+n_c(\r))$. 
Energy derivatives will contain additional terms.
The following contribution must be added to first derivatives, 
Eq.\ \eqref{dE1}:
\begin{equation}
{\partial E^{cc} \over \partial \lambda_i} = \int v_{xc}(n(\r)+n_c(\r))
         {\partial n_c(\r)\over\partial\lambda_i } d\r.
\end{equation}

Let us specialize to the case of monochromatic atomic
perturbations. The following term must be added 
to the screened potential used in the calculation 
of the linear response, Eq.\ \eqref{dVq}:
\begin{equation}
{\partial V_{cc}(\r) \over \partial u^\alpha_s(\q)} = 
 \int  \left . {d v_{xc}(n) \over d n} \right |_{n=n(\r)+n_c(\r)}
 {\partial n_c(\r )\over\partial u^{\alpha}_s(\q)} d\r
\end{equation}
where the core charge $n_c(\r)$ is written as a sum of
ionic terms, as in Eq.\ \eqref{Vion}.
The force constants, 
Eq.\ \eqref{forceconstants}, will contain additional terms :
\begin{eqnarray}
^{cc}\negthinspace\widetilde C^{\alpha\beta}_{st}(\q) & = &
{1\over N_c} \left [
     \int v_{xc}(n(\r)+n_c(\r))
     {\partial^2 n_c(\r)\over\partial u^{*\alpha}_s(\q)
                             \partial u^\beta_t(\q)} d\r \right .
\nonumber \\ & + &
\int \left .
     \left({\partial(n(\r)+n_c(\r))\over\partial u^\alpha_s(\q)}\right)^*
           {\partial V_{cc}(\r) \over \partial u^\beta_t(\q)}  d\r
\right ].
\end{eqnarray}

The matrix elements of the relevant quantities between PW's are
in the appendix.

\subsection{Ultra-soft pseudo-potentials}

In typical bulk semiconductors (e.g. Si, Ge, GaAs, AlAs) at
equilibrium volume 100-150 PW's per atom are sufficient for most 
applications. However, many
atoms---transition metals, first-row elements like F, O,  and to
a lesser extent N and C---require rather {\em hard} PP's to
ensure transferability, and their treatment demands
impractically large PW basis sets. One can try to 
exploit the many degrees of freedom which are present in PP 
generation to obtain softer PP's. 
Several recipes have been proposed to get optimally smooth PP's
(for example by acting on the form of pseudo-wave-functions in the 
core region). Simple and effective recipes have been described 
by \citeasnoun{Vanderbilt}, \citeasnoun{RRKJ} and \citeasnoun{TM}.

A more radical approach to the challenge posed by hard PP's
has been proposed by \citeasnoun{ultrasoftPP} who introduced
{\it ultrasoft} pseudo-potentials.
In this scheme, the orbitals are allowed to be as soft as possible
in the core regions so that their PW's expansion converges
rapidly, at the price of giving up the norm conservation and the
standard ortho-normality constraint of atomic orbitals.
Orthonormality is recovered by introducing a generalized
overlap operator which depends on the ionic positions. 
The full electron density is obtained by adding to the square 
modulus of the orbitals an augmentation charge localized 
in the core regions. 
Despite these technical complications, this approach
has proved to be extremely successful in treating large-scale
electronic structure problems. 

In the ultrasoft scheme, the atomic PP is separated into 
a local $V_{loc}$ and a non-local $V_{\scriptscriptstyle NL}$ part. 
The non-local potential is 
written in the separable form, as a sum of projectors. 
The ionic potential is
written as a sum over all the atoms $I$ of projectors:
\begin{equation}
V_{\scriptscriptstyle NL}({\bf r},{\bf r}')=\sum_{I} \sum_{ij} D^{(0)I}_{ij}
\beta^{I}_i({\bf r}-{\bf R}_{I})\beta^{*I}_j({\bf r}'-
{\bf R}_{I})
\label{eq:nlpot}
\end{equation}
where the functions $\beta^{I}_i({\bf r})$ and the coefficients
$D^{(0)I}_{ij}$ are computed in an atomic calculation,
as described by \citeasnoun{ultrasoftPP} and ~\citeasnoun{laasonen}.

The charge density is computed by augmenting the square modulus of the
orbitals with a term which recovers the full valence charge density:
\begin{eqnarray}
n({\bf r})&=& 2\sum_{n=1}^{N/2} |\psi_n({\bf r})|^2 \nonumber \\ 
& & \mbox{}+ 2\sum_{n=1}^{N/2}\sum_{I}\sum_{ij} Q^{I}_{ij}({\bf r}-{\bf R}_{I})
\langle \psi_n | \beta^{I}_i\rangle \langle
\beta^{I}_j | \psi_n\rangle  \nonumber  \\
&=&2 \sum_{n=1}^{N/2} \langle \psi_n|
\Lambda({\bf r})|\psi_n\rangle.
\end{eqnarray}
Consistently with this definition the orbitals are 
supposed to obey generalized orthogonality constraints
$\langle \psi_{n} | S | \psi_{m}\rangle=\delta_{nm}$,
with an overlap operator $S$ given by:
\begin{eqnarray}
S({\bf r},{\bf r'})& = & \delta({\bf r}-{\bf r'}) \nonumber \\
& & \mbox{} + \sum_{I} \sum_{ij}
q^{I}_{ij}
\beta^{I}_i({\bf r}-{\bf R}_{I})\beta^{*I}_i({\bf r}'-{\bf R}_{I})
\end{eqnarray}
where $q^{I}_{ij}=\int d{\bf r}\ Q^{I}_{ij}({\bf r})$.
The $S$ operator depends on the atomic positions and so do the constraints
obeyed by the orbitals.

The orbitals are determined by minimizing the total energy with the above 
constraints. This yields a generalized Kohn-Sham equation:
\begin{equation}
H_\SCF |\psi_n\rangle = \epsilon_n S|\psi_n\rangle
\end{equation}
where $H_\SCF$ is the KS hamiltonian:
\begin{equation}
H_\SCF= -{\hbar^2\over 2m}{\partial^2\over\partial\r^2} + V_{\KS},
\end{equation}
$V_{\KS}$ is the KS potential:
$V_{\KS}
= \tilde V_{\NL}+V_{loc}+V_{\scriptscriptstyle Hxc}$,
where $V_{\scriptscriptstyle Hxc}$ is the Hartree and exchange 
correlation potential, and
$\tilde V_{\NL}$ is the non local part, Eq.\ \eqref{nlpot},
in which the atomic coefficients $D^{(0)I}_{ij}$ are replaced
by screened coefficients $D^{I}_{ij}$ \cite{laasonen}:
\begin{equation}
D^{I}_{ij}=D^{(0)I}_{ij}+\int Q^{I}_{ij}({\bf r}) 
            V_{eff}
({\bf r}) d{\bf r}.
\end{equation}
where $V_{eff}=V_{loc}+V_{\scriptscriptstyle Hxc}$.
The forces acting on the atoms are obtained as in 
Sec.~\secref{linear-response}, 
differentiating the total energy with respect to the atomic displacements and
using the HF theorem.
In the ultrasoft PP case, however, the 
orthogonality constraints change as the atoms move thus giving rise to
additional terms in the forces.
Differentiating the generalized orthogonality constraints 
with respect to atomic displacements ${u}_s^\alpha$ for
the $s$-th atom we obtain the following relationship: 
\begin{eqnarray}
\left \langle  {\partial \psi_n \over \partial {u}_s^\alpha} \left|S
\right|\psi_m\right\rangle
&+& \left \langle  \psi_n\left| S \right|{\partial \psi_m \over
\partial u_s^\alpha} \right \rangle \nonumber \\
&=&- \left \langle \psi_n \left|
{\partial S \over \partial u_s^\alpha} \right|\psi_m\right \rangle,
\label{eq:cons}
\end{eqnarray}
which, used in the expression of the forces gives:
\begin{equation}
{F}_s^\alpha=2\sum_{n=1}^{N/2}\left \langle\psi_{n}\left |
{\partial V_{\KS}\over\partial u_s^\alpha} - \epsilon_n 
{\partial S \over \partial u_s^\alpha} \right | \psi_{n} \right \rangle,
\label{eq:forceUS}
\end{equation}
where the partial derivative of $V_{\KS}$ is:
\begin{eqnarray}
{\partial V_{\KS}({\bf r}_1,{\bf r}_2)\over \partial u_s^\alpha}&=&
{\partial V_{\NL}({\bf r}_1,{\bf r}_2)\over \partial u_s^\alpha}  
\nonumber \\ & &\mbox{}+
\int d {\bf r}_3\
{\partial V_{loc}({\bf r}_3)\over \partial 
u_s^\alpha }
\Lambda({\bf r}_3;{\bf r}_1,{\bf r}_2) 
 \nonumber \\ & &\mbox{}+\int d {\bf r}_3\
V_{ eff}({\bf r}_3) 
{\partial \Lambda({\bf r}_3;{\bf r}_1,{\bf r}_2)\over
\partial {u}_s^\alpha },
\end{eqnarray}
with no derivative of the Hartree and exchange and correlation
potential.

In order to compute the interatomic force constant and hence the dynamical
matrix we need the mixed second derivatives of the total energy 
with respect to the displacements $u_s^\alpha$
and $u_{t}^\beta$ of the atoms at sites $s$ and $t$. These expressions
have been derived by~\citeasnoun{USphonon}.
Here we report only the final results. 
Taking the derivative of the HF forces,
one finds that the electronic contribution to $C_{st}^{\alpha\beta}$
contains four terms. The first
one, ${C}^{(1)\alpha\beta}_{st}$, corresponds to the expectation value of the
second derivative of the electron-ion potential, and the additional
second derivative of the overlap matrix:
\begin{eqnarray}
{C}^{(1)\alpha\beta}_{st}=
 2\sum_{n=1}^{N/2} 
\Bigg \langle \psi_n\Bigg| &\Bigg[& {\partial^2  \left( \tilde V_{\NL}
+ V_{loc} \right) \over
\partial {u}_{s}^\alpha
\partial {u}_{t}^\beta } \nonumber \\
&-&\epsilon_n
{\partial^2 S \over \partial {u}_s^\alpha
\partial {u}_{t}^\beta }\Bigg]
 \Bigg| \psi_{n} \Bigg \rangle,
\label{eq:secder}
\end{eqnarray}
where the second derivative of $\tilde V_{\NL}$
is performed at fixed charge density as before.
The second term ${C}^{(2)\alpha\beta}_{st}$ is:
\begin{equation}
{C}^{(2)\alpha\beta}_{st}=
2\sum_{n=1}^{N/2} \bigg[ \left \langle
{\partial \psi_{n}\over \partial {u}_s^\alpha}\left| P_c^+
\right|\phi_{t,n}^\beta\right \rangle +
h.c. \bigg],
\end{equation}
where
$P^+_c=1-\sum_{m=1}^{N} S|\psi_m\rangle\langle \psi_m|$
is the projector on the conduction-band subspace,  $h.c.$ indicates
the hermitian conjugate with respect to the $s\alpha,t\beta$ indices, 
$\phi$ is defined as
\begin{equation}
|\phi_{t,n}^\beta\rangle=\Bigg[{\partial
\left(\tilde V_{\NL}
+V_{loc}\right) \over \partial {u}_{t}^\beta } -
\epsilon_n {\partial S\over \partial
{u}_{t}^\beta }\Bigg] \bigg |\psi_{n} \bigg\rangle,
\end{equation}
and again the derivative of $\tilde V_{\NL}$ is performed at
fixed density.
In the norm-conserving PP scheme, the electronic contribution to
${C}_{st}^{\alpha\beta}$ is simply given by the sum
of ${C}^{(1)\alpha\beta}_{st}$ and
${C}^{(2)\alpha\beta}_{st}$ calculated for $S=1$ and
$\Lambda({\bf r})=|{\bf r}\rangle
\langle {\bf r}|$.
In the ultrasoft PP scheme one must consider two additional
contributions to ${C}_{st}^{\alpha\beta}$ which
have no counterparts in the norm-conserving scheme.
${C}^{(3)\alpha\beta}_{st}$ is the interaction between the change of
the augmentation charge $\Delta^{t\beta}n({\bf r})$ due to
the atomic displacement ${u}_{t}^\beta$ (see Eq.\ \eqref{deltarho}
below) and
the change of $V_{\scriptscriptstyle Hxc}$ due to the displacement 
${u}_s^\alpha$
(see Eq.\ \eqref{dVscf}): 
\begin{equation}
C^{(3)\alpha\beta}_{st}=
{1\over 2}\int \ \bigg[
{\partial V_{\scriptscriptstyle Hxc}({\bf r})\over \partial {u}_s^\alpha }
\Delta^{t\beta}n({\bf r}) d{\bf r}
+h.c.\bigg].
\end{equation}
Finally, ${C}^{(4)\alpha\beta}_{st}$ is analogous to 
${C}^{(2)\alpha\beta}_{st}$
but with the projector on the conduction-states subspace replaced by
that on the valence-state subspace. Since the perturbation formalism
provides explicitly only $P_c|{\partial \psi_i
\over \partial {u}_s^\alpha}\rangle$,
the valence-state component must be derived from the constraints
imposed by
the orthogonality condition, Eq.\ \eqref{cons}. One finally
obtains: \refstepcounter{equation} \label{eq:qua} $$ \displaylines{
\quad
{C}^{(4)\alpha\beta}_{st}= \hfill \cr \hfill
- 2\sum_{n,m=1}^{N/2}\left
\langle \psi_n\left | {\partial S \over \partial {u}_s^\alpha }
\right |\psi_{m}\right \rangle \langle
\psi_{m}|\phi_{t,n}^\beta\rangle
+h.c. \quad (\theequation) } $$
We note that in the norm-conserving scheme, the left hand side of
Eq.\ \eqref{cons} vanishes since $S=1$, and Eq.\ \eqref{cons} 
allows one to show that
the contribution to ${C}_{st}^{\alpha\beta}$ from the
valence-states component
of $|{\partial \psi_i \over \partial {u}_s^\alpha}\rangle $ is zero.
In the ultrasoft case, Eq.\ \eqref{cons} is used to evaluate
such a component in terms of the unperturbed orbitals,
as given in Eq.\ \eqref{qua}.

The key ingredient to evaluate the dynamical matrix is
$P_c |{\partial \psi_n \over \partial {u}_s^\alpha}\rangle $, which can be
determined, within first-order perturbation theory, by solving the
linear system: \refstepcounter{equation} \label{eq:ls} $$ \displaylines{
\quad
(H_\SCF-\epsilon_n S) P_c \bigg|{\partial \psi_n \over \partial 
{u}_s^\alpha}\bigg \rangle= \hfill \cr \hfill 
-P_c^+\left[{d V_{\KS} \over d {u}_s^\alpha} - \epsilon_n
{\partial S \over \partial {u}_s^\alpha}\right] 
\bigg| \psi_n \bigg \rangle, \quad (\theequation) } $$
where
\begin{eqnarray}
\left[{d V_{\KS} \over d {u}_s^\alpha} - \epsilon_n
{\partial S \over \partial {u}_s^\alpha}\right] \bigg | \psi_n 
\bigg \rangle &=&
|\phi_{s,n}^\alpha\rangle \nonumber \\  &+&
\int {\partial V_{\scriptscriptstyle Hxc}({\bf r}) \over 
\partial {u}_s^\alpha}
\Lambda({\bf r}) |\psi_n\rangle d{\bf r}.
\end{eqnarray}
Eq.\ \eqref{ls} is the generalization to the ultrasoft case of the
self-consistent linear system given in Eq.\ \eqref{linear}.
The perturbing term depends on the variation of the charge density
${\partial n ({\bf r}) / \partial {u}_s^\alpha}$
through ${\partial V_{\scriptscriptstyle Hxc}({\bf r}) 
/\partial u_s^\alpha}$.
${\partial n ({\bf r}) / \partial {u}_s^\alpha}$
is a functional of
$P_c |{\partial \psi_n / \partial {u}_s^\alpha}\rangle$:
\begin{equation}
{\partial n ({\bf r}) \over \partial {u}_s^\alpha}
 = 4 \sum_{n=1}^{N/2} \left \langle {\partial \psi_n
\over \partial {u}_s^\alpha } 
| P_c^+ \Lambda({\bf r}) |\psi_n \right \rangle
+ \Delta^{s\alpha} n({\bf r}).
\label{eq:deltarho}
\end{equation}
The term $\Delta^{s\alpha} n({\bf r})$, peculiar to the ultrasoft
scheme, has two contributions: $\Delta^{s\alpha} n({\bf r})
=\delta^{s\alpha} n({\bf r})
+ \delta^{s \alpha} n_{ortho}({\bf r})$. The former term,
\begin{equation}
\delta^{s\alpha} n({\bf r}) =
2 \sum_{n=1}^{N/2} \left \langle \psi_n \left| 
{\partial \Lambda({\bf r}) \over
\partial {u}_s^\alpha }\right |\psi_n \right \rangle, 
\end{equation}
accounts for the displacement of the augmentation charge
at fixed orbitals, whereas the latter, \refstepcounter{equation} 
$$ \displaylines{ \quad
\delta^{s \alpha} n_{ortho}({\bf r})= \hfill \cr \hfill
- 2\sum_{n,m=1}^{N/2}
\left \langle \psi_n \left | {\partial S \over \partial 
{u}_s^\alpha} 
\right |\psi_m \right \rangle
\langle \psi_m | \Lambda({\bf r}) |\psi_n\rangle,
\quad (\theequation) } $$
appears because of the orthogonality constraints,
similar to the ${C}^{(4)\alpha\beta}_{st}$ term in the interatomic
force constants.

The generalization of the above formalism to metallic systems can be
done along the same lines as described in Sec.~\secref{metals}.
The presence of the fractional occupation numbers modifies the definition of
the valence-states subspace and
the terms $\delta^{s\alpha} n_{ortho}({\bf r})$ and
${C}^{(4)\alpha\beta}_{st}$ must be modified accordingly.
For instance $\delta^{s\alpha} n_{ortho}({\bf r})$ becomes:
\begin{eqnarray}
\delta^{s\alpha} n_{ortho}({\bf r}) & = &\mbox{}- 2\sum_{n,m=1}^{N/2} 
\left[
\tilde \theta_{F,n} \tilde \theta_{n,m}+
\tilde \theta_{F,m} \tilde \theta_{m,n}\right] \nonumber \\ & &
\mbox{}\times\left \langle \psi_n \left | {\partial S \over \partial 
{u}_s^\alpha} 
\right |\psi_m\right \rangle
\langle \psi_m  | \Lambda({\bf r}) |\psi_n\rangle.
\end{eqnarray}

\subsection{Localized basis sets, all-electron}

All-electron implementations of DFPT based on 
localized basis sets exist for 
both the Linearized Muffin-Tin Orbitals (LMTO)
method \cite{LMTO_LR1,LMTO_LR2} and for the Linearized
Augmented Plane Waves (LAPW) method~\cite{LAPW_LR}. LMTO and LAPW
are amongst the most popular all-electron methods in DFT
calculations. Their extension to DFPT calculations is especially
useful for systems containing transition metals (like e.g.
high-temperature super-conductors and most ferroelectrics)
for which the PW-PP approach is not very practical. 
An earlier implementation using a mixed-basis set
(localized atomic-like states 
plus plane waves) in a PP formalism \cite{Zein3}
and a more recent one \cite{mixedbasis} are also known.

Localized-basis-sets and mixed-basis implementations 
are more complex than PW implementations. 
Part of the additional complexity
arises from Pulay terms in derivatives. The origin of such
terms is easily understood. The first derivative of the 
energy functional (see Eq.\ \eqref{E2}) contains the 
HF term, as in Eq.\ \eqref{dE1}, plus a term
$\widetilde F$, coming from implicit dependence through
the wave-functions:
\begin{equation}
{\partial E\over\partial\lambda} = 
\int n(\r) {\partial V(\r)\over\partial\lambda} d\r + \widetilde F
\end{equation}
where
\begin{equation}
\widetilde F = 2\sum_{n=1}^{N/2}\int 
     {\partial \psi^*_n(\r)\over\partial\lambda} (H_\SCF-\epsilon_n) 
      \psi_n(\r) d\r + c.c.
\label{eq:Pulay}
\end{equation}
vanishes if the wave-functions are {\em exact} KS orbitals. This is not
always true if the wave-functions are {\em approximate} KS orbitals.
Let us expand the wave-functions into a basis set $\phi_n$, taken to
be orthonormal for simplicity: 
\begin{equation}
 \psi_n(\r) = \sum_j c_j^{(n)}\phi_j(\r).
\end{equation}
The solution of the KS equations reduces to the solution
of a secular equation 
\begin{equation}
 \sum_k (H_{jk}-\epsilon_n) c_k^{(n)}=0
\label{eq:secular}
\end{equation}
where
\begin{equation}
  H_{jk}=\int\phi^*_j(\r)H_\SCF\phi_k(\r)d\r.
\end{equation}
By inserting the expansion of the KS orbitals into Eq.\ \eqref{Pulay}
one finds
\begin{eqnarray}
\widetilde F & = & 2\sum_{n=1}^{N/2} \sum_{jk} 
      \left( c_j^{*(n)} c_k^{(n)}\int
      {\partial \phi^*_j(\r)\over\partial\lambda}(H_\SCF-\epsilon_n)
                \phi_k(\r) d\r \right . \nonumber \\
      & & \left. + 
      {\partial c_j^{*(n)}\over\partial\lambda} (H_{jk}-\epsilon_n) 
       c_k^{(n)} \right) + c.c..
\end{eqnarray}
The second term vanish exactly (see Eq.\ \eqref{secular}). 
The first term does not vanish if the basis set is not complete 
and if the basis set depend explicitly on $\lambda$. 
In realistic calculations using atomic-centered basis sets,
the Pulay contribution cannot be neglected.
Accurate and reliable calculations of forces and of the force 
constants require a very careful account of the Pulay terms, 
that are instead absent if a PW basis set is used.

LMTO-based linear-response techniques have been used 
to calculate phonon
spectra and electron-phonon couplings in several elemental metals
\cite{LMTO_e-ph,LMTO_metals} and more recently in doped 
BaBiO$_3$ \cite{BaBiO3} and CaCuO$_2$ \cite{CaCuO2}.
An extension of the method to the calculation of spin
fluctuations \cite{LMTO_spin} has been recently done.

LAPW-based  linear-response techniques have been first used 
in the study of 
the lattice dynamics of CuCl \cite{CuCl} and later applied
to several materials: SiC \cite{LAPW_SiC}, ferroelectrics KNbO$_3$
\cite{KNbO3,KNbO3_H,KNbO3_Z} and SrTiO$_3$ \cite{SrTiO3}, high-T$_c$
super-conductor La$_2$CuO$_4$ \cite{La2CuO4}.

The mixed-basis approach has been tested by calculating the
phonon dispersions for simple metals containing localized 
$4d$ electrons: Ag and Y \cite{mixedbasis}, Ru \cite{rutenio},
and on sapphire ($\alpha-$Al$_2$O$_3$) \cite{zaffiro}.

\section{Other approaches} \label{sec:other}
\subsection{Dielectric approach}

Historically, the microscopic theory of lattice dynamics was first
formulated in terms of {\em dielectric matrices} \cite{PCM}. The basic
ingredient is the inverse dielectric matrix, $\epsilon^{-1}(\r,\r')$,
relating, in the linear regime, the external perturbation $\Delta V$ to
the total electrostatic potential experienced by an external test charge:
\begin{equation}
\Delta V_{test}(\r) = \int \epsilon^{-1}(\r,\r') \Delta V(\r') d\r'.
\end{equation}
As an alternative the theory can be formulated in terms of 
the electron polarizability, $\chi(\r,\r')$,
which gives the charge-density linear response to the external perturbation:
\begin{equation}
\Delta n(\r) = \int \chi(\r,\r') \Delta V(\r') d\r'.
\label{eq:chidef}
\end{equation}

These two response functions are simply related as:
\begin{equation}
\epsilon^{-1}(\r,\r') =
\delta(\r-\r') - \int \frac{e^2}{|\r-\r_1|} \chi(\r_1,\r) d\r_1.
\end{equation}

Within DFT, one can also define the independent-electron polarizability,
$\chi_0(\r,\r')$, as the charge-density response to variation of 
the total KS potential:
\begin{equation}
\Delta n(\r) = \int \chi_0(\r,\r') \Delta V_\SCF(\r') d\r'.
\label{eq:chi0def}
\end{equation}
The expression for $\chi_0(\r,\r')$ in terms of KS orbitals has the
well known form:
\begin{equation}
\chi_0(\r,\r') = 
\sum_{n,m} \frac{f_n-f_m}{\epsilon_n-\epsilon_m}
              \;\;  \psi^*_n(\r) \psi_m(\r) \psi^*_m(\r') \psi_n(\r')   
\label{eq:chi0}
\end{equation}
where $f_n$ is the occupancy of the state 
($f_n=\tilde\theta\left({(\varepsilon_F-\varepsilon_n)/\sigma}\right)$ 
in the notations of Sec.\secref{metals}), and
the sums over $n$ and $m$ extend to both occupied and empty
states. As shown in
Sec.\secref{metals}, only terms involving virtual
transitions from occupied or partially occupied to empty or
partially empty states contribute.

Combining Eqs.\ \eqref{chidef} and \eqref{chi0def} and recalling
the relationship between the bare and the KS self-consistent perturbing
potential, Eqs.\ \eqref{DeltaVvsDeltapsi} and \eqref{Kappa}:
\begin{equation}
     \Delta V_\SCF(\r) =
        \Delta V(\r) + \int K(\r,\r')\Delta n(\r') d\r',
\label{eq:DeltaV}
\end{equation}
one gets an integral equation for $\chi$:
\begin{eqnarray}
 \chi(\r,\r')&=&\chi_0(\r,\r') \nonumber \\
             & &\mbox{} +
                \int \chi_0(\r,\r_1)K(\r_1,\r_2) \chi(\r_2,\r') d\r_1d\r_2,
\end{eqnarray}
or equivalently
\begin{equation}
 \chi^{-1}(\r,\r') = \chi_0^{-1}(\r,\r') - K(\r,\r').
\end{equation}
This equation, projected onto a PW basis set, becomes a matrix
equation, one for each \q-point in the BZ, that can be solved by
matrix inversion.

The original dielectric matrix approach has the major drawback that the
perturbation must be described by a local potential and thus it cannot
be applied to the lattice dynamics problem if modern {\it non-local}
pseudo-potentials are employed to describe electron-ion interaction.
In fact in this case, not only the unperturbed external potential,
but also the perturbation itself is described by a non local operator 
and Eq.\ \eqref{chidef} is not appropriate.
For these reasons the calculation of dielectric matrix has been of limited
utility for the study of vibrational properties---see for some early
examples the empirical pseudo-potential calculations by \citeasnoun{BBCT},
\citeasnoun{RRAB} and \citeasnoun{RR83}---while it has been successfully
employed in the study of the macroscopic dielectric properties of simple
materials \cite{Hx1} and, more generally, is an essential ingredient
in quasi-particle GW calculations. In these latter cases, in fact,
even if the unperturbed external potential is described by non local
pseudo-potentials, the perturbation of interest is only local and
Eqs.\ \eqref{chidef} and the following ones still apply.

A modification of the dielectric matrix approach that removes
its limitation to local pseudo-potentials has been devised by
\citeasnoun{Quong_LR}. The response to the bare potential is stored
into a bare response $\Delta n_b(\r)$:
\begin{eqnarray}
\label{eq:nbare}
\Delta n_b(\r)&=&\int \chi_0(\r,\r') \Delta V(\r') d\r' \\
              &=&\sum_{n,m } 
                 \frac{f_n-f_m} {\epsilon_n-\epsilon_m} 
                 \psi_n^*(\r) \psi_m(\r)
                 \langle \psi_m | \Delta V | \psi_n \rangle,\nonumber
\end{eqnarray}
that can be calculated once for all for non local potentials as well. Then
Eq.\ \eqref{chi0def} is applied together with Eq.\ \eqref{DeltaV} to yield 
$\Delta n(\r)$:
\begin{eqnarray}
\Delta n(\r)&=&\Delta n_b(\r)  \nonumber\\
 & &\mbox{} + \int \chi_0(\r,\r_1) K(\r_1,\r_2) \Delta n(\r_2) d\r_1d\r_2.
\end{eqnarray}
This equation is solved for $\Delta n(\r)$ by inverting the kernel of
the above integral equation (a matrix in a PW basis):
\begin{equation}
\left (\delta(\r-\r') - \int \chi_0(\r,\r_1) K(\r_1,\r') d\r_1\right).
\label{eq:kernel}
\end{equation}
The second-order change in the energy can then be calculated as usual.

This approach has been used for the calculations of force constants in
Au \cite{Quong_gold}, of electron-phonon coupling \cite{Quong_e-ph} and
of thermal expansion in metals \cite{Quong_metals}, and of the
structural stability and electron-phonon coupling in Li
\cite{PhasesLi}.

This modified dielectric matrix approach is conceptually rather
similar to the original DFPT. The main difference with respect to DFPT
is the replacement of the self-consistency cycle needed in DFPT with
the construction and inversion of kernel in Eq.\ \eqref{kernel}. This
operation is rather time-consuming, since it requires the inversion of
large matrices and an expensive sum over unoccupied bands, but must be
done only once for any given point in the BZ where the vibrations are
computed (as opposed to the self-consistency in DFPT that must be
performed for each phonon mode).

For small systems the overall computational workload of the two approaches
is similar, but the size of the systems that can be treated by the
dielectric approach is limited by the growing dimension of the kernel
operator, Eq.\ \eqref{kernel}.

\subsection{Frozen phonons}

The frequencies of selected phonon modes can be calculated from the
energy differences---or from the forces acting on atoms---produced by
finite, periodic, displacements of a few atoms in an otherwise perfect
crystal at equilibrium. The first such, so called {\em frozen-phonon},
LDA calculations were performed already in the early 80's (see for
instance \citeasnoun{Yin}). A frozen-phonon calculation for lattice
vibrations at a generic \q-vector requires a super-cell having \q\ as a
reciprocal-lattice vector and whose linear dimensions must be
therefore at least of the order $\sim 2\pi / | \q |$. In practice, the
size of the super-cell that one can afford to deal with has
traditionally limited the application of this technique to
zone-center or selected zone-boundary phonon modes in relatively
simple materials. However, zone-center phonons are also the best
characterized because they may be Raman- or infrared-active, so that
they do not require neutron spectroscopy to be detected. Moreover, the
frozen-phonon and linear-response techniques may be combined to study
anharmonic effects that would be otherwise difficult to calculate
directly from perturbation theory \cite{raman,anharm4}.

Phonon dispersions along high-symmetry lines in simple materials are
determined by the so called {\it inter-planar force constants} (i.e. by
the forces acting on planes perpendicular to the phonon wave-vector
when another such plane is rigidly moved from equilibrium). Lattice
vibrations along some high-symmetry lines in cubic semiconductors have
been calculated in this way using reasonably sized super-cells
\cite{interplanar}. More recently several authors
\cite{Wei1,Wei2,Frank,Ackland,Parlinski-2} have started to calculate
phonon dispersions from the {\em interatomic force constants}
determined in real space using the frozen-phonon approach. 

The main advantage of the frozen-phonon approach is that it does not
require any specialized computer code, as DFPT does. This technique
can in fact be straightforwardly implemented using any standard
total-energy and force code, and just some care is needed in the
evaluation of numerical derivatives. The principal limitation is the
unfavorable scaling of the computational workload with the range of
\def\range{{\cal R}_{IFC}} the IFC's, $\range$. In fact, the
calculation of IFC's within the frozen-phonon approach requires the
use of super-cells whose linear dimensions must be larger than
$\range$, thus containing a number of atoms $N_{at}^{SC} \sim
\range^3$. As the computer workload of standard DFT calculations
scales as the cube of the number of atoms in the unit cell, the cost
of a complete IFC calculation will scale as $3 N_{at} \times
\range^9$, where $N_{at}$ is the number of (inequivalent) atoms in the
elementary unit cell (the factor 3 accounts for the three, generally
independent, phonon polarizations). The calculation of IFC's using
DFPT requires instead the evaluation of the dynamical matrices on a
regular grid of wave-vectors in the Brillouin zone, whose spacing must
be chosen of the order of the inverse of the range of the IFC's:
$\Delta q \sim 2 \pi / \range$ (see Sec. \secref{force-constants}).
The number of \q-points in such a grid is of the order of
$\range^3$. As the computational cost for the calculation of each
column of the dynamical matrix is of the order of $N_{at}^3$---and the
number of such columns is $3 N_{at}$---the total cost for the
calculation of the IFC's (and, hence, of complete phonon dispersions)
using DFPT is of the order of $\range^3 \times 3 N_{at}^4$. 

Another problem strictly related to these considerations is that of
the calculation of phonon dispersions in polar materials. In
Secs. \secref{polar} and \secref{force-constants} we saw that the
long-range character of the dipole-dipole interactions in polar
insulators determines a non analytic behavior of the dynamical
matrices as functions of the wave-vector in the long wave-length
limit. The real-space counterpart of this property is that IFC's are
long ranged as they decay with the inverse cube of the distance.  The
interpolation of the dynamical matrices in reciprocal space as well as
the calculation of the long-range tails of the IFC's in real space is
made difficult by this problem. Within DFPT, the standard remedy is
the separate treatment of the non-analytic part of the dynamical
matrix, using information on the ionic effective charges and crystal
dielectric constant, as explained in Secs. \secref{polar} and
\secref{force-constants}. Frozen-phonon super-cell calculations,
instead, do not directly provide such information,\footnote{This is a
fact occasionally overlooked in the literature: see e.g.\
\citeasnoun{ZrO2wrong} and the comment by \citeasnoun{ZrO2}} that must
instead be extracted from the limiting behavior of inter-planar force
constants \cite{frozenIFC}, supplied by a separate calculation using
the Berry's phase approach \cite{BerryPhase,P},\footnote{Note that the
Berry's phase approach cannot be used to
calculate the dielectric constant} borrowed from DFPT
calculations \cite{Parlinski-1}, or fitted to the experiment
\cite{Parlinski-2}.
\subsection{Vibrational properties from molecular dynamics}
\label{sec:MD} 

All the methods described so far are static, zero-temperature,
methods. In the last 15 years the combined use of molecular dynamics
(MD) and of DFT \cite{CP} has become a very powerful tool for the ab
initio study of condensed-matter systems at finite temperature. In MD
simulations atomic trajectories are generated from
the classical equations. 
Equilibrium properties are then estimated as time
averages over the trajectories which also contain information on the
dynamics of the system, {\em i.e.} on phonon modes. In fact, the
vibrational density of states---which exhibits peaks at the phonon
frequencies---can in principle be computed by Fourier transforming the
atomic velocity auto-correlation function~\cite{autocorr}. Ab initio MD
simulations are usually performed using super-cells which contain a
small number of atoms (from a few tens to a few hundreds) and periodic
boundary conditions. Because of this, only phonons which are
zone-center phonons of the super-cell are accessible to the simulation.

The straightforward estimation of phonon frequencies from MD
simulations suffers from three types of problems. First, at low
temperature all the systems are strongly harmonic and, hence, poorly
ergodic. The time necessary to reach equilibrium may thus be
impractically long. In these cases, bringing the system to thermal
equilibrium may require technical tricks such as coupling to
Nos\'e-Hoover thermostats \cite{Nose}.

The second problem is that the simulation time necessary to attain a
frequency resolution, $\Delta\omega$, cannot be shorter than $ \tau
\gtrsim 2\pi/ \Delta\omega $. In practice, this time may be too
long for first-principles MD simulations. This problem can be overcome
by using more sophisticated spectral estimation methods, such as the
Multiple Signal Classification (MUSIC) algorithm \cite{music}, to
extract information on phonon modes from relatively short MD
runs. MUSIC exploits the harmonic character of phonon modes and has
proven to be very accurate and useful in simple situations. In more
complicated cases, where the frequencies are too many and too close
with respect to the inverse length of the run, MUSIC suffers from
instabilities that may prevent the full determination of the
spectrum. An improved self-consistent MUSIC algorithm has been
developed \cite{koh1,koh2} to cope with instabilities and to extract
information on eigen-vectors as well. In this algorithm, a first
estimation of the frequencies is followed by the determination of the
eigen-vectors through a least squares fit of the trajectory including
orthogonality constraints.  Then, the trajectory is projected onto
each of the normal modes. At this point each projected trajectory
contains mainly one frequency component. This component is
re-estimated using MUSIC, and the scheme is iterated until
self-consistency in the frequencies is achieved.

Finally---when MD simulations are used to predict the temperature
dependence of individual vibrational modes and the thermal behavior of
properties which depend on them---the results may depend on the size
of the simulation (super-) cell. In fact, even though in the harmonic
approximation a mode commensurate with the simulation cell is strictly
decoupled from those which are not, this is not the case at high
temperature when anharmonic effects are important. As a consequence,
the neglect of modes which are not commensurate with the simulation
cell may affect the evaluation of the frequencies of commensurate
modes which, in the harmonic approximation, would be directly
accessible to the simulation.

MD simulations are in some sense complementary to lattice-dynamical
calculations, in the sense that the latter are better suited at low
temperatures where the former are subject to ergodicity
problems. Lattice-dynamics is by definition limited to the (quasi-)
harmonic regime, while MD naturally accounts for all the anharmonic
effects occurring at high temperature, provided the size of the
simulation cell is large enough to allow a proper description of the
relevant phonon-phonon interactions.

\section{Applications} \label{sec:applic}
\subsection{Phonons in bulk crystals}

Phonon dispersions in crystals have been for a long time calculated
using model force constants where some form of interatomic potentials is
assumed and the parameters of the model are adjusted so as to reproduce
some known experimental results. Although this approach works reasonably
well in some cases it has serious drawbacks that in many situations call
for more predictive methods. The typical experimental data used to fix the
parameters in the model are the phonon frequencies themselves. Model
force constants can be seen as a compact way to encode the available
experimental input, with very limited predictive power when applied to
other properties. In many cases only a few selected frequencies (usually
of infrared (IR) and Raman-active modes) are known and the results of model
calculations in the rest of the BZ are questionable. Even when the entire
dispersion spectra is known (usually by neutron diffraction measurements),
the knowledge of phonon frequencies alone is not sufficient to determine
completely the force constants: the knowledge of the phonon displacement
patterns would be needed as well. The latter experimental information
is seldom available and, in the few cases in which the pattern were
measured, the comparison with the results of even the best models is
sometimes very poor.

Using DFPT the interatomic force constants are computed from
first-principles and, as they are usually accurate, both good frequencies
{\it and} good displacement patterns are obtained, without need of
experimental inputs. 

Most, if not all, calculations described in the following sections are
done at the LDA level that usually provides very good results. A test of
the performance of generalized gradient corrections (GGA) in the
flavor proposed by \citeasnoun{PBE}, has been performed in 
\citeasnoun{GGAph}. It is found that GGA systematically lowers 
the frequencies of phonon branches with positive Gr\"uneisen parameters. 
This effect is correlated with the GGA's expansion of the lattice constant,
since GGA phonon frequencies computed at the experimental lattice constants 
are higher than the corresponding LDA ones. 
A similar trend is found also for magnetic Fe and Ni \cite{DCdG}.
In this case GGA equilibrium geometries are much superior to LDA ones
and phonon dispersions are correspondingly closer to the experimental results
\cite{DCdG}.
In Diamond, Al, and Cu, LDA and GGA equilibrium geometries and phonon 
dispersions
have similar accuracy with respect to the experimental data. Si is an
exception since the LDA phonon dispersions are already in very good
agreement with experiment and GGA slightly worsens the comparison
\cite{GGAph}.

In most applications phonon dispersions are computed at the theoretical
equilibrium geometry (lattice parameters and internal coordinates).
This choice is mandatory when the experimental geometry is poorly
known, but it is
also, in our opinion, the most consistent one when comparing with
experimental data at low temperature. Inclusion of thermal expansion
may become necessary in some cases when comparing with room and higher
temperature data. See \secref{QHA} for the treatment of thermal effects.

\subsubsection{Simple semiconductors}

\paragraph{Elemental and III-V semiconductors}
\label{sec:III-V}

\begin{figure*}[p]
\centerline{\psfig{figure=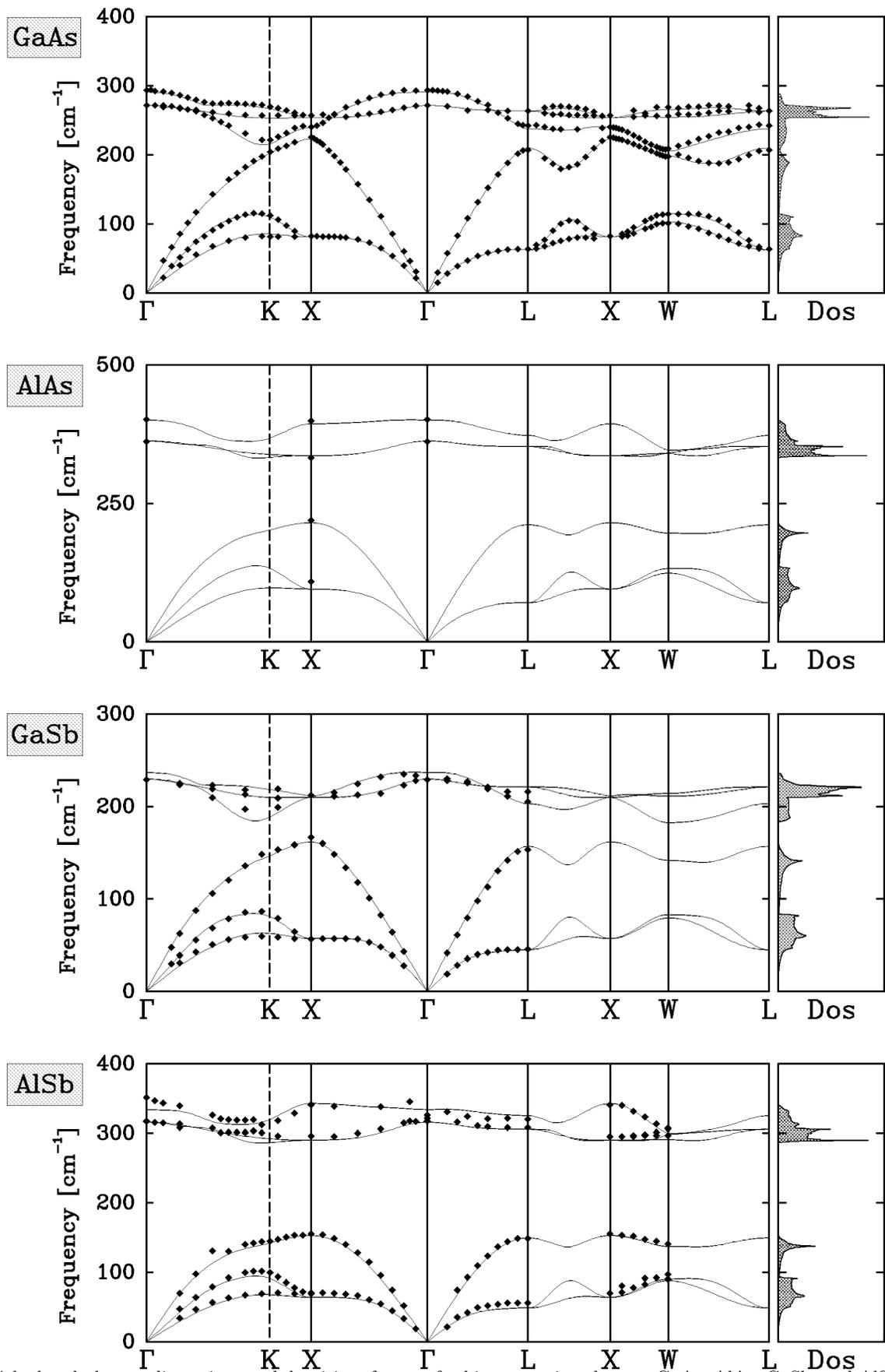,width=15.0truecm}}
\caption{Calculated phonon dispersions and densities of states for
binary semiconductors GaAs, AlAs, GaSb, and AlSb. Experimental data
are denoted by diamonds. (Reproduced from \protect\citeasnoun{bulkoni}.)}
\label{fig1}
\end{figure*}

The phonon spectra and effective charges of group IV semiconductors Si and
Ge (diamond structure) and of zincblende structure III-V semiconductors
GaAs, GaSb, AlAs, AlSb were calculated in \citeasnoun{bulkoni}.
The calculated phonon dispersions and densities of states for the
latter four compounds are shown in Fig.\ \ref{fig1}, together with
experimental data.
Zone-center phonons, effective charges, and dielectric constants of nine
III-V zincblende semiconductors were computed, along with their
piezoelectric constants, by \citeasnoun{piezoSdG}. The phonon dispersions
of Si were calculated as well by \citeasnoun{LMTO_LR2} as a test of the 
LMTO implementation of DFPT. Dispersions for InP appear in a paper 
devoted to the (110) surface phonons of InP \cite{InP}; dispersions for
both GaP and InP were published in a study of phonons in GaInP$_2$
alloys \cite{GaInP}. For all these materials, phonon spectra and effective
charges are in very good agreement with experiments, where available.
For AlAs---for which experimental data are very scarce---these calculations
provide the only reliable prediction of the entire phonon
dispersions. For Si, the
calculated phonon displacement patterns compare favorably to those
extracted from inelastic-neutron-scattering experiments \cite{Siexp}.

In all these materials the interatomic force constants turn out to be
quite long-ranged along the (110) direction. This feature had already been
observed in early calculations \cite{Kane,Fleszar-Resta} and is
related to the
peculiar topology of diamond and zinc-blende lattices, with bond
chains propagating along the (110) directions.

The force constants of GaAs and of AlAs are especially interesting in
view of their use in complex GaAlAs systems such as super-lattices,
disordered super-lattices, and alloys. While the phonon dispersions
in GaAs are experimentally well characterized, bulk samples of AlAs
of good quality are not available and little experimental information
on its vibrational modes have been collected. Since several years it
has been assumed that the force constants of GaAs and those of AlAs
are very similar and that one can obtain the dynamical properties
of AlAs using the force constants of GaAs and the masses of AlAs
({\em mass approximation}) \cite{massapprox}. The DFPT calculations
provided convincing evidence that the mass approximation
holds to a very good extent between GaAs and AlAs \cite{bulkoni}.
This 
transferability of the
force constants
makes it possible to calculate 
rather easily and
accurately
the vibrational spectra of complex GaAlAs systems
\cite{GaAlAs_sl1,GaAlAs_sl2,GaAlAs_all,wires}. Somewhat surprisingly, the
mass approximation does not seem to be valid when the interatomic force
constants for a well-known and widely used model, the Bond-Charge Model
(BCM), are employed. A six-parameter BCM for GaAs that gives dispersions
that compare favorably with experiments and {\it ab initio} calculations,
yields, when used in the mass approximation, AlAs dispersions that are
quite different from first-principles results. This clearly shows that
information on the vibrational frequencies alone is not sufficient to
fully determine the force constants, even when complete phonon dispersions
are experimentally available. In order to obtain more realistic dispersions
for AlAs in the mass approximation, one has to fit the BCM for GaAs to
{\em both} frequencies {\em and} at least a few selected eigenvectors
\cite{BCMrefit}.

\paragraph{II-VI semiconductors}

The II-VI zincblende semiconductors ZnSe, ZnTe, CdSe, CdTe present some
additional difficulties in a PP-PW framework with respect to their III-V
or group IV counterparts. The cation $d$ states are close in energy
to the $s$ valence states so that the $d$ electrons should be included 
among the valence electrons.
Phonon calculations performed several years ago, when the inclusion of
localized $d$ states in the pseudo-potential was difficult,
showed that the effects of cation $d$ electrons could be accounted
for also by including the $d$ states in the core and by using the 
nonlinear core correction approximation. The results showed an accuracy 
comparable to that previously achieved for 
elemental and III-V semiconductors \cite{II-VI}. 
Similar calculations have been more recently performed for hexagonal
(wurtzite
structure) CdS \cite{CdS,CdSiso} and CdSe \cite{CdSe} and compared
with the results of inelastic neutron scattering experiments.

\paragraph{Diamond and Graphite}

The phonon dispersions of Diamond, together with the
internal-strain parameter, thermal expansion coefficient
in the Quasi-Harmonic Approximation, and the mode Gr\"uneisen 
parameter dispersion curves, were calculated by
\citeasnoun{Diamond}. A unique feature found in Diamond is the 
presence of an {\em over-bending} of the uppermost phonon branch,
whose frequencies have a minimum at the BZ center
(instead of a maximum as in the other elemental semiconductors
Si and Ge). This feature has important consequences for the 
second-order Raman spectra (see paragraph \ref{2ndRaman}).

High-pressure spectra, up to 1000 GPa, for Diamond are reported
in \citeasnoun{CompressedDiamond}. Under pressure, the phonon 
frequencies of the $X_4$ and $L'_3$ modes gradually go higher 
than those of $X_1$ and $L'_2$, respectively.
The over-bending of the uppermost 
phonon branch decreases with the increase of pressure.

The phonon dispersions of Graphite along the $A-\Gamma-K-M$ line
were calculated in \citeasnoun{Graphite}. The dispersions exhibit
an over-bending similar to that of Diamond for the in-plane 
dispersion. The dispersions between graphitic planes is very 
flat. The peculiar behavior of low-frequency branches along the
$\Gamma-K$ line can be related to the long extent of IFC's along the
zigzag chains in the graphitic planes.

\paragraph{Silicon Carbide}

Silicon Carbide (SiC) may crystallize in a large variety of 
tetrahedrally coordinated polymorphs.
Phonon dispersion curves were calculated for the
3C (cubic zincblende), 2H (wurtzite) and 4H hexagonal structures
\cite{SiC}. For the 3C structure, elastic and Gr\"uneisen 
constants were calculated as well \cite{SiC}. The behavior 
under pressure of phonon dispersions, effective charges, and 
of the dielectric tensor was studied by several authors
\cite{LAPW_SiC,SiCunderP,SiCunderP2}. The interest for the 
dynamical properties of SiC under pressure was prompted by 
a report \cite{SiCexp} that the splitting between LO and transverse
optic (TO) modes of 6H SiC increases 
with increasing pressure until $P=60$GPa, then it decreases.
This was attributed to a decrease of the effective charges.
This interpretation was not confirmed by any of the above 
theoretical studies, which pointed out instead that an incorrect 
volume dependency for the dielectric tensor was assumed in the
analysis of experimental data.

\paragraph{Nitrides}

The group-III nitrides are very interesting materials for
optoelectronic applications at short wavelengths, as well as
in high-frequency and high-temperature electronic devices.
Several groups performed calculations of lattice-dynamical
properties for zinc-blende and wurtzite GaN \cite{GaN}, 
BN and AlN \cite{BNandAlN}, wurtzite AlN, GaN and InN
\cite{ph_Nitrides}, wurtzite GaN \cite{GaN_PP} and 
AlN \cite{AlN_PP}. All calculations yield good agreement with 
available experiments, not only for mode frequencies
but also for displacement patterns. In particular, the 
calculated eigenvector for the Raman-active $E_2$ mode 
in wurtzite GaN compares well with the eigenvector obtained 
from the study of the isotope effect \cite{GaN_Raman}.
The comparison with recent High-Resolution Inelastic X-Ray 
scattering measurement in wurtzite GaN \cite{GaN_PP} and 
AlN \cite{AlN_PP} shows good agreement with scattering 
intensities, thus validating the correctness of computed 
eigenvectors.

The phonon dispersions in BN and AlN are considerably different
from those of III-V semiconductors not containing first-row
elements: phonon dispersions for BN are similar to those of 
Diamond, while AlN dispersions are close to those of SiC.
Furthermore, the difference between BN and AlN phonon dispersions
cannot be explained by a simple mass approximation but derives
from the quite different degree of ionicity and covalent strength 
of the two materials. The marked ionicity of AlN bonding yields 
a pronounced structural and dielectric anisotropy in the wurtzite 
structure, larger than that of wurtzite BN and SiC \cite{BNandAlN}.
The three-phonon decay of the LO phonon in two acoustic phonons 
is not allowed in GaN and InN, since the LO frequency is much 
larger than the acoustic frequencies over the entire frequency 
spectrum \citeasnoun{ph_Nitrides}.

The pressure dependence of the dielectric and lattice-dynamical
properties of both zincblende and wurtzite GaN and AlN have been
recently calculated in \citeasnoun{GaNAlN_P}.

\paragraph{Other semiconductors}

Phonon dispersions of zincblende semiconductor CuCl were
calculated using the LAPW method \cite{CuCl}. CuCl exhibits large anharmonic 
effects: in particular, many peaks in neutron scattering measurements 
disappear at temperatures as low as room temperature. The calculated
phonon dispersions however agree well with low temperature experimental 
results.

The phonon dispersions for bulk layered semiconductor
$\epsilon$-GaSe were calculated by \citeasnoun{GaSe}. 
The bulk dispersions agree well both with neutron scattering 
results and with surface phonon measurements with inelastic
Helium-atom scattering. The calculation of (0001) surface 
phonons at the $\bar K $ point yields small differences 
with respect to the corresponding bulk phonons, as expected
for a layered material. This rules out previous assumptions
of anomalous surface phonon dispersions.
In a subsequent study \cite{GaSeInSe} the phonon dispersions 
of the similar material InSe was calculated.
The bulk dispersions for both GaSe and InSe compare favorably 
with experiments on (001) thin films epitaxially grown on 
hydrogen-terminated Si(111) $(1 \times1)$ surfaces.

The face-centered orthorhombic inter-metallic semiconductor
Al$_2$Ru exhibits strong far-IR absorption by optical phonons. 
Zone-center phonon frequencies, effective charges and dielectric 
tensor were calculated in \citeasnoun{Al2Ru}, showing 
anomalously large Born effective charges, in agreement with 
experiments. The analysis of the valence charge density shows
instead that the static ionic charges of Al and Ru are negligible.
Hybridization is proposed as the origin of both of the 
semiconducting gap and the anomalous Born effective charges.

\paragraph{Second-order Raman spectra in simple semiconductors}
\label{2ndRaman} 

Neglecting matrix element effects, second-order Raman spectra are
approximately given by the overtone density of states: 
$I(\omega)=g(\omega/2)$, where $g(\omega)$ is the vibrational DOS.
A more accurate description is obtained by combining the ab-initio
phonon spectra with phenomenological polarizability coefficients.
This technique was applied to solve the long-standing controversy
on the sharp peak in the spectrum of Diamond near the two-phonon cutoff
\cite{DiamondRaman}. As already observed in \citeasnoun{Diamond},
the sharp peak is due to a maximum in the vibrational density of 
states, originating from the over-bending of the uppermost phonon branch.
Such over-bending, and the maximum in the vibrational density of
states, are absent in the other elemental semiconductors Si and Ge.
Neither two-phonon bound states nor polarizability matrix element
effects are needed to explain the peak.

A similar technique was applied to the calculation of second-order 
Raman spectrum of AlSb in the $\Gamma_1$ symmetry \cite{AlSbRaman}.
\begin{figure}
\centerline{\psfig{figure=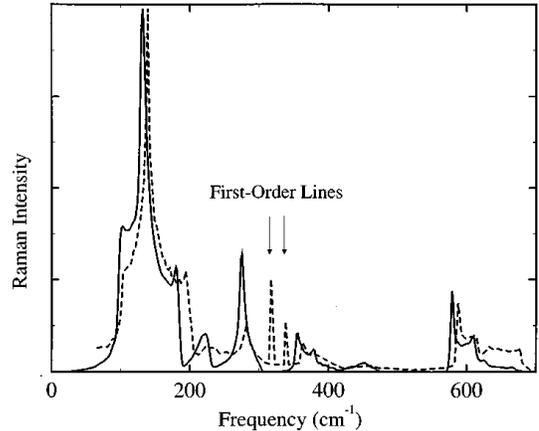,width=8.5truecm}}
\caption{Second-order Raman spectrum of AlSb in the $\Gamma_1$
representation at room temperature; the solid line represents the
theoretical calculation, the dashed line is the experimental spectrum.
(Reproduced from \protect\citeasnoun{AlSbRaman}.)}
\label{fig2}
\end{figure}
The calculated spectrum, shown in Fig.\ \ref{fig2},
agrees well with recent experimental data.
Both theory and experiment give clear evidence for the existence of 
an over-bending in the highest frequency branch of the phonon dispersion 
with a saddle point at the $\Gamma$ point as predicted 
by \citeasnoun{bulkoni}. This evidence is derived from a general
discussion of the critical-point behavior. Furthermore, the
over-bending in AlSb is explained as an effect of the very different
masses of the Al and Sb atoms in contrast to the over-bending in
Diamond, whose origin lies in a peculiarity of the force constants. 

In the above cases, a one-to-one correspondence between the overtone 
density of states and the second-order Raman spectrum is quite visible.
This is not the case for SiC \cite{SiCRaman}. The $\Gamma_1$ spectrum 
of SiC exhibits three distinct peaks at 1302, 1400, and 1619  cm$^{-1}$ 
which occur in the gaps of the overtone density of states. 
This exemplifies the importance of taking into account the coupling 
matrix elements.

\paragraph{Piezo-electricity in binary semiconductors}

DFPT is used to evaluate the linear response
of wave-functions to a macroscopic electric field, from which---using
the stress theorem of \citeasnoun{stress}---one finds 
the induced stress. Due to the presence of large cancellations between
contributions of opposite sign, the
results are very sensitive to the overall accuracy of the
calculation. Well-converged 
calculations yield results in good agreement with available 
experimental data in III-V \cite{piezoSdG} and in II-VI compounds,
with the exception of ZnTe \cite{piezoSdG2}. Nonlinear
piezoelectricity in CdTe was studied as well \cite{piezoAdC}. It was
found that piezoelectricity is linear over a wide range of
volume-conserving strains, while it displays strong 
nonlinearity  whenever the strain is not volume conserving. This
implies that the observed nonlinear effects can be accurately
accounted for by the linear piezoelectric response of the cubic system
at the strained volume.

\subsubsection{Simple metals and super-conductors}

An early calculation of a phonon dispersions in Nb and Mo metals
was performed by \cite{Zein3}, using a mixed basis set. The aim
of the calculation was to explain the presence of a dip in in 
the $(\zeta00)$ branch of Nb and its absence in Mo.

\begin{figure}[t]
\centerline{\psfig{figure=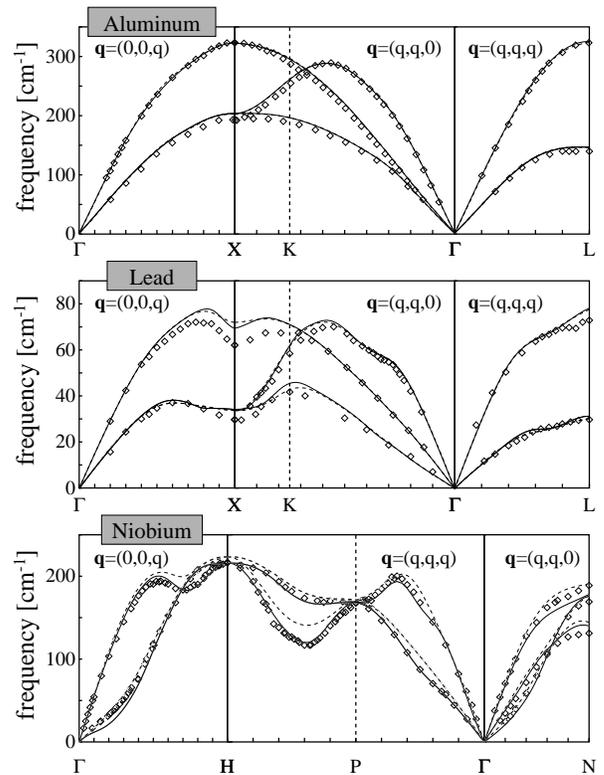,width=8.5truecm}}
\caption{Calculated phonon dispersions for fcc simple metal Al and Pb
and for the bcc transition metal Nb. Solid and dashed lines refer to
different smearing widths (0.3 and 0.7 eV, respectively). Experimental
data are denoted by diamonds. (Reproduced from 
\protect\citeasnoun{SdGmetalli}.)} \label{fig3}
\end{figure}

The accuracy of DFPT for metals (described in Sec. \secref{metals})
was demonstrated by \citeasnoun{SdGmetalli} for three test cases: 
face-centered cubic (fcc) Al, fcc Pb, and body-centered cubic (bcc) 
Nb (see Fig.\ \ref{fig3}).
\begin{figure}[tb]
\centerline{\psfig{figure=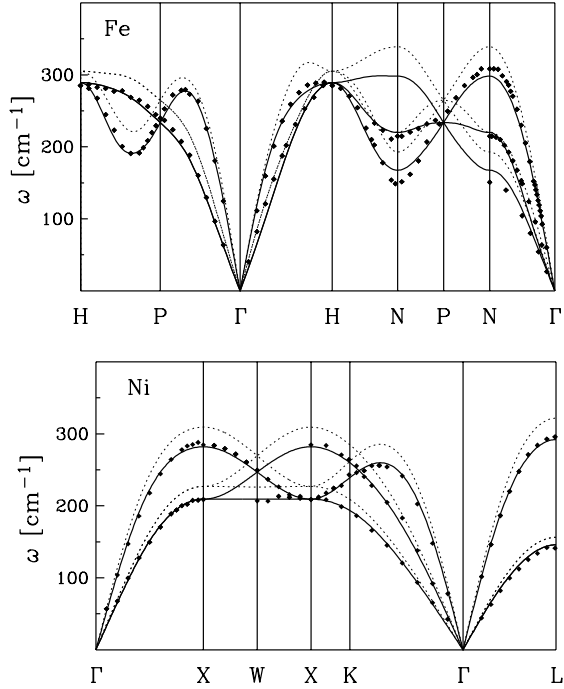,width=8.5truecm}}
\caption{Upper panel: calculated GGA phonon dispersions (solid lines)
for magnetic bcc Fe compared to inelastic neutron scattering data
(solid diamond) and to dispersions calculated within 
local spin density approximation (LSDA) (dotted lines).
Lower panel: calculated GGA phonon dispersions (solid lines) for magnetic
fcc Ni compared to inelastic neutron scattering data (solid diamond)
and to calculated LSDA dispersions (dotted lines). 
(Reproduced from \protect\citeasnoun{DCdG}.)} \label{fig4}
\end{figure}
In all cases the calculated dispersion 
curves are in good agreement with experiments if an appropriate 
smearing technique is used.
Phonons in fcc Cu, Ag, Au were calculated as a test case
for the use of ultrasoft PP's~\cite{USphonon}.
The alternative linear response technique of \citeasnoun{Quong_LR}
was applied to the phonon dispersions and interatomic force
constants of fcc Al.
Phonons of magnetic bcc Fe and fcc Ni have been recently calculated
by~\citeasnoun{DCdG}.
For these metals, good agreement with experiments is obtained 
using a spin-polarized GGA functional for the exchange and 
correlation energy
(see Fig.\ \ref{fig4}).
Phonons in hexagonal close-packing (hcp) Ru were calculated 
and measured with inelastic neutron scattering \cite{rutenio}. 
Many phonon anomalies have been found. Phonons in fcc Ag and 
in hcp Y were calculated as a test of mixed basis set
technique \cite{mixedbasis}.

\begin{figure}[bt]
\centerline{\psfig{figure=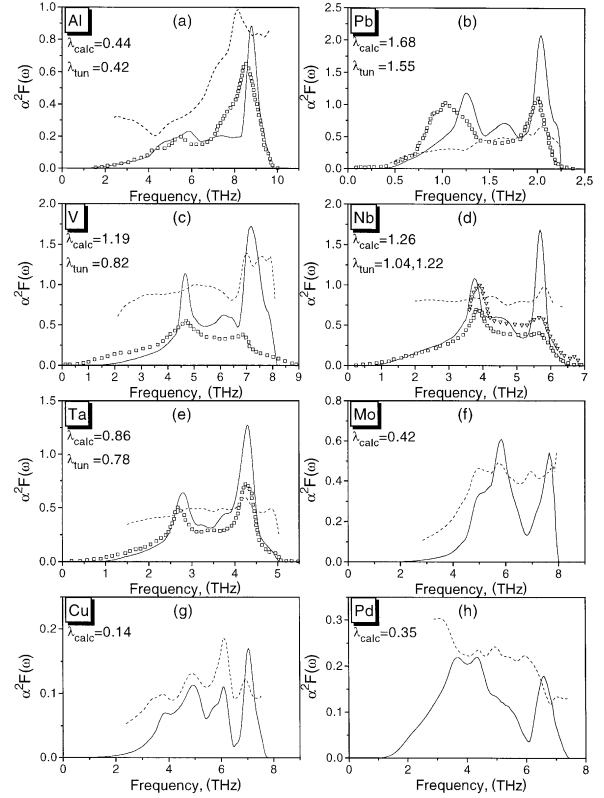,width=8.5truecm}}
\caption{(a)-(h) Calculated spectral functions $\alpha^2 F(\omega)$ of
the electron-phonon interaction (full lines) for the eight elemental
metals considered in \protect\citeasnoun{LMTO_metals}. The behavior
of the electron-phonon prefactor $\alpha^2(\omega)$ [defined simply
as the ratio $\alpha^2 F(\omega)/ F(\omega)$] is shown by dashed
lines. Symbols plots present the results of available tunneling
experiments. (Reproduced from \protect\citeasnoun{LMTO_metals}.)}
\label{fig5}
\end{figure}

The main interest of phonon calculations in metals is possibly
related to transport properties, and notably super-conductivity. 
The availability of accurate phonon dispersions in the entire 
BZ allows the calculation of the electron-phonon 
(Eliashberg) spectral function $\alpha^2 F(\omega$) and of 
the mass enhancement parameter $\lambda$ that enters the
MacMillan equation for the transition temperature $T_c$ to 
super-conductivity. Other important quantities that can be
calculated are the transport spectral function
$\alpha_{tr}^2 F(\omega$) and the $\lambda_{tr}$ coefficients,
that determine the electrical and thermal resistivity
in the normal state.
In simple metals calculations of $\lambda$ and $\alpha^2 F(\omega$)
were performed in Al, Pb, Li \cite{Quong_e-ph},
in Al, Cu, Mo, Nb, Pb, Pd, Ta, V \cite{LMTO_metals},
in Al, Au, Na, and Nb \cite{el-ph}.
Transport spectral functions and coefficients
\cite{LMTO_metals,el-ph} and phonon line-widths 
due to the electron-phonon coupling \cite{el-ph}
were also calculated.
Fig.\ \ref{fig5} shows the results of \citeasnoun{LMTO_metals}
for $\alpha^2 F(\omega$).

The calculation of electron-phonon coefficients found
a remarkable application, beyond simple metals, in the study 
of the behavior of molecular solids S, Se, Te under 
pressure. With increasing pressure these transform first to 
a base-centered orthorhombic (bco) super-conducting structure, 
followed by a rhombohedral $\beta$-Po phase, and finally for 
Se and Te by a bcc phase.

At the phase transition between the $\beta$-Po and the bcc phase,
a jump is observed in $T_c$ in Te. The origin of this jump was
clarified \cite{Te} through the study of phonon dispersions and of 
the electron-phonon interactions. The phonon contribution to the
free energy is shown to be responsible for the difference in 
the structural transition pressure observed in low and room
temperature experiments.

In S the $\beta$-Po phase is predicted to be followed by a
simple-cubic phase that is stable over a wide range of pressures 
(280 to 540 GPa), contrary to what is observed in Se and Te.
The calculated phonon spectrum and electron-phonon coupling 
strength \cite{CompressedSulfur} for the lower-pressure 
$\beta$-Po phase is consistent with the measured super-conducting 
transition temperature of 17 K at 160 GPa. The transition temperature 
is calculated to drop below 10 K upon transformation to the predicted 
simple-cubic phase.

\subsubsection{Oxides}

Oxides present a special interest and a special challenge for
anyone interested in phonon physics. On the one hand, many
very interesting materials, such as ferroelectrics and high-$T_c$
super-conductors, are oxides. On the other hand, good-quality 
calculations on oxides are usually nontrivial, both for technical
and for more fundamental problems. In a straightforward PW-PP 
framework, the hard PP of Oxygen makes calculations expensive:
the use of ultrasoft PP's is generally advantageous. LDA is
known to be not accurate enough in many cases (and sometimes the
entire band-structure approach is questionable in oxides). Many
oxides have complex structural arrangements. In spite of all
these problems, there have been several calculations of
phonon-related properties in oxides from first principles.
These calculations are described in the remaining of this 
paragraph, with the exception of work on phase transitions
that is deferred to Sec. \secref{appl_soft}.

\paragraph{Insulators}

In alkaline-earth oxides MgO, CaO, and SrO in the rocksalt structure,
LDA yields good agreement with available experimental data for lattice
vibrations \cite{MgO}.
The investigation of the phonon-induced charge-density fluctuations
of MgO and CaO at the L point of the BZ partially
supports the breathing-shell model of lattice dynamics and rules out
the charge-transfer model for this class of materials. Moreover, the
calculations show that the breathing-like charge-density response is
more pronounced for oxygen than for the cations in these compounds. 

\begin{figure}[tb]
\centerline{\psfig{figure=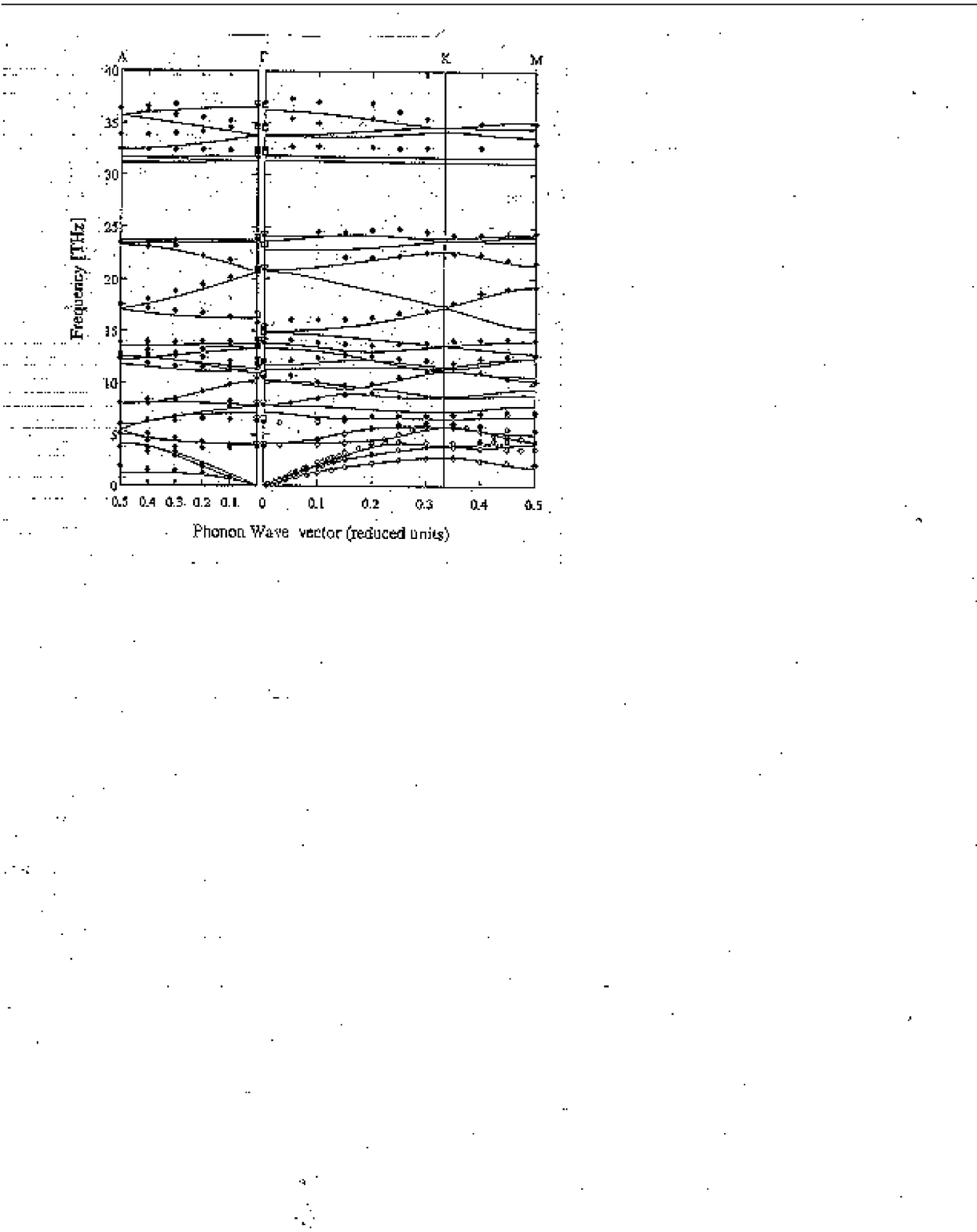,width=8.5truecm}}
\caption{Phonon band structure of $\alpha$-quartz along selected
directions. Symbols: experimental data; 
lines: theoretical results. (Reproduced from 
\protect\citeasnoun{GonzeFC1}.)}\label{fig6}
\end{figure}

Silicon Dioxide (SiO$_2$) is a much-studied prototypical
tetrahedrally coordinated compound, existing in a large
variety of different structures. At ambient conditions
the ground state structure of SiO$_2$ is $\alpha-$quartz,
whose phonon dispersions, dielectric tensor, and effective 
charges were studied by \citeasnoun{GonzeCG}. Effective 
charges in $\alpha-$quartz are anisotropic (here calculated
for the first time in DFPT) and those for O exhibit some 
anomalous character (see following section).
Fig.\ \ref{fig6} shows the calculated phonon dispersions for
$\alpha-$quartz. IFC's in $\alpha-$quartz were
calculated in a subsequent paper \cite{GonzeFC1}.
The availability of force constants is important because it 
allows to extensively test semiempirical interatomic potentials
that are used in molecular-dynamics simulations of silica.
Phonons, dielectric tensor, and effective charges of the 
high-pressure, sixfold-coordinated phase stishovite were
studied by \citeasnoun{stishovite}.

The phonon dispersions of $\alpha-$Al$_2$O$_3$ (sapphire)
have been recently calculated in \citeasnoun{zaffiro}, 
using LDA and a mixed
basis set. Sapphire has a rhombohedral unit cell containing 2 formula
units (10 atoms). A weak anisotropy in the dielectric tensor and in
the effective charges is found.

\paragraph{Ferroelectrics}

The phonon frequencies at $\Gamma$, dielectric tensor and 
effective charges of Titanium dioxide (TiO$_2$) in the rutile
structure were calculated in \citeasnoun{TiO2}.
Rutile TiO$_2$ is an {\em incipient ferroelectric}: the
frequency of the TO mode A$_{2u}$ decreases with temperature
but never goes to zero. It is found that the Born effective 
charges of TiO$_2$ rutile are much larger than the nominal 
ionic charges of Ti (4+) and O $(2-)$ ions (and much larger 
than those of stishovite in spite of the similar structure).
This may sound rather counterintuitive but it is typical 
of all ABO$_3$ ferroelectrics in the perovskite 
structure \cite{ABO3_Z}, whose prototypical material
is Barium Titanate (BaTiO$_3$). The effective charges of 
BaTiO$_3$ and of similar compounds exhibit both values
that are definitely larger ({\em anomalous}) than the ionic 
charge, and a strong anisotropy (for O only in the cubic 
structure: the effective charge is anomalous for 
displacements parallel to the Ti-O bond, {\em normal},
that is close to the ionic value, in the orthogonal
direction). 
By performing an appropriate band-by-band decomposition \cite{bandbyband}
of contributions, this effect can be tracked
to the dynamical change of hybridization, mainly between 
O $2p$ and Ti $3d$ orbitals \cite{BaTiO3_Z}. 
Born effective charges for cubic WO$_3$ in the defect-perovskite 
structure \cite{WO3} and for KNbO$_3$ \cite{KNbO3_Z} follow the 
same pattern.
The important role of covalence in determining the anomalous polarization
was demonstrated also by~\citeasnoun{PRB}, by computing the 
effective charges of a fake material, similar to KNbO$_3$, 
where covalence was artificially removed by an additional potential.

Phonons at $\Gamma$ for the cubic (ideal perovskite) and for 
the rhombohedral phases of BaTiO$_3$ were calculated in 
\citeasnoun{BaTiO3_G}. The complete phonon dispersions for the 
cubic structure, together with an analysis of the IFC's, can 
be found in \citeasnoun{BaTiO3_ph}.

\paragraph{High-$T_c$ super-conductors}

Although the microscopic mechanism which gives rise to
super-conductivity in high-$T_c$ oxides is still under active debate,
accurate phonon dispersions and electron-phonon coefficients constitute an
important piece of information for understanding the properties of these
materials. Calculations were performed---at the LDA level---for
CaCuO$_2$ \cite{CaCuO2}, Ba$_{0.6}$K$_{0.4}$BiO$_3$ \cite{BaBiO3}, and
La$_2$CuO$_4$ \cite{La2CuO4}.

In hole-doped ($n=0.24$) CaCuO$_2$, the phonon dispersions and 
electron-phonon coupling, for both $s-$ and $d-$wave pairing,
were calculated using LMTO linear-response techniques. 
The resulting values of 
$\lambda\simeq 0.3$ for $d_{x^2-y^2}$ symmetry and 
$\lambda \simeq 0.4$ for $s$ symmetry suggest that the 
electron-phonon mechanism alone is insufficient to explain the high 
$T_c$ but could enhance another $d-$wave pairing mechanism \cite{CaCuO2}.

Similar calculations \cite{BaBiO3} were performed for
Ba$_{0.6}$K$_{0.4}$BiO$_3$, using the virtual crystal 
and mass approximations. The $\lambda$ parameter,
including also anharmonic contributions, is found to be 
$\lambda=0.34$, a too small value to explain high-$T_c$
super-conductivity in this system within the standard mechanism.

In tetragonal La$_2$CuO$_4$, the phonon frequencies and eigenvectors 
were calculated using LAPW linear response techniques \cite{La2CuO4}. 
\begin{figure}[tb]
\centerline{\psfig{figure=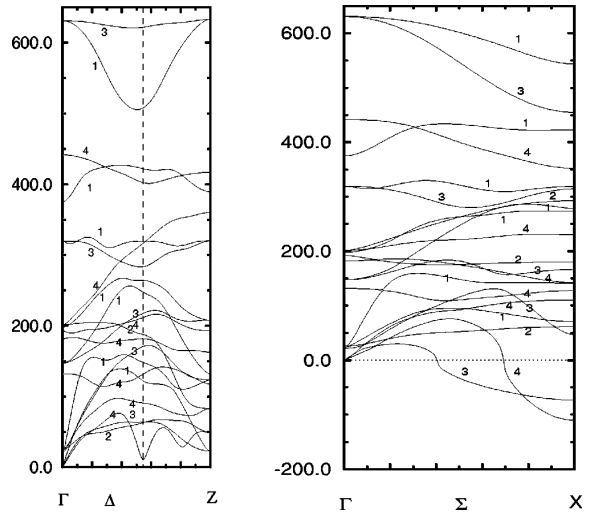,width=8.5truecm}}
\caption{Calculated phonon dispersion of tetragonal La$_2$CuO$_4$ along
the $(\xi,\xi,0)$ ($\Sigma$) and $(\xi,0,0)$ ($\Delta$) directions. The
frequencies are in cm$^{-1}$ and the imaginary frequencies are
represented as negative numbers. The vertical dashed line corresponds
to the boundary of the first BZ. (Reproduced from
\protect\citeasnoun{La2CuO4}.)}
\label{fig7}
\end{figure}
The results (see Fig.\ \ref{fig7}) are 
generally in good agreement with experiments, with the exception 
of lowest-lying branches involving anharmonic motion of the apical 
oxygen atoms parallel to the CuO$_2$ planes. The octahedral tilt 
mode at the $X$ point is found to be the most unstable mode throughout 
the BZ, consistent with the observed phase transition
to the orthorhombic structure at low temperature. The calculated 
dispersion of the highest frequency $\Sigma_3$ branch is in good 
agreement with experiment, showing that a proposed large 
renormalization of the phonon spectrum by a Jahn-Teller electron-phonon 
interaction is unlikely. 

\subsubsection{Other materials}

Fullerene C$_{60}$ forms a molecular solid that can be 
{\em doped} with up to six alkali atoms (K, Rb, Cs) per
C$_{60}$. The alkalis lose their electrons
that fill the conduction band of C$_{60}$,
originated by molecular t$_{1g}$ states.
Sizable frequency shifts
and a large enhancement in the intensity of the four 
IR-active modes of molecular C$_{60}$ is observed
when K (or Rb, or Cs) is added to solid C$_{60}$.
The ten Raman-active modes of molecular C$_{60}$
also exhibit large shifts.
In insulating bcc K$_6$C$_{60}$ the phonons at $\Gamma$
and the effective charges were calculated \cite{K6C60}.
The frequencies of Raman- and IR-active modes are
reported in Table I.
It is found that the structural relaxation of the C$_{60}$
molecule is primarily responsible for the frequency changes,
while the change in IR relative intensities is a 
consequence of the electron transfer. 
The potassium vibrations are found to lie within the range
68--125~cm$^{-1}$ and are well decoupled from the C$_{60}$
intramolecular modes. The DFPT results do not support
the so called {\em charged phonon} model.

\begin{table} 
\caption{
K$_6$C$_{60}$: Theoretical and experimental frequencies $\omega$ of the
optically C$_{60}$-like active modes and shifts $\Delta \omega$ with respect to
C$_{60}$. We use the labeling of the I$_h$ group. The H$_{g}$ modes
split into a triplet T$_g$ (left) and a doublet E$_g$ (right).
The experimental values are ascribed according to the calculated
ordering. Their shifts refer to the centers of gravity.
Units are cm$^{-1}$ \protect\cite{K6C60}.} 
\label{tab2}
\begin{tabular}{l|rrr|rr}
&\multicolumn{3}{c|}{$\omega$} & \multicolumn{2}{c}{$\Delta\omega$}\\\cline{1-6}
Mode & Theory & Expt.\tablenotemark[1]& Expt.\tablenotemark[2] & 
Theory & Expt.\tablenotemark[1]\tablenotemark[2]\\
\tableline
 A$_{g}(1)$  &  507      &   502    &   501     &  12   &   9,~~6\\
 A$_{g}(2)$  & 1469      &  1432    &  1431     & $-$35 & $-$36,$-$36\\
 H$_{g}(1)$  &  274,~266 &  281,~269&  280,~268 &  11   &   6,~~5\\
 H$_{g}(2)$  &  412,~419 &  419,~427&  419,~427 & $-$12 & $-$8,~$-$9\\
 H$_{g}(3)$  &  660,~660 &  656,~676&  656,~676 & $-$51 & $-$44,$-$47\\
 H$_{g}(4)$  &  770,~769 &       761&  760,~728 & $-$13 & $-$11,$-$25\\
 H$_{g}(5)$  & 1107,1109 & 1094,1120& 1093,1122 & $-$12 &   5,~~6\\
 H$_{g}(6)$  & 1264,1268 & 1237     & 1232,1237 & $-$15 & $-$11,$-$14\\
 H$_{g}(7)$  & 1423,1414 & 1383     & 1384      & $-$31 & $-$43,$-$38\\
 H$_{g}(8)$  & 1508,1498 & 1476     & 1481,1474 & $-$74 & $-$97,$-$95\\
 \tableline
      &  & Expt.\tablenotemark[3]& Expt.\tablenotemark[4]& & Expt.\ 
\tablenotemark[3]\tablenotemark[4]\\
 T$_{1u}(1)$ &  466      &   467    &  467      & $-$61 & $-$61,$-$60\\
 T$_{1u}(2)$ &  571      &   565    &  564      & $-$15 & $-$12,$-$12\\
 T$_{1u}(3)$ & 1215      &  1182    & 1183      &  $-$3 &     0,~~0\\
 T$_{1u}(4)$ & 1395      &  1340    & 1341      & $-$67 & $-$86,$-$88\\
\end{tabular}
\tablenotetext[1]{Data from \citeasnoun{Dresselhaus}}
\tablenotetext[2]{Data from \citeasnoun{kuzrev}}
\tablenotetext[3]{Data from \citeasnoun{MCMartin}}
\tablenotetext[4]{Data from \citeasnoun{kuz94}}
\end{table}

Elemental Boron and Boron-rich solids tends to form complex structures
formed by assembly of icosahedral B$_{12}$ units. The exact structure
and vibrational properties of such materials are not well known.
The comparison of accurate phonon calculations with IR and Raman 
measurements is of great help in determining the atomic structure.

Icosahedral Boron presents a very sharp peak at 525 cm$^{-1}$
whose vibrational character has been for a long time denied. 
New Raman scattering experiments under pressure were compared 
with ab initio lattice dynamics calculations \cite{Boron}. The very 
good agreement of the mode frequencies and their pressure coefficients 
yields unambiguous assignment of all observed features, including the 
525 cm$^{-1}$ line which is a highly harmonic librational mode of the 
icosahedron and mainly involves bond bending. 
This mode is also identified in the Raman spectrum of other 
icosahedral boron-rich solids \cite{Boron}.

Boron Carbide, B$_4$C, is the third hardest material after Diamond and
cubic BN. The building blocks of the crystal are distorted icosahedral
B$_{11}$C units, but their precise arrangement is still experimentally
unknown. The structure of icosahedral B$_4$C boron carbide was
theoretically determined by comparing existing IR and Raman spectra
with accurate ab initio lattice-dynamical calculations, performed
for different structural models \cite{B4C}. The examination of the
inter- and intra-icosahedral contributions to the stiffness shows that
intraicosahedral bonds are harder than intericosahedral ones, contrary
to previous conjectures \cite{B4C}.

Phonons in solid Cl--a typical molecular solid--were calculated
in \citeasnoun{Cl}. The calculated intramolecular distance is
too large by 8\%, maybe due to the pseudo-potentials used,
and as a consequence the internal phonon frequencies are too low.
Intermolecular distances are in good agreement with
experiment\footnote{The agreement is possibly fortuitous since LDA
does 
not treat correctly van der Waals interactions and materials held
together by them.} and
so are most of the calculated external phonons (calculated at
$\Gamma$ and $Y$) \cite{Cl}.

The phonon dispersions of transition-metal carbide NbC were studied 
as a sample application in a technical paper \cite{LMTO_LR2}. 
NbC presents several peculiar phonon anomalies that are well 
reproduced by the LMTO calculations.

Zone-center phonons and dielectric properties ($\epsilon_\infty$,
$Z^*$) of cubic rocksalt alkali hydrides LiH and NaH were
calculated in \citeasnoun{Zein2}. More complete phonon dispersions 
for LiH and LiD appeared in \citeasnoun{Guido}.

\subsection{Phonons in semiconductor alloys and super-lattices}

The calculation of phonon dispersions in systems described by large
unit cells or lacking periodicity altogether presents a special 
challenge. Disordered systems (such as amorphous materials or
substitutionally disordered alloys) can be described in a PW-PP
framework by periodically repeated fictitious {\em
super-cells}. However the needed computational effort quickly grows
with the size of the unit cell or super-cell 
(as $\propto N^\alpha$, where $N$ is the number of atoms, 
$\alpha\sim 3\div 4$ in practical calculations) 
so that even with the best algorithms and the best computers 
available one is limited to systems having at most $\sim 100$ atoms.
Such size may or may not be adequate for the physical system under 
investigation. If it is not, the brute-force approach must be
supplemented by a more targeted approach. Typically, accurate
first-principles calculations in suitably chosen small systems
are used to set up a computationally manageable model for the 
large system.

\subsubsection{GaAs/AlAs super-lattices}

GaAs/AlAs super-lattices and alloys are a very successful
example of how first-principles calculations supplemented by
an appropriate model can lead to an accurate description
of the dynamical properties of real systems. As already
mentioned in Sec.\secref{III-V}, the mass approximation 
is the key for obtaining accurate dynamical matrices for
large systems at a modest computational cost, once the
interatomic force constants in real space for GaAs
(or for the Ga$_x$Al$_{1-x}$As virtual crystal) were
obtained.

The interest for phonons in GaAs/AlAs super-lattices
is a consequence of the progress in epitaxial techniques
that allowed the growth of ultrathin super-lattices
(with a period of $<10$ atomic layers), in particular 
along the (100) direction. Owing to the large difference
between the cationic masses, GaAs and AlAs optic mode
occur in different frequency ranges. In the super-lattice, 
optical vibrations are confined in one or in the other of 
the materials. In perfectly ordered super-lattices, the 
relation between confined modes and bulk dispersions of 
the component materials is given by the {\em unfolding}
model \cite{unfolding}. 

In ultrathin super-lattice it was not clear however how
valid the unfolding model was, how much disorder at the 
interfaces (cation intermixing) was present, and what 
the effect of disorder would be. The simulation of the 
dynamical properties of ordered and of partially disordered 
super-lattices, using the mass approximation and large
super-cells \cite{GaAlAs_sl1,GaAlAs_sl2} allowed
to clarify the problem. It was found that the unfolding
model is very well verified in ultrathin super-lattices,
but that a sizable amount of cation intermixing is
needed in order to explain the details of the Raman
spectra. These findings later received independent 
confirmations \cite{GaAsAlAs_confexp,GaAsAlAs_confth}.

It should be remarked that the ability to obtain such results
critically depends on the quality of the modelization used.
Empirical models, even good ones like the BCM, are too crude
and do not adequately describe the dispersions of AlAs, and as
a consequence, the details of the spectrum.

Calculations of the phonon spectra in more exotic GaAs/AlAs
systems proceeds quite in the same way. For instance, the 
vibrational properties of an array of GaAs thin wires embedded
in AlAs were calculated in \citeasnoun{wires}.

\subsubsection{GaAs/AlAs alloys}

\begin{figure*}[t]
\centerline{\psfig{figure=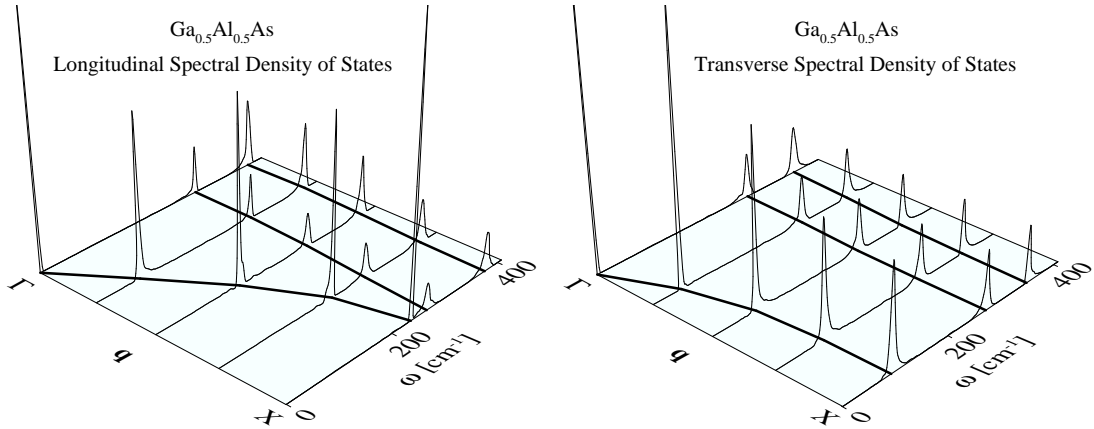,angle=90,width=15.0truecm}}
\caption{Spectral densities of states of Ga$_{0.5}$Al$_{0.5}$As along
the $\Gamma-X$ direction. The position of the peaks are indicated by
the solid lines in the $\omega$-{\bf q} plane. (Reproduced from 
\protect\citeasnoun{GaAlAs_all}.)}
\label{fig8}
\end{figure*}

The same approach that was used for GaAs/AlAs super-lattice
was also used to clarify the results of Raman measurements
in GaAs/AlAs homogeneous alloys. The Raman spectra has two distinct
peaks, corresponding to the vibrations of each cationic species
separately. This behavior is called {\em two-mode} and is typical 
of all $A_xB_{1-x}C$ III-V alloys, with the exception of
Ga$_x$In$_{1-x}$P. The peaks are shifted and asymmetrically broadened 
with respect to the pure materials. This asymmetry was
interpreted assuming that Raman-active phonon modes are localized 
on a scale $\sim 100 ~\rm\Angstrom$, but such assumption was challenged
by other experimental results indicating that phonons have 
well-defined momentum and are coherent over distances $> 700 ~\rm\Angstrom$.
The results of simulations using a 512-atom super-cell (Fig.\ \ref{fig8}) 
clearly 
indicate that the latter picture is the correct one \cite{GaAlAs_all}.

\subsubsection{Si/Ge super-lattices and alloys}

Si/Ge super-lattices and alloys are more difficult not only to grow,
but also as a subject of theoretical study, than GaAs/AlAs systems.
The mass approximation in Si/Ge systems is quite poor, yielding 
errors of as much as 20 cm$^{-1}$ for optic phonons. Moreover
in Si/Ge systems the lattice parameters of the two components 
differ by as much as $4\%$, thus giving raise to sizable strain
and atomic relaxations that must be taken into account. GaAs/AlAs 
systems instead are almost perfectly matched (their lattice 
mismatch is a modest $0.2\%$).

In order to achieve the same level of accuracy for Si/Ge systems as
in GaAs/AlAs systems, one has to supplement the mass approximation
with a correction that takes into account the effect of strain and of
atomic relaxation. This goal is achieved by introducing higher-order
interatomic force constants that are fitted to first-principles results
for a few selected configurations \cite{SiGe_2}.  More complex systems
can then be simulated by suitable super-cells as in GaAs/AlAs. In this
way the vibrational properties of Si$_x$Ge$_{1-x}$ were studied
\cite{SiGe_2}. In particular it was shown that this approach is able
to reproduce the Raman spectra in Si$_{0.5}$Ge$_{0.5}$, including the
details due to the local arrangement of atoms, while the mean-field
approach (Coherent Potential Approximation, or CPA) badly fails in
reproducing the three-mode character of the spectra \cite{SiGe_1}.
The higher-order interatomic force constants were also used to
study the vibrational properties of ideal and realistically intermixed
Si/Ge super-lattices \cite{SiGe-sl1,SiGe-sl2,SiGe-sl3}.

\subsubsection{AlGaN alloys}

The zone-center vibrational properties of wurtzite Al$_x$Ga$_{1-x}$N
alloys, over the entire range of composition from pure GaN to pure
AlN, were studied by \citeasnoun{AlGaN_all} using the mass 
approximation and the arithmetic average of the interatomic
force constants of the two pure materials previously calculated by
\citeasnoun{ph_Nitrides}.  While some of the alloy modes display a
two-mode like behavior, do not preserve a well-defined symmetry and have
a large broadening, the LO modes, instead, display a one-mode behavior
and have a well defined symmetry, small broadening, and a pronounced
dependence of the frequency upon alloy composition. Therefore these modes
are proposed as the best candidates for the compositional characterization
of the alloy \cite{AlGaN_all}.

\subsubsection{GaP/InP alloys}

A different approach to study phonons in semiconductor alloys 
uses suitably chosen small super-cells, the {\em special quasi-random
structures} \cite{SQS}, to simulate a disordered system. 
The evolution of 
the vibrational properties in GaP/InP systems with long-range order 
was studied using such approach (with a 16-atom cell)
to calculate the phonon spectra of random Ga$_{0.5}$In$_{0.5}$P 
\cite{GaInP}. The phonon spectra of pure GaP, InP, and of CuPt-type
ordered GaInP$_2$ were calculated for comparisons.
It was found that ordered GaInP$_2$ and Ga$_{0.5}$In$_{0.5}$P
have qualitatively different phonon spectra: ordered
GaInP$_2$ exhibits a two-mode behavior, with two GaP-like and 
two InP-like phonon modes, while disordered Ga$_{0.5}$In$_{0.5}$P 
exhibits a {\em pseudo-one-mode} behavior: two LO modes, one of GaP 
and another of mixed GaP/InP character, appear, while the TO modes 
of GaP and InP have merged into a single alloy mode. This is in 
remarkable agreement with experiments \cite{GaInP}.

\subsubsection{Localized vibrations at defects}

Localized vibrational modes of impurities contain a wealth of
information on the local structure of the defect. Their analysis
require an accurate knowledge of the phonon spectra of the host
crystal. The interpretation of the isotopic fine structure of
substitutional impurities in III-V semiconductors was performed 
using DFPT for the bulk crystal and a Green's function technique, 
with results that are far superior to those obtained using model
calculations \cite{fine1}. With these techniques, the host isotope
fine structure of $^{12}$C:As and $^{11}$B:As local modes in GaAs
\cite{fine1}, of the As:P gap mode \cite{fine2} and of
B:Ga gap and local modes in GaP \cite{fine3} were
successfully analyzed.

\subsection{Lattice vibrations at surfaces}

\begin{figure}[t]
\centerline{\psfig{figure=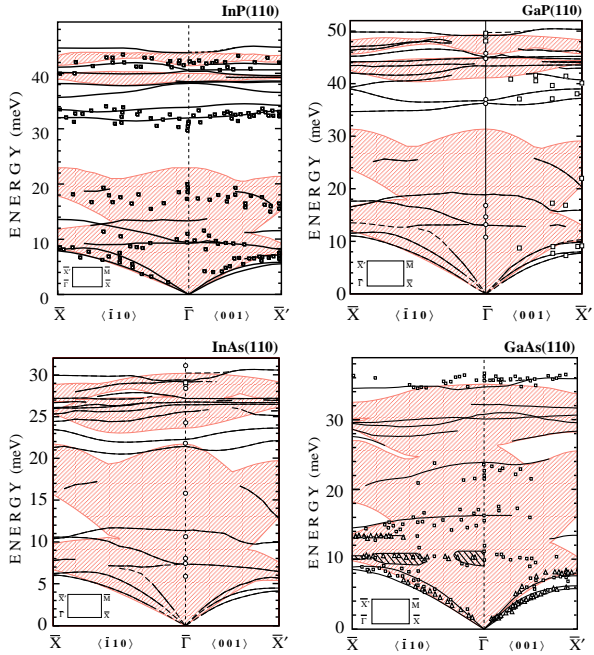,width=8.5truecm}}
\caption{ Phonon dispersion of InP, GaP, InAs, and GaAs(110). The
large shaded area represents the surface projected bulk
states. Surface-localized and resonant modes are indicated by {\it solid}
and {\it dashed} lines. The squares and triangles represent data from
HREELS and HAS, respectively. Earlier LDA-results 
are indicated by circles. (Reproduced from \protect\citeasnoun{Eckl}.)}
\label{fig9}
\end{figure}

Most DFPT calculations on surface phonons focussed on
semiconductor surfaces \cite{review,Eckl,001}. 
Calculations were performed for GaAs(110) \cite{GaAs110}, 
InP(110) \cite{InP}, GaP(110) and InAs(110) \cite{Eckl2}, 
InSb(110) \cite{InSb110}; for Si(111) \cite{Ancilotto}; 
for Si(001) \cite{Si001} and Ge (001) \cite{Ge001}; 
for H-covered (110) surfaces of GaAs, InP \cite{HonSurf},
and of GaP, InAs \cite{Eckl}; for H- and As-covered Si(111) 
$(1\times 1)$ surfaces \cite{Eckl,HonSi111}; for Ga- and 
B-covered  Si(111) $(\sqrt{3}\times\sqrt{3})$ R30 \cite{Eckl};
for Ge on GaAs(110)~\cite{GeonGaAs}; for As-covered (110)
surfaces of GaAs, GaP, InAs, InP \cite{As-covered-III-V}.
The work on semiconductor surface phonons has been described 
in a recent extensive review article \cite{review}.
In Fig.\ \ref{fig9} we show an example of the accuracy that can
be reached by these calculations in III-V surfaces.

Other DFPT calculations were performed for H-covered and 
clean W (110) surfaces \cite{H:W110}, for Be(0001) \cite{Be0001} 
and for Ag(111) \cite{Ag111} surfaces. The description of the two 
latter works is deferred to Sec.\secref{appl-thermal}. 

The hydrogen-covered (110) surface of W exhibits phonon anomalies,
clearly caused by H adsorption, whose nature is still unclear. 
An anomaly in the upper Rayleigh phonon branch along the (001) 
direction  is observed both in Helium-atom scattering (HAS) and 
in electron energy-loss (EELS) spectra. In the lower branch,
a similar anomaly is observed by EELS, whereas a much deeper one
is only detected by HAS. DFPT calculations \cite{H:W110} yield 
excellent agreement with both HAS and EELS for the upper anomalous
branch, and with EELS for the lower branch, provided that a careful 
sampling of the surface BZ is performed. Such anomalies
are interpreted as due to Fermi-surface nesting (Kohn anomaly). 
However the calculations do not predict the deep anomaly in the 
lower branch observed by HAS, whose nature is still not clear. As a
byproduct of these calculations, the phonon
dispersions for bulk bcc W and for the clean (110) surface are
calculated, and both are found
in excellent agreement with experiments \cite{H:W110}.
\subsection{Soft phonons and pressure-induced lattice transformations}
\label{sec:appl_soft}

Many phase transitions, both induced by pressure and by temperature,
are driven by a lattice instability. This may be an elastic
instability (leading to a change of shape of the unit cell) or 
a phonon instability (a {\em soft phonon}, whose energy goes to
zero). When soft phonons have a nonzero ${\bf k}$ vector (usually at
the border of the BZ) the distortion that sets up causes an increase
of the dimension of the unit cell.
The identification of the soft phonon responsible for the phase
transition is only the first step in understanding the phase
transition. The next step is usually the construction of a realistic
model that takes into account 
all the relevant anharmonic interactions responsible for the
stabilization of the low-symmetry structure (these may include
coupling with the strain, 
multiple soft modes, and so on). 

The earliest calculation using DFPT of a phonon instability was
performed on narrow-gap semiconductors GeTe, SnTe, PbTe
\cite{Zein4}. The first two exhibit a transition for a 
high-temperature cubic rocksalt phase to a low-temperature
rhombohedral phase, respectively at $T\sim 700K$ and $T\sim 140K$.
The calculation of the dielectric and zone-center phonons
in the cubic phase yields negative values for $\omega_{TO}$
in GeTe and SnTe and an anomalously low value in PbTe,
consistently with experimental observations.

\subsubsection{Ferroelectrics}

The ferroelectric transition in perovskite materials, whose most
famous example is BaTiO$_3$, is closely related to a 
lattice instability. In ferroelectrics (and in general in 
temperature-induced phase transitions) anharmonicity plays
a fundamental role: harmonic calculations generally yield
a negative frequency for the soft mode at zero temperature.
At higher temperature, anharmonicity stabilizes the soft mode.
Accurate phonon calculations are in any case the starting point
to  construct an effective Hamiltonian for the ferroelectric 
transition through the use of a localized, symmetrized basis set of
{\em lattice Wannier functions} \cite{LatticeWannier}. These are the
phonon analogous of electronic Wannier functions for electrons.
Furthermore, the mapping of the phonon instabilities in the full BZ gives
a real space picture of the ferroelectric instability, even when it
involves coordinated atomic displacements in several unit cells.

Ferroelectric phase transitions were studied using DFPT 
in Barium Titanate, BaTiO$_3$ \cite{BaTiO3},
Potassium Niobate, KNbO$_3$ \cite{KNbO3,KNbO3_Z},
Strontium Titanate, SrTiO$_3$ \cite{SrTiO3fer},
Lead Titanate, PbTiO$_3$ \cite{PbTiO3}, Lead Zirconate, PbZrO$_3$
\cite{BaPbTiZrO3}.

In KNbO$_3$ phonons at the $\Gamma$ point, effective charges 
and dielectric tensor were calculated for the cubic, 
tetragonal and rhombohedral perovskite structure.
In the hypothetical cubic structure soft modes are present, 
one of these mode is stabilized in the experimentally observed tetragonal 
structure, and all of them are stable in the rhombohedral structure 
that turns out to be the most stable one. 
An effective Hamiltonian is constructed with lattice Wannier function 
by \cite{KNbO3_H}.
For the cubic structure a complete mapping of the phonon dispersion in
the Brillouin zone has been computed. The results show a soft mode 
dispersion that exhibits an instability of a pronounced two 
dimensional nature in reciprocal space and suggest a one-dimensional chain
type instability oriented along the (100) direction
of displaced Nb atoms\cite{KNbO3}.
Similar phonon dispersions and chain type instability were
also found in BaTiO$_3$~\cite{BaTiO3_ph}.

SrTiO$_3$ exhibits both ferroelectric and anti-ferrodistortive instabilities.
In the cubic structure, phase space of the ferroelectric 
instability is greatly reduced compared to KNbO$_3$.
Anti-ferrodistortive instabilities exist in one-dimensional 
cylindrical tubes extending along the entire R$-$M$-$R line in the
BZ. The one-dimensional character of these tubes corresponds to
real space planar instabilities characterized by rotations of 
oxygen octahedra~\cite{SrTiO3fer}.

In PbTiO$_3$, phonons at $\Gamma, R, X, M$ for the cubic structure are
used to construct lattice Wannier functions for an effective 
Hamiltonian. In contrast with the results for BaTiO$_3$ and 
KNbO$_3$, a significant involvement of the Pb atom in the lattice 
instability is found.  Monte Carlo simulations for this Hamiltonian
show a first-order cubic-tetragonal transition at 660 K. The resulting 
temperature dependence of spontaneous polarization, $c/a$ ratio, and 
unit-cell volume near the transition are in good agreement with
experiment. Both coupling with strain and fluctuations are necessary
to produce the first-order character of this transition \cite{PbTiO3}.

\begin{figure*}[p]
\centerline{\psfig{figure=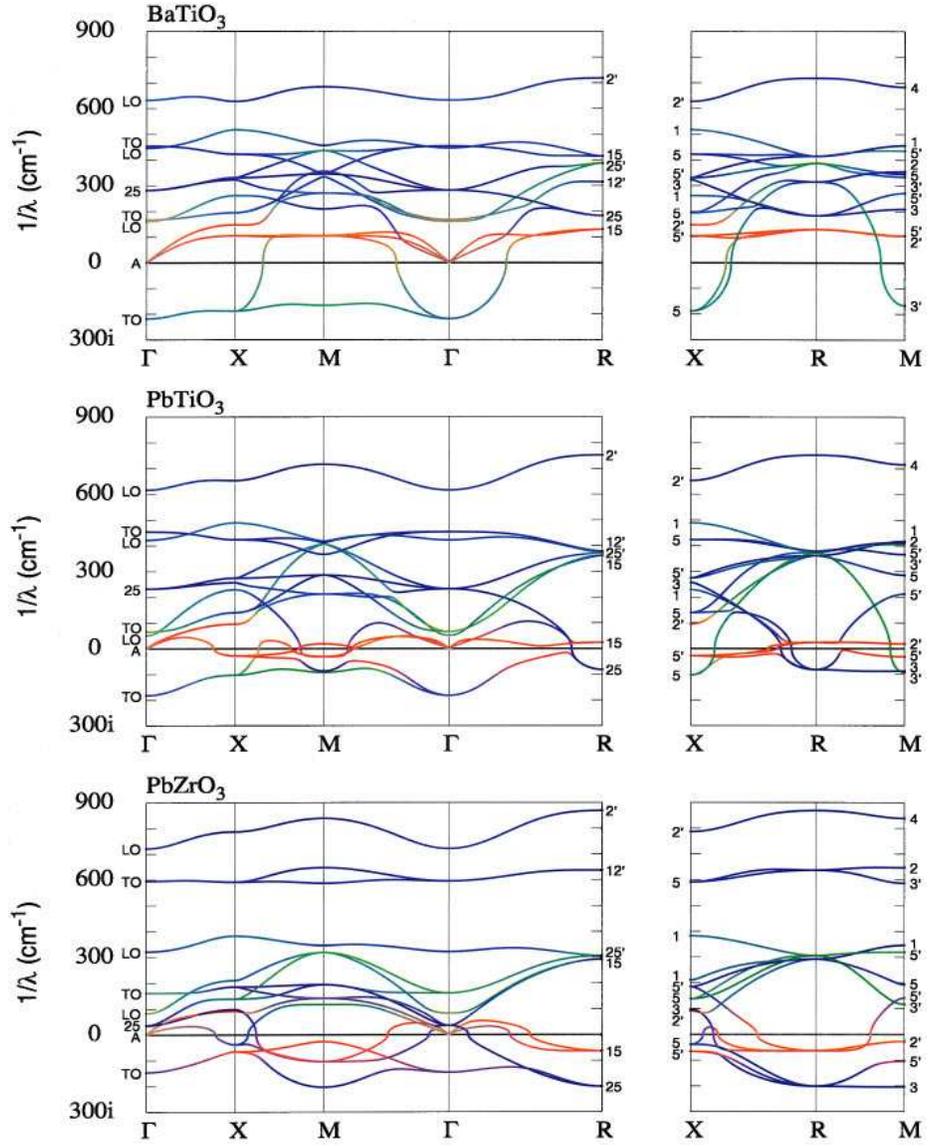,width=15.0truecm}}
\caption{(Color) Calculated phonon dispersion relations of BaTiO$_3$,
PbTiO$_3$, and PbZrO$_3$ along various high symmetry lines in the simple
cubic BZ. A color has been assigned to each point based on
the contribution of each kind of atom to the associated dynamical matrix
eigenvector (red for the A atom, green for the Batom, and blue for the
oxygen). (Reproduced from \protect\citeasnoun{BaPbTiZrO3}.)}
\label{fig10}
\end{figure*}

The full phonon dispersion relations of PbTiO$_3$ and of PbZrO$_3$
in the cubic perovskite structure were computed and compared 
with previous results for Barium Titanate in \citeasnoun{BaPbTiZrO3}.
The comparison (see Fig.\ \ref{fig10})
shows that the change of a single constituent has a 
deep effect on the character and dispersion of unstable modes, with 
significant implications for the nature of the phase transitions 
and the dielectric and  piezoelectric responses of the compounds. 
The unstable localized ferroelectric mode of PbTiO$_3$ 
has a much weaker dispersion with respect to BaTiO$_3$. As
a consequence the ferroelectric distortion is almost isotropic in
real space. Furthermore, there is an anti-ferrodistortive instability at
the R point, not present in BaTiO$_3$ or KNbO$_3$.
In PbZrO$_3$ the ground state is anti-ferroelectric and is obtained
by freezing mainly modes at R and $\Sigma$. The phonon dispersions
show therefore even more pronounced instabilities.
The unstable branches are dominated by Pb and O displacements.
Examination of the interatomic force constants in real space for
the three structures PbTiO$_3$, PbZrO$_3$ and BaTiO$_3$ shows 
that while most are very similar, it is the difference in a 
few key interactions which produces the observed changes in the 
phonon dispersions. This suggests the possibility of using the
transferability of force constants to predict the lattice
dynamics of perovskite solid solutions.

\subsubsection{Pressure-induced phase transitions}

The subject of pressure-induced phase transitions has become
increasingly important since the invention of the diamond
anvil cell. Beyond a fundamental interest, the behavior of 
matter at very high pressure (such as those found in the 
interiors of the earth and of other planets) is relevant in
geology and astronomy.

Contrary to what happens in temperature-induced phase transitions,
anharmonicity does not necessarily play an important role in
pressure-induced phase transitions. One or more harmonic 
frequencies may become soft as a consequence of the changes 
in volume and in the atomic positions caused by applied pressure. 
It is therefore important to determine if phonon instabilities 
occur and if they occur at lower or higher pressure
with respect to other possible instabilities.

\paragraph{Cesium Halides}

Cesium Halides---CsI, CsBr, CsCl---crystallize in the cubic B2
structure at low 
pressure. Under high pressure, CsI makes a continuous transition 
to a lower-symmetry phase, whose onset is as low as 
$P\simeq 15$ GPa. The lower-symmetry phase was originally thought
to be tetragonal, but it was later identified as an orthorhombic
phase, approaching an hexagonal closed-packed
structure with increasing pressure \cite{CsIexp1,CsIexp2}. CsBr has
also a phase transition around $P=53$ GPa, while it is not clear if
such a transition
is present in CsCl.

\begin{figure}
\centerline{\psfig{figure=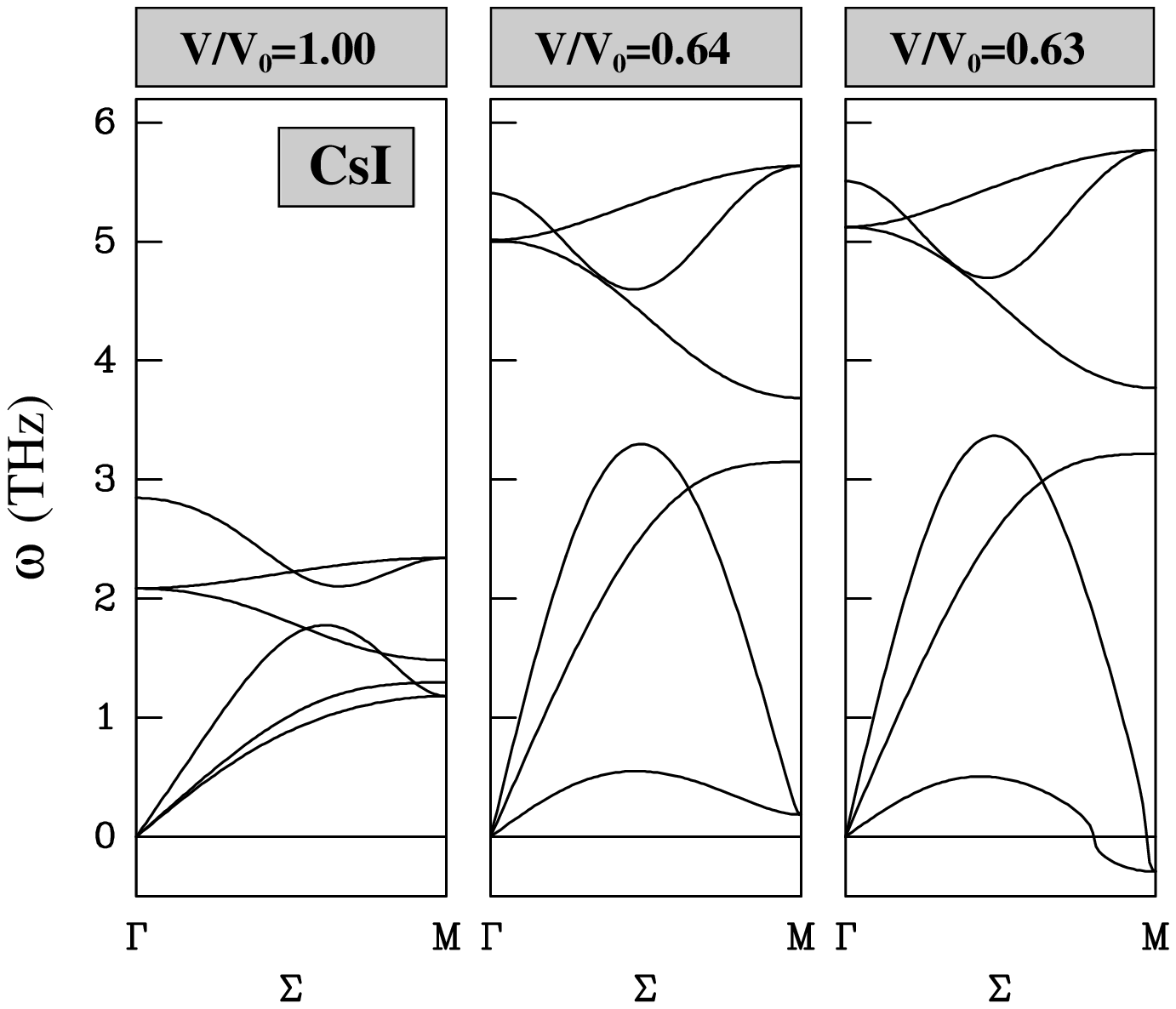,width=8.5truecm}}
\caption{Phonon dispersion relation along the $\Sigma$ (110) for CsI at
equilibrium volume and just above and below the softening pressure of the
$M^-_5$ acoustic mode. {\em Negative} frequencies actually mean
{\em imaginary} ({\it i.e.} negative squared frequencies). 
(Reproduced from \protect\citeasnoun{CsI}.)}
\label{fig11}
\end{figure}

A detailed study of the phonon spectra and of the elastic stability
of CsI (see Fig.\ \ref{fig11})
reveals that an elastic instability leading to a cubic to 
tetragonal transition is in competition with a phonon instability 
of zone-boundary $M^-_5$ modes \cite{CsI}.
In the framework of Landau theory, a phenomenological model 
for the free energy around the transition can be set up using 
the results from first-principles calculations. Group theory allows
one to restrict the search for minimum-energy structures to those 
whose symmetry group is a subgroup of the group of the
undistorted structure.
The resulting model is still quite complex: the order parameter 
(the amplitude of the soft phonons) is six-dimensional.
The coupling between phonons and strain plays a crucial role 
in favoring in CsI the transition to an orthorhombic structure,
driven by a soft phonon, with respect to the tetragonal one. 
In CsBr, instead, the elastic instability leading to the cubic 
to tetragonal transition may occur before the phonon instability.
Finally, in CsCl no softness of the zone-boundary phonons and a 
very weak tendency towards the elastic instability is observed
\cite{CsI}. 

\paragraph{Cesium Hydride}

Like most alkali hydrides, CsH crystallizes in the rocksalt (cubic B1) 
structure and undergoes a transition to the CsCl (cubic B2) structure 
under moderate pressure. A second transition from the B2 structure to 
a new orthorhombic phase (assigned to a CrB structure, space group 
$D^{17}_{2h}$) has been observed in CsH at $P\sim 17$ GPa \cite{CsHexp}.
In \citeasnoun{CsH} it is shown that this first-order transition 
is intimately related to a displacive second-order transition 
(driven by a soft phonon at the $M$ point of the BZ,
$M_2^-$) which would occur upon application of a shear 
strain to the (110) planes.

\paragraph{Silicon Dioxide}

\begin{figure}
\centerline{\psfig{figure=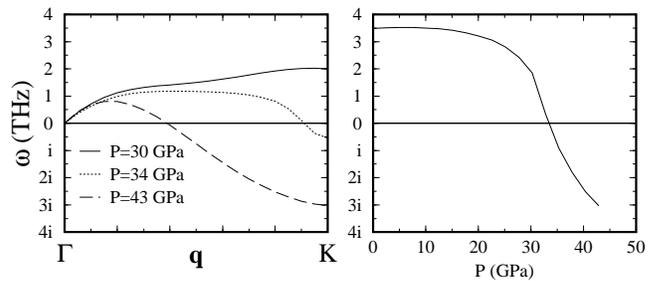,width=8.5truecm}}
\caption{Left panel: $\alpha$-quartz phonon spectra along the $\Gamma-L$
line for the soft mode at P=30 GPa (solid), P=34 GPa (dotted), P=43 GPa
(dashed). Right panel: dependence of the soft-mode frequency at the $K$
point upon pressure. (Reproduced from \protect\citeasnoun{amorph}.)}
\label{fig12}
\end{figure}

Silicon Dioxide exists in a large number of different phases,
both crystalline and amorphous. A surprising and still not fully understood
phenomenon (observed in many other materials as well)
is pressure-induced amorphization, taking place for SiO$_2$
at room temperature with an onset around 15-25 GPa 
\cite{amorph1,amorph2}. DFPT calculations (Fig.\ \ref{fig12})
show that a zone-boundary phonon, 
at the $K$ point of the BZ, becomes soft at 
$P\sim 30$ GPa \cite{amorph}. The relevance of this fact
to pressure-induced amorphization is still
unclear. \citeasnoun{amorph} suggested that the extreme flatness of
the acoustic phonon band whose edge goes soft may be related to the
strange behavior of amorphization.
The iso-structural compound AlPO$_4$ berlinite, which undergoes
a similar pressure-induced amorphization, has also a soft
phonon at the $K$ point of the BZ \cite{belinite}, and
displays a similar---although less pronounced---flattening of the
acoustic band just before amorphization.

The stability of stishovite (tetragonal rutile structure)
under high pressure was studied in \citeasnoun{stishoP}. 
Both a ferro-elastic instability to the orthorhombic CaCl$_2$ 
phase and a softening of the $B_{1g}$ phonon mode at $\Gamma$ 
were found. The former precedes the latter, at $P=64$ GPa, 
thus leading to a second-order transition to the CaCl$_2$ structure.

\paragraph{Semiconductors}

In $A^NB^{8-N}$ octet semiconductors, one would expect a sequence 
of transitions with increasing pressure into the structures of
more ionic binary compounds: 
zincblende$\rightarrow$NaCl$\rightarrow\beta-$tin. 
Earlier experimental and theoretical data supported this picture. 
However newer and 
more accurate angle-dispersive x-ray measurements have revealed
that the high-pressure structures of most III-V semiconductors 
are more complex than expected on the basis of this simple picture. 
The reason is that the NaCl and $\beta-$tin structures
may become unstable with respect to lattice instabilities
(soft phonons). In \citeasnoun{Zunger1}
the systematic absence of the NaCl phase for all covalent 
compounds below a critical ionicity value is explained 
in terms of the instability of a zone-boundary transverse acoustic
(TA) phonon
at the $X$ point, leading to a $Cmcm$ phase. The $\beta-$tin structure
turns out to be unstable for all but the most covalent compound
due to the presence of a phonon anomaly along the $c-$axis
direction in the LO branch \cite{Zunger1}.
A similar approach \cite{Zunger2} shows that in most III-V 
semiconductors the CsCl structure, that was believed to
follow the $\beta-$tin structure at very high pressure, 
is actually unstable for GaP, GaAs, InP, and InAs, due to
the softening of TA phonons at the $M$ point. A Landau 
analysis of the phase transition leads to two candidate
high-pressure structures, respectively the InBi and the AuCd 
structure.

The softening of zone boundary TA modes in Germanium was
studied both experimentally with inelastic neutron scattering
and theoretically with DFPT calculations in \citeasnoun{Gesoft}.
The softening is prevented by an unrelated first order transition 
to the $\beta-$tin structure occurring at $P=9.7$ GPa.

\paragraph{Solid Hydrogen}

The nature of the high-pressure phases of solid hydrogen is a fascinating
subject. In the gas and diluted solid phases, quantum effects are
very important and the homo-polar $H_2$ molecules, interacting via
electric quadrupole-quadrupole interactions, freely rotate even at
low temperature. Upon compression solid Hydrogen undergoes a first
transition to a broken-symmetry phase (phase II) where the molecule
rotation is frozen. At $P\sim 150$ GPa Hydrogen undergoes a second transition
from phase II to a phase III, not clearly identified, experimentally
characterized by a strong increase in IR activity in the vibronic range.
A theoretical study combining constant-pressure ab initio molecular
dynamics and density-functional perturbation calculations \cite{koh3}
addressed this transition and found that phase II (identified as a
quadrupolar $Pca2_1$ hcp phase) becomes unstable due to a zone boundary
soft phonon. The resulting candidate structures for phase III have larger
unit cells, accounting for the many libronic peaks experimentally observed,
and much larger effective charges, leading to a strongly increased IR
activity, as observed experimentally.

\paragraph{Metals}

The lattice dynamics of the bcc and fcc phases of W under pressure
was studied in \citeasnoun{WunderP}. The bcc phase is stable
at zero pressure. Under applied pressure, the bcc phase develops
phonon softening anomalies for $P \sim 1200$ GPa. At this pressure,
however, the fcc and hcp phases have a lower enthalpy than the
bcc phase. The fcc phase of W has elastic instabilities at zero
pressure that stabilize with increasing pressure before its 
enthalpy becomes lower than that of the bcc phase.

Phonon instabilities in bcc Sc, Ti, La, Hf were studied in
\citeasnoun{bccmetals}. These metals exhibit hexagonal (Sc, Ti, Hf)
or double hexagonal (La) closed-packed structure at low temperature,
while at high temperature they become bcc.

\subsubsection{Other phase transitions}

In the substitutionally disordered narrow-gap semiconductor 
Pb$_{1-x}$Ge$_x$Te, a finite-temperature cubic-to-rhombohedral 
transition appears above a critical concentration $x\sim 0.005$.
A hypothetical ordered cubic Pb$_3$GeTe$_4$ super-cell is studied 
as a model for such alloy \cite{Pb3GeTe4}.
Unstable lattice modes are found, dominated by off-centering of the 
Ge ions coupled with displacements of their neighboring Te ions. 
A model Hamiltonian for this system (using the lattice Wannier 
function formalism) is constructed and studied via Monte Carlo
Simulations. A transition temperature of $\sim 620$ K 
is found for the cubic model, compared to the experimental 
value of $\sim 350$ K for the alloy.

\subsection{Thermal properties of crystals and surfaces}
\label{sec:QHA}

The knowledge of the entire phonon spectrum granted by DFPT enables
the calculation of several important thermodynamical quantities and of
the relative stability of different phases as functions of
temperature. The first calculation of a thermal property 
(expansion coefficients in Si) using DFPT dates back to 1989 
\cite{firstSith}.

The thermodynamical properties of a system are determined by the
appropriate thermodynamical potential relevant to the given
ensemble. In the ensemble where the sample volume and temperature are
independent variables, the relevant potential is the Helmholtz free
energy, $F=E-TS$. For a solid in the adiabatic approximation
the free energy can be
written as the sum of an electronic and a vibrational term. The
electronic entropy contribution is easily evaluated in metals,
although usually neglected, whereas it is
totally negligible for insulators: $F_{el} \simeq E_{el}$. 
The key quantity to calculate in order to have access to the
thermal properties and to the phase stability is the vibrational 
free energy $F_{ph}$.

Far from the melting point, the vibrational free energy, $F_{ph}$, can
be conveniently calculated within the {\em quasi-harmonic
approximation} (QHA).  
This consists in calculating $F_{ph}$ in the harmonic 
approximation, retaining only the implicit volume dependence 
through the frequencies:
\begin{equation}
F_{ph}(T,V)=-k_B T\log
   \left(\mbox{Tr }e^{-{\cal H}_{ph}(V)/k_BT}\right),
\end{equation}
where ${\cal H}_{ph}(V)$ is the phonon hamiltonian at a given volume.
In terms of the phonon spectra, $F_{ph}$ can be written as
\begin{equation}
F_{ph}(T,V) = -k_B T\sum_{i,\k} \log
             \left(
                   \sum_n e^{-(n+1/2)\hbar\omega_{i\k}(V)/k_B T} 
             \right).
\end{equation}
Once the sum over occupation numbers $n$ is performed, one gets 
the final formula:
\begin{equation}
F_{ph}(T,V) = k_B T\sum_{i,\k} \log
             \left[
  	           2 \sinh\left(\hbar\omega_{i\k}(V)/2k_B T\right) 
             \right].
\end{equation}
In practical calculations, the force constants are calculated
at a few volumes and interpolated in between to get the volume dependence.
Once the phonon spectra on the entire BZ is available, the 
calculation of $F_{ph}$ reduces to a straightforward integration 
over the BZ.

\begin{figure}[t]
\centerline{\psfig{figure=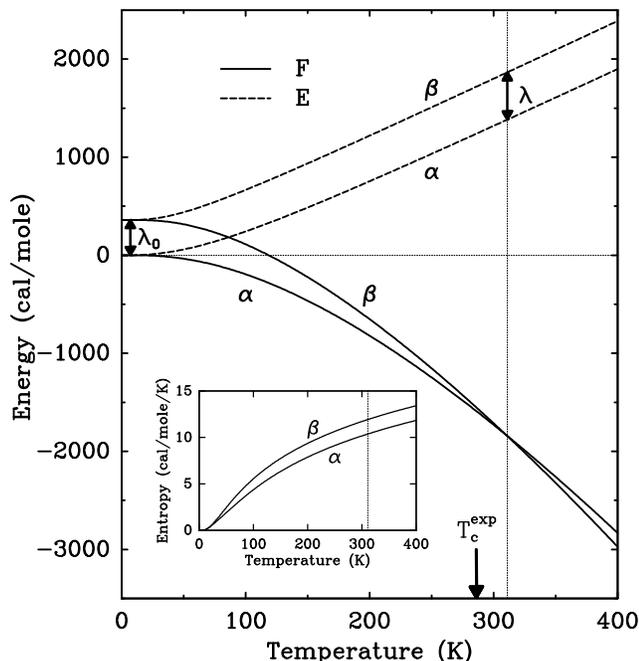,width=8.5truecm}}
\caption{
Lower panel: Zero-pressure free-energy (solid lines) and
internal-energy (dashed lines) curves for the $\alpha$ and $\beta$
phases of Tin as functions of temperature. The thin vertical dotted
line indicates the theoretical transition temperature, while the
experimental value for $T_c$ is shown by the arrow. $\lambda_0=359$
cal/mole is the $T=0$ K free-energy difference---including the
zero-point contribution---while $\lambda=482$ cal/mole indicates and
the latent heat adsorbed in the $\alpha\leftrightarrow\beta$
transition. Finally, the inset displays the temperature dependence of
the vibrational entropies of the two phases. (reproduced from 
\protect\citeasnoun{Tin}.)}
\label{fig13}
\end{figure}

The QHA accounts only partially for the
effects of the anharmonicity, through the volume dependence of the
phonon spectra (this is clearly an anharmonic effect: the perfect
harmonic crystal would  have no volume expansion with the temperature)
but it turns out to be a  very good approximation at temperatures not
too close to the melting point.

The quantities that can be calculated in the QHA include equilibrium
lattice parameters and  
elastic constants, specific heat, thermal expansion coefficients, 
as a function of the temperature. Corrections due to quantum 
fluctuations ({\em zero-point motion}) at zero temperature can
be estimated as well. The comparison of the free energies
of different phases (not related by symmetry relations,
unlike those considered in \secref{appl_soft}) yields
the relative stability as a function of the pressure and 
of the temperature.
Most calculations in this field have been performed
on simple systems, but there are a few examples of
applications to surfaces, notably to anomalous thermal
expansion.

\subsubsection{Metals}

The equilibrium lattice parameter and thermal expansion 
coefficients for bcc Li and fcc Al and Na were computed in 
\citeasnoun{Quong_metals}. The relative stability of the various
polymorphs of Li (bcc, fcc, hcp, 9R) was examined in
\citeasnoun{PhasesLi}. It is found that the transformation
from the 9R structure at low temperatures to the bcc phase
upon heating is driven by the large vibrational entropy 
associated with low-energy phonon modes in bcc Li. 
The strength of the electron-phonon interaction in Li 
is calculated and found to be significantly reduced in 
the low-temperature 9R phases as compared to the bcc
phase, consistently with the observed lack of a 
super-conducting transition in Li \cite{PhasesLi}. 

The thermal properties of fcc Ag, plus the Gr\"uneisen parameters,
were calculated in \citeasnoun{Silver}.

The $\alpha-\beta$ phase transition in Tin was studied in
\citeasnoun{Tin}. At $T=0$ K the free energy of the $\beta$ phase 
lies $\sim 359$ cal/mole above that of the $\alpha$ structure.
The narrower frequency range spanned by the vibrational
band in the $\beta$ phase makes its entropy larger at high
temperature. As a consequence, the free energies of the two phases,
shown in Fig.\ \ref{fig13},
equal each other at a temperature of $\sim 38$ C, in close agreement
with the observed transition temperature $T_c=13$ C \cite{Tin}.

\subsubsection{Semiconductors and Insulators}

Calculations were reported of the thermal properties, 
plus the Gr\"uneisen parameters, of Si \cite{Sith};
of the thermal expansion coefficients of Diamond \cite{Diamond};
of the thermal expansion coefficient and specific heat of cubic
SiC in the 3C structure \cite{SiC}.

In Diamond, thermal properties were calculated at high pressure,
up to 1000 GPa \cite{CompressedDiamond}. The P-V-T equation of states
has been calculated from the Helmholtz free energy. The thermal 
expansion coefficient is found to decreases with the increase
of pressure, and at ultrahigh pressure (700  GPa), Diamond exhibits 
a negative thermal expansion coefficient at low temperatures. 

The temperature dependence of the diamond-$\beta$-Tin phase 
transition in Si and Ge, that occurs under pressure at 
$\sim$10 GPa, was calculated in \citeasnoun{SiGeunderP}
using the QHA.

Thermal properties (specific heat, entropy, phonon contribution to 
the free energy, atomic temperature factors) for SiO$_2$ in the
$\alpha-$quartz phase and in the high-pressure stishovite phase
were calculated in \citeasnoun{SiO2th}.

The ability of the QHA to yield
thermodynamical properties of materials over a considerable
pressure-temperature regime was asserted in recent papers
\cite{MgOunderP1,MgOunderP2}. In these works the thermoelastic properties 
of MgO was calculated over a wide range of pressure and temperature.
Thermodynamical potentials and several derived quantities (such as
the temperature dependence of elastic constants at high pressures)
were computed and successfully compared with experimental 
data \cite{MgOunderP1,MgOunderP2}.

\subsubsection{Surfaces}
\label{sec:appl-thermal}

The thermal expansion of surfaces was calculated for
Ag(111) \cite{Ag111} and for Be(0001) \cite{Be0001}.

In Ag(111) the top-layer relaxation changes from an inward 
contraction (-0.8\%) to an outward expansion (+6.3\%) as the
temperature increases from T = 0 K to 1150 K, in agreement 
with experimental findings. The calculated surface phonon 
dispersion curves at room temperature are in good agreement 
with Helium-scattering measurements \cite{Ag111}.

At the Be(0001) surface, an anomalously large thermal expansion 
was recently observed in low-energy electron diffraction experiments.
The calculations were tested in bulk Be, where they describe very well 
the thermal expansion, and checked against first-principles 
molecular dynamics simulations for the surface. The large
thermal expansion is not found. The discrepancy could be 
explained assuming that the actual surface is less ideal
than assumed \cite{Be0001}.

\subsubsection{Alloys}

The vibrational contribution to the free energy is known to
affect the phase stability of alloys. The importance of such
effect was examined in two metallic alloys, the Cu-Au 
system \cite{CuAu} and the Re-W systems \cite{ReW}.

Cu-Au systems were studied using a combination of DFPT
and of cluster expansion methods. The vibrational free energy
of the alloy is calculated by a cluster expansion over a small 
set of representative ordered structures having small super-cells,
quite in the same way as the configurational free energy is
calculated. Anharmonic effects are taken into account through 
the QHA. The results indicate that the 
vibrational free energy contributes significantly to the phase 
stabilities and thermodynamic functions of the CuAu system.
In particular it tends to stabilize the compounds and alloys 
with respect to the phase-separated state, and lowers the 
order-disorder transition temperatures. It is found out
that the vibrational free energy, not the vibrational entropy,
is the relevant quantity, due to the larger thermal expansion 
coefficients of the alloy with respect to the ordered ground 
states \cite{CuAu}.

A somewhat similar approach was applied to the study of the
dynamical and thermodynamical stability of the bcc and fcc disordered 
Re$_x$W$_{1-x}$ system. As a byproduct, the phonon dispersion curves
for fcc and bcc Re were calculated. Fcc Re is dynamically 
stable but has pronounced phonon anomalies; bcc Re exhibits phonon 
instabilities in large parts of the BZ, similar to those 
found in fcc W. Due to the instabilities in bcc Re and fcc W 
the vibrational entropy, and therefore the free energy, is undefined. 
The problem is circumvented by using the virtual crystal approximation
to calculate the phonon dispersions of disordered Re$_x$W$_{1-x}$
and by applying a concentration-dependent nonlinear interpolation 
to the force constants. A region where the bcc phase would become 
thermodynamically unstable towards a phase decomposition into 
disordered bcc and fcc phases is found \cite{ReW}.

\subsection{Anharmonic effects}
\label{sec:anharmonic}
\begin{figure}
\centerline{\psfig{figure=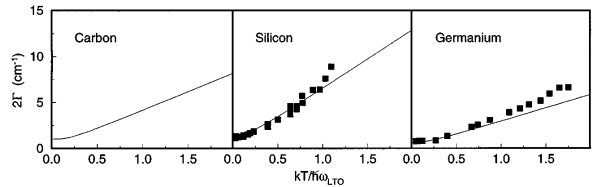,width=8.5truecm}}
\caption{temperature dependence of the full width at half maximum,
$2\Gamma$, of the zone-center optical phonon in Diamond, Si, and Ge.
Solid lines are the result
of DFPT calculation; squares represents experimental data. 
(Reproduced from  \protect\citeasnoun{anharm1}.)}
\label{fig14}
\end{figure}

Third-order energy derivatives can be calculated
directly from the linear response using the $2n+1$ 
theorem (See Sec.~\secref{higher}).
The first calculation of anharmonic force constants
using DFPT and the $2n+1$ theorem was performed for
the lifetimes of zone-center optical phonons in Diamond, Si, and Ge 
\cite{anharm1} due to its anharmonic decay into two 
phonons of lower frequency. This is the main contribution to
the line-width  of such Raman-active modes, once isotopic and 
other inhomogeneous broadening contributions are subtracted.
The temperature and pressure dependences of anharmonic 
lifetimes were calculated as well (see Fig.\ \ref{fig14}
for the former). The results are in good agreement with
experimental data. 
The microscopic mechanisms responsible for the decay 
are found to be different for different materials and
to depend sensitively on the applied pressure \cite{anharm1}.

Subsequent work was performed in zinc-blende semiconductors 
GaAs, AlAs, GaP, InP \cite{anharm2} and in SiC \cite{anharm3}.
In III-V semiconductors the line-width of Raman-active modes,
both transverse and longitudinal, and their temperature 
dependence were computed. For longitudinal phonons a 
simple approximation consisting in neglecting the effect
of the macroscopic electric field in the anharmonic terms
yields good results \cite{anharm2}. 
In 3C (cubic zincblende) SiC the pressure dependence, 
up to 35 GPa, of the line-widths of the LO and TO modes 
at the BZ center was calculated. 
An anomalous behavior is found:
the line-width of the transverse mode changes very little 
with pressure, while the longitudinal mode shows a monotonic 
increase up to 26 GPa, decreasing abruptly above 
this pressure. The results are in good agreement with
new experimental data, up to 15 GPa \cite{anharm3}.

A different approach to anharmonicity combines DFPT to
calculate harmonic force constants and frozen phonons to 
calculate higher-order force constants through numerical 
differentiation. This  approach is less elegant and more
sensitive to numerical accuracy than the use of the $2n+1$ 
theorem. However it allows to calculate quartic
anharmonic terms and
those third-order energy derivatives whose calculation 
with the $2n+1$ theorem is hindered by technical difficulties, 
such as Raman cross sections in the nonresonant 
limit \cite{C60}.
Such approach was first used to calculate the contribution of 
quantum fluctuations (zero-point motion) to the bulk modulus of Diamond,
to the dielectric constant of Diamond, Si, Ge, and to calculate the 
temperature dependence of these quantities \cite{quantum}. 
More recently, the anharmonic shift of the Raman frequency of 
Diamond and Si, in which both cubic and quartic anharmonic terms are 
equally important, have been calculated \cite{anharm4}. 
The temperature dependence of the Raman frequency and the 
contribution of zero-point motion
are calculated as well as the Raman line-width. The same quantities 
were calculated in Ge using the $2n+1$ theorem for the cubic 
terms and assuming that the quartic anharmonic force constants may 
be approximated by those of silicon \cite{lineshift}.

\subsection{Isotopic broadening of Raman lines}

In the harmonic approximation, the effect of isotopic disorder or of
isotopic substitution is only reflected in the change of masses in
the dynamical matrix. The frequency shift caused by isotopic 
substitution depends on phonon displacements pattern and may be
used as a probe of the latter. DFPT results were applied
to the interpretation of Raman experiments in isotopically
substituted C$_{60}$ \cite{isoC60}, in GaN\cite{GaN_Raman},
and in SiC \cite{isoSiC}.

Isotopic disorder is also responsible for a broadening and further
shift of the Raman lines, beyond the virtual-crystal approximation
which replaces the masses of each individual atom with their
composition-averaged value. Disorder effects are usually treated
within some kind of mean-field approximation (such as the Coherent
Potential Approximation, or CPA). In \citeasnoun{IsoRaman} both 
the CPA and a super-cell approach were used to calculate the
broadening of Raman peaks in Si, Ge, and $\alpha$-Tin, using IFC's
calculated from first principles.

In a recent work \cite{isobroad} a new method to study the effects 
of isotopic composition on the Raman spectra of crystals beyond
mean-field approximation was presented. The Raman cross section 
is expressed as a diagonal element of the vibrational Green's function, 
which is accurately and efficiently calculated using the recursion
technique. The method was applied to Diamond---where isotopic effects 
dominate over the anharmonic ones---as well as to Ge, where anharmonic
effects are larger. In both case the results are in very good agreement
with experiments \cite{isobroad}.

Other effects of isotopic substitution appear beyond harmonic 
approximation. The molecular volume of crystals depends on 
isotopic masses through the zero-point motion. This is a tiny
but measurable effect that can be calculated using the 
quasi-harmonic approximation (Sec. \secref{QHA}). The dependence 
of the crystal lattice constant on isotopic composition 
was calculated for elemental semiconductors Diamond, Si, and Ge 
\cite{isoSemicon} and for compound zincblende semiconductors
GaAs and ZnSe \cite{isoZnSe}. In the latter case, the temperature 
dependence of the derivatives of the lattice constant with respect
to both the anion and the cation mass was computed, together
with the linear thermal expansion coefficients and mode
Gr\"uneisen parameters. 

The dependence of the Raman line-width in 3C-SiC on both isotopic
composition and temperature for longitudinal and 
transverse modes is calculated in \citeasnoun{isoSiC2}. The 
line-widths exhibit a marked and nontrivial dependence on isotopic 
composition.
\subsection{Vibrational broadening of electronic core levels}

Phonons cause a temperature-dependent broadening of core-level
spectra. The calculation of such broadening requires an 
accurate description of both phonon spectra and of core-excited 
electronic states. A theoretical framework to deal with such 
problem has been provided by \citeasnoun{coreC}, together with 
an application to the $1s$ core exciton of Diamond, for which
a strong vibronic coupling is present. The {\em sudden approximation} 
(optical absorption and emission taking place in much shorter times
than typical phonon response times) is assumed, leading to a 
Franck-Condon picture. The electronic and vibrational degrees
of freedom are consistently treated using DFT. A suitable
pseudo-potential, generated on the excited atomic configuration,
allows to simulate the excited electronic states. Due to the 
large lattice distortion, anharmonicity cannot be neglected. 
In the limit of infinite core-hole lifetime, anharmonic effects 
are included within a self-consistent phonon approach. 
The Stokes shift for the $1s$ exciton level is found to be 
about 3 eV (the harmonic approximation would overestimate it
by about 1.4 eV) and the phonon broadening is about 2 eV
\cite{coreC} .
\section{Conclusions and perspectives} \label{sec:conclusions}

The results reviewed in this paper witness the blossoming of ab-initio
lattice-dynamical calculations in solids, based on density-functional
(perturbation) theory. Our ability to predict from first-principles
phonon-related properties of materials depends both on the accuracy of
the ab-initio calculation of lattice vibrations, and on the quality of
the approximations that are needed to relate these calculations to the
specific property one is interested in (such as, {\em e.g.}, the
electric conductivity or the temperature dependence of the crystal
volume). The accuracy of the calculations can be appraised by
comparing the calculated frequencies with infrared, Raman, or
neutron-diffraction experiments. It is fair to state that lattice
dynamics is the one field of solid-state physics where the accuracy of
ab-initio calculations can compete with absorption or diffraction
spectroscopies. Of course, it would be vain to put much effort in the
calculation of quantities that can be measured with a comparable or
better accuracy. The real value of first-principles
calculations stems from their ability to provide unbiased predictions
for those materials and in those cases which are not easily accessible
to the experiment.

Even though necessarily over-simplified, the physical conditions of
the {\em sample} being studied numerically are under our total control
and can therefore be varied at will. This allows to assess to quality
and validity of models that relate the atomistic and electronic
structure of materials, which is usually unknown, to their
macroscopic, and experimentally accessible, properties. Once the
accuracy of the calculated phonon frequencies has been assessed, the
agreement of the predictions for derived quantities gives indications
on the validity of the approximations used to derive them. In the case
of the dependence of inter-atomic distances upon temperature, for
instance, their comparison with experiment gives indications on the
validity of the quasi-harmonic approximation used to calculate
them. Rather surprisingly, it turns out that in a variety of cases,
this approximation gives precise results up to near the melting
temperature.

We conclude by arguing that the field of lattice-dynamical
calculations based on density-functional perturbation theory is ripe
enough to allow a systematic application of DFPT to systems and
materials of increasing complexity. The availability to a larger
community of scientists of the software necessary to perform such
calculations will make them a routine ingredient of current research,
much in the same way as it has occurred for standard DFT calculations
over the past five years or so. Among the most promising fields of
application, we mention the characterization of materials through the
prediction of the relation existing between their atomistic structure
and experimentally detectable spectroscopic properties; the study of
the structural (in)stability of materials at extreme pressure
conditions; and the prediction of the thermal dependence of different
materials properties using the quasi-harmonic approximation.
\acknowledgments

Finally, we would like to thank all those collaborators and friends of
ours, mainly but not exclusively former SISSA students, who have
greatly contributed to the development of this field and without whom
this review would be much shorter or would never have been coinceived
at all. Among them, we are pleased to mention Marco Buongiorno
Nardelli, Claudia Bungaro, Alberto Debernardi, Michele Lazzeri, Kurt
M\"ader, Pasquale Pavone, Andrea Testa, Raffaele Resta, and Nathalie
Vast.

This work has been partially funded by Italian MURST through a COFIN99
project and {\it Progetti di ricerca di rilevante interesse
nazionale}, and by INFM through {\it Iniziativa Trasversale Calcolo
Parallelo}. 

\appendix

\section{Plane-wave pseudo-potential implementation}

\subsection{Pseudo-potentials}

Norm-Conserving PP's are relatively
smooth functions, whose long-range tail goes like 
$-Z_v e^2/r$ where $Z_v$ is the number of valence electrons.
There is a different PP for each atomic angular momentum $l$:
\begin{equation}
\widehat V = V_{loc}(r) + \sum_l V_l(r) \widehat P_l,
\end{equation}
where $\widehat P_l = \sum_{m=-l}^l|lm\rangle\langle lm| $
is the projection operator on states of angular momentum
$l$, $|lm\rangle$ being the angular state with $(l,m)$ angular
quantum numbers. Usually $ V_{loc}(r)\simeq -Z_v e^2/r$ for 
large $r$ so that the $V_l(r)$ are short-ranged.
PP's in this form are called {\em semi-local}:
\begin{eqnarray}
\label{eq:semilocal}
V(\r,\r') & =& V_{loc}(r)\delta(\r-\r') \\ 
          &\mbox{} &\qquad + \sum_{lm} Y_{lm}(\r) V_l(r) 
                   \delta(r-r') Y_{lm}^*(\r'),\nonumber
\end{eqnarray}
where the $Y_{lm}(\r)= \langle \r | lm \rangle $ are the usual spherical
harmonics.

PP's are usually recast into the \citeasnoun{KB} {\em separable} form. 
Each PP is projected
on to the atomic reference pseudo-wave-functions $\phi^{ps}_l$:
\begin{equation}
\label{eq:separable}
\widehat V = V_{loc} + V_L + \sum_l { |V'_l \phi_l^{ps}\rangle 
\langle V'_l \phi_l^{ps}| \over 
\langle\phi_l^{ps}| V'_l|\phi_l^{ps}\rangle }
\end{equation}
where $V_L(r)$ is an arbitrary function,
$V'_l(r) = V_l(r) - V_L(r)$. By construction
the original PP and the projected PP $\widehat V$ 
have the same eigenvalues and eigenvectors
on the reference states $\phi^{ps}_l$.
The separable form may badly fail in some cases due to the 
appearance of spurious {\em ghost} states. Some recipes 
have been devised to avoid such problem \cite{GonzePP1,GonzePP2}.

\subsection{Matrix elements}

A PW basis set is defined as 
\begin{equation}
\langle\r|\k+\G\rangle={1\over V} e^{i(\k+\G)\cdot\r}, 
\quad {\hbar^2\over 2m} |\k+\G|^2\le E_{cut},
\end{equation}
where the \G's are reciprocal lattice vectors,
\k\ is the wave-vector in the BZ, 
$V$ is the crystal volume, and $ E_{cut}$ is the
cutoff on the kinetic energy of the PW.

In the PW-PP implementation, Sec.\secref{implem_pw}, we assume that the
electron-ion potential $V_{ion}(\r,\r')$ is written as 
\begin{equation}
V_{ion}(\r,\r') = \sum_{ls} v_s(\r-\R_l-\boldtau_s,\r'-\R_l-\boldtau_s)
\end{equation}
where $v_s$ is the PP for the $s$-th atomic species, whose general
form is
\begin{equation}
 v_{s}(\r,\r') = v_{s,loc}(r)\delta(\r-\r') + \sum_l v_{s,l}(\r,\r').
\end{equation}

The plane-wave matrix
elements of the above operator are given by:
\begin{eqnarray}
\langle\k+\G | v_s | \k+\G' \rangle& = &\widetilde v_{s,loc}(\G-\G')\\
 & \mbox{} &  + \sum_l\widetilde v_{s,l}(\k+\G,\k+\G'),\nonumber
 \end{eqnarray}
where:
\begin{equation}
\widetilde v_{s,loc}(\G)  = {1 \over \Omega} \int v_{s,loc}(r)
e^{-i{\bf G\cdot\r}} d\r,
\label{eq:ft}
\end{equation}
and
\begin{equation}
\widetilde v_{s,l}(\k_1,\k_2) = {1 \over \Omega} \int e^{-i\k_1\cdot\r}
v_{s,l}(\r,\r') e^{i\k_2\cdot\r'} d\r d\r'.
\end{equation}
For PP's in the semi-local form, Eq.\ \eqref{semilocal}:
\begin{equation}
v_{s,l}(\r,\r')=\widehat P_l f_{s,l}(r),
\end{equation}
and
\begin{eqnarray}
\widetilde v_{s,l}(\k_1,\k_2) & = &
{4\pi \over \Omega} (2l+1) P_l(\hat\k_1\cdot\hat\k_2) \\
& & \quad \times \int_0^\infty r^2 j_l(k_1 r)j_l(k_2 r) f_{s,l}(r) dr,
\nonumber
\end{eqnarray}
where the $j_l$'s are spherical Bessel functions, the $P_l$'s are
Legendre polynomials of degree $l$.

For PP's in the separable form, Eq.\ \eqref{separable}:
\begin{equation}
v_{s,l}(\r,\r')=c_{s,l} \beta^*_{s,l}(\r)\beta_{s,l}(\r'),
\end{equation}
and
\begin{equation}
\widetilde v_{s,l}(\k_1,\k_2)=
c_{s,l} \widetilde \beta^*_{s,l}(\k_1) \widetilde \beta_{s,l}(\k_2)
\end{equation}
where the $\widetilde \beta(\k)$'s are the Fourier transform of 
$\beta(\r)$, as in Eq.\ \eqref{ft}.

The matrix elements between PW's of the derivatives of the
ionic potential, Eq.\ \eqref{dVion}, are
\begin{eqnarray}
&&\left\langle {\bf k+q+G} \left | {\partial V_{ion}\over \partial u^\alpha
_s(\q)} \right | {\bf k+G'} \right \rangle = \qquad\qquad \\
&&\qquad\mbox{} - i(q_\alpha + G_\alpha -G_\alpha') 
e^{-i(\q+\G-\G')\cdot\boldtau_s }  \nonumber \\
&&\qquad\qquad\times\mbox{} \biggl ( \widetilde v_{s,loc}({\bf q+G-G'})  
\nonumber \\ &&\qquad\qquad\qquad\mbox{}
         + \sum_l  \widetilde v_{s,l}({\bf k+q+G,k+G'}) 
\biggr ). \nonumber
\end{eqnarray}
The screening contribution to  $ \partial V_{\SCF} / \partial u^\alpha_s(\q)$,
which is a local potential in DFT, can be advantageously evaluated in real
space and transformed back to reciprocal space by FFT techniques.

The matrix elements of the commutator $[H_\SCF,\r]$ between PW's are:
\begin{eqnarray}
& & \langle \k_1| [H_\SCF,\r] |\k_2\rangle 
    = {\hbar^2\over m}\k_1\delta_{\k_1\k_2} \\
& & \qquad\mbox{} - i \sum_{s,l} e^{-i(\k_1-\k_2)\cdot\boldtau_s }
 \left(  { \partial \over \partial \k_1 }
       + { \partial \over \partial \k_2 }\right )
   \widetilde v_{s,l}(\k_1,\k_2) ,\nonumber
\end{eqnarray}
where $\delta_{\k_1\k_2}=1$ if $\k_1=\k_2$, $\delta_{\k_1\k_2}=0$
otherwise.

The matrix elements of the second derivative of the 
electron-ion interaction potential appearing in the
second term of Eq.\ \eqref{forceconstants} are given by
\begin{eqnarray}
&& \left \langle {\bf k+G} \left|
{\partial^2 V_{ion}\over\partial u^\alpha_s(\q=0)\partial u^\beta_s(\q=0)}
\right| \k+\G'\right\rangle = \\
& & \qquad\mbox{} - (G_\alpha - G_\alpha')(G_\beta - G_\beta')
 e^{-i(\G-\G')\cdot\boldtau_s } \nonumber \\
& & \qquad  \times \biggl ( \widetilde v_s({\bf G-G'}) + \sum_l
\widetilde v_{s,l}({\bf k+G,k+G'}) \biggr ) .
\nonumber
\end{eqnarray} 

\section{Force Constants, ionic term}

The total energy of a crystal contains a divergent ion-ion 
energy term, Eq.\ \eqref{ion-ion}, that combines with the
divergent $\G=0$ terms in the electron-ion energy (the 
second term in Eq.\ \eqref{toten-eigv}) to yield the 
{\em Ewald} term $E_{Ew}$:
\begin{eqnarray}
E_{Ew} & = & {4\pi N_c \over \Omega} {e^2 \over 2}
 \sum_{\G\neq 0} {e^{-G^2/4\eta} \over G^2}
 \left| \sum_s Z_s e^{i\G\cdot\boldtau_s} \right|^2
  \\ & + &
    {N_c e^2\over 2} \sum_{s,t} \sum_\R
{ Z_s Z_t \mbox{erfc}(\sqrt{\eta}|\boldtau_s-\boldtau_t-\R|)
 \over |\boldtau_s-\boldtau_t-\R|}\nonumber
\\ & - &
  {\pi N_c e^2\over \Omega\eta} \left(\sum_s Z_s\right)^2
 - N_ce^2 \sqrt{\eta\over\pi} \sum_s Z_s^2, \nonumber
\end{eqnarray}
where $\mbox{erfc}(x)=1-\mbox{erf}(x)$ ($\mbox{erf}$ is the 
error function); the sum over \R-space excludes
$\boldtau_s-\boldtau_t-\R=0$; $N_c$ is the number of
unit cells in the crystal; $Z_s$ indicates the bare ionic 
charges for the $s$-th atom (pseudo-charges in a PP-PW framework);
$\eta$ is an arbitrary parameter, whose value ensures good 
convergence of both sums over \G- and \R-space.

The second derivative of the Ewald term yields the ionic
contribution to the force constants:
\begin{eqnarray}
\label{eq:Ewald}
^{ion}\negthinspace\widetilde C^{\alpha\beta}_{st}(\q) & = &
  {4\pi e^2\over \Omega}
 \sum_{\G}{e^{-(\q+\G)^2/4\eta} \over (\q+\G)^2}
 Z_s Z_t  \nonumber\\
 & & \mbox{} \times e^{i(\q+\G)\cdot(\boldtau_s-\boldtau_t)}
 (q_\alpha+G_\alpha) (q_\beta + G_\beta )
 \nonumber\\& - &
  {2\pi e^2\over \Omega} \sum_{\G \neq 0 }{e^{-G^2/4\eta} \over G^2}
  \nonumber \\&\mbox{} &  \times \biggl[
  Z_s \sum_l Z_l e^{i\G\cdot(\boldtau_s-\boldtau_l)}
 G_\alpha G_\beta + c.c.
 \biggr] \delta_{st}
 \nonumber\\& + &
  e^2 \sum_\R  Z_s Z_t e^{i\q\cdot\R} \nonumber \\
      &\mbox{}& \times \biggl[ \delta_{\alpha\beta} f_2(x)
      + f_1(x)x_\alpha x_\beta \biggr]_{{\bf x}=\boldtau_s-\boldtau_t-\R}
\nonumber
\\& - &
  e^2 \delta_{st} \sum_\R \sum_l Z_s Z_l \nonumber \\ 
      &\mbox{}& \times \biggl[ \delta_{\alpha\beta} f_2(x)
      + f_1(x) x_\alpha x_\beta \biggr]_{{\bf x}=\boldtau_s-\boldtau_l-\R}
\end{eqnarray}
where the sum over \G-space excludes $\q+\G=0$, the sums \R-space 
exclude $\boldtau_s-\boldtau_t-\R=0$, and the functions $f_1$ and
$f_2$ are defined as follows:
\begin{equation}
 f_1(r)= { 3\mbox{erfc}(\sqrt{\eta}r) 
           + 2\sqrt{\eta\over\pi} r (3+2\eta r^2)e^{-\eta r^2}
          \over r^5 }
\end{equation}
\begin{equation}
 f_2(r)= { - \mbox{erfc}(\sqrt{\eta}r)
          - 2\sqrt{\eta\over\pi}re^{-\eta r^2} \over r^3 }
\end{equation}

\end{document}